\documentclass[12pt,a4paper]{article}
\pdfoutput=1

\usepackage{a4wide}
\usepackage{array}
\usepackage{epsfig}
\usepackage{amssymb}
\usepackage{graphics,graphpap}
\usepackage{amsmath}
\usepackage{wasysym}
\usepackage{slashed}
\usepackage{mciteplus}

\usepackage{placeins}

\newcommand{\lnp}{\big<\!\!:\!}
\newcommand{\rnp}{\!:\!\!\big>}
\newcommand{\qq}{\lnp\bar qq\rnp^{(0)}}
\newcommand{\Slash}{\not\!}
\newcommand{\qFq}{\lnp g_{s}\bar q\sigma Fq\rnp^{(0)}}
\newcommand{\smvs}{\vbox{\vskip 8mm}}
\newcommand{\be}{\begin{equation}}
\newcommand{\ee}{\end{equation}}
\newcommand{\bea}{\begin{eqnarray}}
\newcommand{\eea}{\end{eqnarray}}
\newcommand{\nn}{\nonumber}
\newcommand{\bi}{\begin{itemize}}
\newcommand{\ei}{\end{itemize}}

\usepackage{cite}

\begin{document}
\begin{titlepage}
\renewcommand{\thefootnote}{\fnsymbol{footnote}}
\begin{flushright}
\begin{tabular}{l}
IPPP/10/31\\
DCPT/10/62\\
TUM--HEP--756/10
\end{tabular}
\end{flushright}
\vskip1.5cm
\begin{center}
{\large \bf \boldmath
Theoretical and Phenomenological Constraints
on \\[0.3em]  Form Factors for Radiative and Semi-Leptonic $B$-Meson Decays}
\vskip1.3cm 
{\sc
Aoife Bharucha\footnote{a.k.m.bharucha@durham.ac.uk}$^{,1}$,
Thorsten Feldmann\footnote{thorsten.feldmann@ph.tum.de}$^{,2}$,
Michael~Wick\footnote{michael.wick@ph.tum.de}$^{,2}$
} \vskip1.5cm
        $^1$ {\em IPPP, Department of Physics,
University of Durham, Durham DH1 3LE, UK}\\
\vskip0.5cm
$^2$ {\em Physik-Department, Technische Universit\"at M\"unchen, D-85748 
Garching, Germany}\\

\vskip1.5cm

{\large\bf Abstract\\[10pt]} \parbox[t]{\textwidth}{
We study transition form factors for radiative and rare semi-leptonic $B$-meson decays into
light pseudoscalar or vector mesons, combining theoretical constraints and phenomenological
information from Lattice QCD, light-cone sum rules, and dispersive bounds.
We pay particular attention to form factor parameterisations which are
based on the so-called series expansion, and study the related
systematic uncertainties on a quantitative level.
In this context, we also provide the NLO corrections to the correlation
function between two flavour-changing tensor currents, which enters the
unitarity constraints for the coefficients in the series expansion.
}

\vfill

\end{center}
\end{titlepage}

 \newcommand\T{\rule{0pt}{2.5 ex}}
 \newcommand\TT{\rule{0pt}{3. ex}}
\newcommand\B{\rule[-1.5ex]{0pt}{0pt}}
\newcommand\BB{\rule[-2.3ex]{0pt}{0pt}}

\tableofcontents

\clearpage

\setcounter{footnote}{0}

\section{Introduction}\label{sec:1}

At the upcoming LHC experiments (notably LHCb), 
exclusive $B$-meson decays will play one of the major roles for
precision tests of the flavour sector in the Standard Model (SM) and its
possible New Physics (NP) extensions \cite{Buchalla:2008jp,Antonelli:2009ws}. 
In order to extract information about
the underlying short-distance flavour transitions,  
hadronic matrix elements will be required as theoretical input,
and the precision to which they can be predicted will be essential for 
the success of the flavour program at LHCb. 
The most prominent example are transition form factors (FFs) for $B$-meson
decays into light mesons, which will be the subject of this work.
Being non-perturbative hadronic quantities, the theoretical calculation of FFs requires
techniques such as Lattice QCD (see e.g.\ \cite{Bailey:2008wp,AlHaydari:2009zr,Bernard:2009ke,Dalgic:2006dt,Bowler:2004zb})
or QCD sum rules on the light cone (LCSR, see e.g.\ \cite{Ball:2004ye,Ball:2004rg},
and \cite{Colangelo:2000dp} and references therein).
The two methods are complementary with respect to the momentum transfer $q^2$ between
the initial and final-state mesons: In Lattice QCD, results are more easily obtainable at high values of $q^2$, as discretisation effects can only be controlled 
for small momenta of the final state in units of the Lattice spacing. 
This is in contrast to the LCSR method, which involves an expansion in inverse powers 
of the energy of the light daughter meson that is valid for low values of $q^2$.

Different FF parameterisations, which can be used to interpolate between the results 
for small and large momentum transfer, have been suggested in the literature;
a good review can be found in \cite{Dalgic:2006dt}.
These include simple pole-type parameterisations, like the Be\'cirevi\'c--Kaidalov (BK)
approach \cite{Becirevic:1999kt}, or variants like the Ball--Zwicky (BZ) parametrisation \cite{Ball:2004ye}. 
Another representation is found from the Omnes solution to the dispersion relation,
see the discussion in  \cite{Flynn:2007ii,Bourrely:2008za}.
In this paper, we will make use of the so-called Series Expansion (SE),
which was advocated in Refs.~\cite{Arnesen:2005ez,Boyd:1994tt,Boyd:1997qw,
Caprini:1997mu,Becher:2005bg}. 
Here, one can make use of dispersive bounds to obtain additional theoretical constraints
on the expansion coefficients. A simplified version of the SE (SSE) was recently suggested 
in \cite{Bourrely:2008za}.
The aim of this work is to use the SE/SSE to describe the transition FFs 
on the basis of recent Lattice and LCSR results,
including a detailed analysis of systematic errors. 
We will focus in particular on the FFs
entering  $B \to V\gamma$, $B \to L\,\ell^+\ell^-$, $B\to L \nu \bar \nu$
decays, where $L=P,V$ is a light vector or pseudoscalar meson. 
We will give numerical results for $B\to \rho,K,K^*$ and $B_s \to \phi$
transitions, which are particularly interesting with respect to NP studies,
see e.g.\ \cite{Buchalla:2008jp,Antonelli:2009ws,Beneke:2004dp,Bobeth:2007dw,Egede:2008uy,Altmannshofer:2008dz,Bartsch:2009qp}. (Detailed phenomenological studies for FFs relevant
for the determination of the CKM elements $|V_{ub}|$ and $|V_{cb}|$
from semi-leptonic $B$ decays can be found in the recent literature
\cite{Ball:2006jz,Bourrely:2008za,Bailey:2008wp,AlHaydari:2009zr,Bernard:2009ke}.)
For the discussion of dispersive bounds for tensor FFs,
we will include the result of a precise calculation of the tensor 
current two-point correlator at NLO in the QCD coupling constant, including 
the leading non-perturbative corrections from quark and gluon condensates.

Our paper is organized as follows.
In Section~\ref{sec:FFdefs}, we provide convenient definitions for
the $B$-meson FFs and introduce the idea of the SE/SSE. 
In Section~\ref{sec:DB}, we review the derivation of dispersive bounds from
current-correlation functions and summarize the results for
the profile functions obtained from the operator-product expansion.
We apply our formalism to $B\to K$, $B\to \rho$, $B \to K^*$ and $B_s \to \phi$
FFs, by fitting the (truncated) SE/SSE to theoretical
``data'' from Lattice QCD and/or LCSRs in Section~\ref{sec:fit}. Our conclusions
are presented in Section~\ref{sec:concl}, and some technical details are 
given in the Appendix.

\section{Form Factors}\label{sec:FFdefs}

In this section, we will first provide the definitions for the 
various $B$-meson FFs in question, fix the notation to be used in the 
subsequent discussion, and introduce the SE/SSE.

\subsection{Definition of Form Factors and Helicity Amplitudes}

The hadronic matrix elements
for transitions between a pseudoscalar $B$ meson 
and a generic (light) pseudoscalar meson 
are usually written in terms of three FFs 
$f_0(q^2)$, $f_+(q^2)$ and $f_T(q^2)$, 
which depend on the momentum transfer $q^2=(p-k)^2$,  
\begin{eqnarray}  
\langle P(k)|\bar q \gamma_\mu b| B(p)\rangle&=&\left(p_\mu + k_\mu - q_\mu\,\frac{m_B^2-m_P^2}{q^2}\right)f_+(q^2)
+\frac{m_B^2-m_P^2}{q^2}\, q_\mu \,f_0(q^2) \,,
\nonumber \\
\langle P(k)|\bar q \sigma_{\mu\nu}q^\nu b | B(p)\rangle&= &
\frac{i}{m_B+m_P}\left(
q^2(p+k)_\mu-(m_B^2-m_P^2) \, q_\mu\right) f_T(q^2)\,.
\end{eqnarray}
At zero momentum transfer, the additional relation $f_+(0)=f_0(0)$ holds. 

Similarly, the matrix elements for the transition between a $B$-meson and 
a generic vector meson\footnote{Our phase convention for the vector state
differs by a relative factor of $i$ from the convention that is used,
for instance, in \cite{Ball:2004rg}.}
can be written in terms of FFs 
$V(q^2)$, $A_{0-3}(q^2)$, $T_{1-3}(q^2)$,
which are conventionally defined as
\begin{eqnarray}
\langle V(k,\varepsilon)|\bar q\gamma_\mu b |\bar B(p)\rangle &=& i \epsilon_{\mu\nu\rho\sigma} \, \varepsilon^{*\nu}(k) \, p^\rho k^\sigma \, \frac{2V(q^2)}{m_B+m_V}
\,,
\nonumber \\[0.2em]
\langle V(k,\varepsilon)|\bar q\gamma_\mu\gamma_5 b |\bar B(p)\rangle &=& -\varepsilon^*_\mu(k) \, (m_B+m_V) \, A_1(q^2)
+ (p+k)_\mu \, (\varepsilon^*(k) \cdot q) \, \frac{A_2(q^2)}{m_B+m_V}
\nonumber\\
&&+q_\mu \, (\varepsilon^*(k) \cdot q) \,\frac{2m_V}{q^2}\left(A_3(q^2)-A_0(q^2)\right) \,,
\end{eqnarray}
where $A_0(0) =  A_3(0)$.
For transitions involving a tensor current, 
the matrix elements are characterised by the tensor FFs,
\begin{eqnarray}
\langle V(k,\varepsilon) |\bar q \sigma_{\mu\nu} q^\nu  b|
\bar B(p)\rangle&=& i\epsilon_{\mu\nu\rho\sigma} \, \varepsilon^{*\nu} \,
p^\rho k^\sigma\, 2 T_1(q^2)\,, \nonumber 
\\
\langle V(k,\varepsilon) |\bar q \sigma_{\mu\nu} q^\nu \gamma_5 b|
\bar B(p)\rangle&=& T_2(q^2) \left( \varepsilon^*_\mu(k)
  \, (m_B^2-m_V^2) - (\varepsilon^*(k) \cdot q) \, (p+k)_\mu \right)\nonumber\\
&& + T_3(q^2) (\varepsilon^*(k) \cdot q) \left( q_\mu - \frac{q^2}{m_B^2-m_V^2} \, (2p-q)_\mu
\right) \,, \label{eq:pengFF}
\end{eqnarray}
where $T_1(0) = T_2(0)$. 
The equations of motion for the quarks imply the additional constraint
\begin{equation}
 A_3(q^2)=\frac{m_B+m_V}{2m_V}\,A_1(q^2) -\frac{m_B-m_V}{2m_V}\,A_2(q^2) \,,
\end{equation}
and therefore the $B\to V$ transitions 
are characterized by seven independent FFs.

The above constitute the standard definitions for the FFs widely used in the literature. However, for this work, we find it convenient to work with certain linear combinations of these, dubbed helicity amplitudes in Ref.~\cite{Boyd:1997qw}. 
This is primarily to diagonalize the unitarity relations, which shall be used to 
derive the dispersive bounds on certain FF parameterisations. 
The helicity amplitudes also have definite spin-parity quantum numbers, which is useful when considering the contribution of excited states. In addition, they 
have simple relations to the universal FFs, 
appearing in the heavy-quark and/or large-energy limit (see appendix~\ref{sec:newFF}),
and lead to simple expressions for the observables in $B \to L\,\ell^+\ell^-$ decays
in the naive factorization approximation.
To put the contributions to the various correlation functions entering the dispersive
bounds on an equal footing, we also choose a particular normalization convention
and define new $B \to P$ vector FFs via
\begin{align}
\label{eq:AVdef}
\mathcal{A}_{V,\sigma}(q^2 )& =\sqrt{\frac{q^2}{\lambda}} \,
  {\varepsilon_\sigma ^{*\mu}(q)} 
\, \langle P(k)|\bar q \, \gamma_\mu \, b |\bar B(p)\rangle \,.
\end{align}
Here 
\begin{align} 
\lambda &= \left( (m_B-m_P)^2-q^2\right) \left( (m_B+m_P)^2-q^2\right)
\equiv (t_- - q^2)(t_+ - q^2) 
\label{eq:lambdadef}
\end{align}
is a standard kinematic function,
and $\varepsilon_\sigma ^{*\mu}(q)$ are 
transverse ($\sigma=\pm$), longitudinal ($\sigma=0$) or time-like ($\sigma=t$) polarization vectors as defined in (\ref{eq:defeps}) in the appendix.
This implies
\begin{align}
 \mathcal{A}_{V,0}(q^2)  &= f_+(q^2) \,, \qquad  
 \mathcal{A}_{V,t}(q^2)  = \frac{m_B^2  - m_P^2}{\sqrt \lambda} \, f_0(q^2) \,,
\end{align}
while the transverse projections vanish.
Similarly, for the $B\to P$ tensor FF, we define
\begin{equation}
\label{eq:ATdef}
\mathcal{A}_{T,\sigma}(q^2 )= (-i )\sqrt{\frac{1}{\lambda}} \,
{\varepsilon_\sigma^{*\mu}(q)}
\, \langle P(k)|\bar q \, \sigma_{\mu\nu} q^\nu \, b |\bar B(p)\rangle  \,.
\end{equation}
Here, the only non-zero FF is\footnote{The newly 
defined tensor FF $\mathcal{A}_{T,0}(q^2)$ vanishes as $\sqrt{q^2}$,
which might look somewhat artificial at first glance. 
However, the tensor current does not contribute to physical processes at $q^2=0$ anyway.} 
\begin{equation}
\mathcal{A}_{T,0}(q^2 )  = \frac{\sqrt{q^2}}{ m_B + m_P} \, f_T(q^2) \,.
\end{equation}
A similar analysis for the $B \to V$ vector and axial-vector FFs yields
\begin{equation}
\label{eq:BVdef}
\mathcal{B}_{V,\sigma}(q^2 )
= \sqrt{\frac{q^2}{\lambda}} \, \sum\limits_{\varepsilon(k)} 
{\varepsilon_\sigma^{*\mu}(q)} \, \langle V(k,\varepsilon(k))
|\bar q \, \gamma_\mu(1-\gamma^5) \, b |\bar B(p)\rangle
\end{equation}
with
\begin{align}
 \mathcal{B}_{V,0}(q^2)&= \frac{(m_B+m_V)^2 \, (m_B^2-m_V^2-q^2) \, A_1(q^2)
 -\lambda \, A_2(q^2)}{2 m_V \sqrt{\lambda} \, (m_B+m_V)} \,,
\cr 
\mathcal{B}_{V,t}(q^2) & = A_0(q^2) \,,
\cr 
\mathcal{B}_{V,1}(q^2)&\equiv - \frac{\mathcal{B}_{V,-}-\mathcal{B}_{V,+}}{\sqrt{2}} 
 =\frac{\sqrt{2 \, q^2}}{m_B+m_V} \, V(q^2) \,,
\cr 
\mathcal{B}_{V,2}(q^2)&\equiv - \frac{\mathcal{B}_{V,-}+\mathcal{B}_{V,+}}{\sqrt{2}}
=\frac{\sqrt{2 \, q^2 } \, (m_B+m_V)}{\sqrt{\lambda }} \, A_1(q^2) \,.
\end{align}
Finally, the $B\to V$ matrix elements with tensor currents are projected on
\begin{equation}
\label{eq:BTdef}
\mathcal{B}_{T,\sigma}(q^2 )= \sqrt{\frac{1}{\lambda}} \,
\sum\limits_{\varepsilon(k)} \,  {\varepsilon _\sigma ^{*\mu}  (q)} 
\, \langle V(k,\varepsilon(k)) |\bar q \, \sigma_{\mu\alpha} q^\alpha (1+\gamma^5) \, b|\bar B(p)\rangle
\end{equation}
giving rise to the FFs
 \begin{align}
\mathcal{B}_{T,0}(q^2 )&
=
\frac{\sqrt{q^2} \, (m_B^2+3 m_V^2-q^2)}{2 m_V \sqrt{\lambda}} \, T_2(q^2)
-
\frac{\sqrt{q^2 \, \lambda }}{2 m_V \, (m_B^2-m_V^2)} \, T_3(q^2)
 \cr 
\mathcal{B}_{T,1}(q^2 )&=
- \frac{\mathcal{B}_{V,-}-\mathcal{B}_{V,+}}{\sqrt{2}}
=\sqrt{2} \, T_1(q^2 ) \,,
\cr 
\mathcal{B}_{T,2}(q^2 )&=
- \frac{\mathcal{B}_{V,-}+\mathcal{B}_{V,+}}{\sqrt{2}}=
\frac{\sqrt{2} \,(m_B^2-m_V^2)}{\sqrt{\lambda }} \, T_2(q^2) \,.
\end{align}  


\subsection{Series Expansion}

\subsubsection{Resonances}

 An important factor in determining the shape of the FF is the presence of 
low-lying resonances with appropriate  quantum numbers and mass $m_{\rm R}$ in the range $ t_- <m_{\rm R}^2< t_+$. 
Common to most parameterisations is the inclusion of the low-lying resonance 
by a simple pole. The various descriptions differ in the modelling of the continuous part.
In the following, we use the abbreviation $P(q^2) =1-q^2/m_{R}^{2}$.
If multiple resonances are present in the given region, then $P(q^2)$ should be a product of such poles, and if no resonances are present then $P(q^2)=1$. A summary of 
the relevant resonance masses is provided in Table~\ref{tab:reson}.

\begin{table}[tpbh]

\caption{\label{tab:reson} Summary of the masses of low-lying $B_d$ and $B_s$
resonances, using PDG values \cite{Amsler:2008zzb} and/or 
theoretical estimates from heavy-quark/chiral symmetry \cite{Bardeen:2003kt}.
Notice that the mass values for $(0^+,1^+)$ predicted in
\cite{Bardeen:2003kt} have not been confirmed experimentally, yet. Instead
the PDG quotes ``effective'' resonances $B_J^*(5698)$ and $B_{sJ}^*(5853)$ 
with undetermined spin/parity.}

\begin{center}
\begin{tabular}{l|c|c|c|c|c}
\hline
\hline
\T Transition & $J^P$ & Mass (GeV) & $J^P$ & Mass (GeV) & Ref. \\
\hline
\hline
\T $b \to d$& $0^-$ & 5.28 & $1^-$ & 5.33 & \cite{Amsler:2008zzb} \\
         & $0^+$ & 5.63 & $1^+$ & 5.68 & \cite{Bardeen:2003kt} \\
          & $1^+$ & 5.72 & $2^+$ & 5.75 & \cite{Amsler:2008zzb} \\
\hline
\hline
\T $b \to s$ & $0^-$ & 5.37 &  $1^-$ & 5.42 & \cite{Amsler:2008zzb} \\
           & $0^+$ & 5.72 &  $1^+$ & 5.77 & \cite{Bardeen:2003kt} \\
           & $1^+$ & 5.83 &  $2^+$ & 5.84 & \cite{Amsler:2008zzb} \\
\hline
\hline
\end{tabular}
\end{center}
\end{table}

\subsubsection{Series Expansion(SE)}

The SE has its origin in dispersive relations \cite{Arnesen:2005ez,Boyd:1994tt,Boyd:1997qw,Becher:2005bg}. 
The starting point is to extend the FFs defined in the physical range 
(from $q^2=0$ to $t_-=(m_B-m_L)^2$)
 to analytic functions throughout the complex $t=q^2$ plane, except for along the branch cut at the threshold for production of real $BP$/$BV$ pairs at $q^2 \geq t_+=(m_B+m_L)^2$.
If  low-lying resonances are present below $t_+$, they are 
accounted for by the so called Blaschke factor $B(t)$, see below.
 Complex analysis can then be used to map the cut $t$\/--plane onto the unit disc
 in terms of the coordinate $z(t)$. The variable $z(t)$ is found to be an excellent expansion 
parameter for the FFs. Furthermore,
with an appropriately chosen normalization function $\phi_f(t)$, one obtains simple dispersive bounds
on the coefficients of the SE, see below.
 We will discuss the calculation of the functions $\phi_f(t)$ as well as the derivation
of the dispersive bounds in greater detail in Section~\ref{sec:DB}. 
The Series Expansion (SE) then corresponds to the
 following FF parametrisation,
 \begin{equation}
\label{par:SE}
  f(t) = \frac{1}{B(t) \, \phi_f(t)} \, \sum_{k} \alpha_k \, z^k(t) \,,
 \end{equation}
with 
\begin{equation}
\label{eq:zdef}
z(t)\equiv z(t,t_0)= \frac{ \sqrt{t_+-t}- \sqrt{t_+-t_0}}{\sqrt{t_+-t}+ \sqrt{t_+-t_0}} \,.
\end{equation}
Here $0 \leq t_0<t_-$ is a free parameter 
which can be optimised to reduce the maximum value of $|z(t)|$ in 
the physical FF range,
\begin{equation}
 t_0 \big|_{\rm opt.} =  t_+ \left(1 - \sqrt{1 - \frac{t_-}{t_+}} \right)
\,.
\label{eq:t0opt}
\end{equation}
We will later see that with the optimised value for $t_0$,
the FFs can be well described by a SE which is truncated after
the second term proportional to $z(t)$. Other values of $t_0$
(e.g.\ $t_0=0$) are still allowed but sometimes require to go
to higher order in the SE.

As a crucial property, the function $z(t)$ satisfies $|z(t)| \equiv 1$ in
the pair-production region, $t \geq t_+$.
The Blaschke factor is thus chosen as $B(t)=z(t,m_{\rm R}^2)$. 
As for $P(q^2)$ defined above, if multiple resonances are present,
 then $B(t)$ is a product of the corresponding Blaschke factors.
Further discussion about the physical basis for the SE is found in Ref.~\cite{Boyd:1997qw}.

\subsubsection{Simplified Series Expansion (SSE)} 

Another form of the Series Expansion method can also be considered. Instead of the Blaschke factor $B(t)$, one can use a simple pole $P(q^2)$ to account for low-lying resonances. This idea was proposed in Ref.~\cite{Bourrely:2008za}, yielding
\begin{eqnarray}
\label{par:SSE}
f(t) &=& 
 \frac{1}{P(t)} \, \sum_{k} \tilde\alpha_k \, z^k(t,t_0) \,.
\end{eqnarray}
It was found that the dispersive bounds can still be imposed on the coefficients 
$\tilde\alpha_k$ of the SSE. We will discuss this and other issues concerning the validity of the simplifications in the following section.

\section{Dispersive Bounds} \label{sec:DB}

The FFs describe the process $B \to L$ with $L=P,V$ in the decay region 
$ 0 < q^2  < t_-=(m_B  - m_L)^2 $.
Using crossing symmetry, they can also describe the process 
in the pair-production region $ q^2>(m_B  + m_L)^2$.
This can be exploited to obtain a bound on parameters describing the FFs. A detailed derivation of this bound can be found in Refs.~\cite{Caprini:1997mu,Boyd:1994tt}.
Here we provide a brief outline of the argument, in order to introduce our notation
and to extend the method to tensor FFs.

The crucial observation of the idea of dispersive bounds (as it is for QCD sum rules) is the possibility to evaluate the correlator of two flavour-changing currents, 
\begin{equation} 
\Pi^X_{\mu \nu}(q^2)  =  i \, \int d^4 x \, e^{i \, q \cdot x}
 \left\langle 0 \right|{\rm T} \, j^X_{\mu}(x) \, j^{\dag\,X}_{\nu}(0)\left| 0 \right\rangle \,, 
\label{eq:correlator}
\end{equation}
either by an operator product expansion (OPE) or by unitarity considerations. 
Here the relevant currents $j_\mu^X$ are defined as\footnote{In phenomenological
applications, we are only interested in the currents $j_\mu^{T+A_T}$. The connection
to correlators with genuine tensor currents $j_{\mu\nu} = \bar q \sigma_{\mu\nu} q$
is given in Appendix~\ref{app:proj}.}  

\begin{align}
 j_\mu^{V}      &= \bar q\gamma _\mu  b  \,,
&
j_\mu^{V-A}  &= \bar q\gamma _\mu  (1 - \gamma^5  )b \,,
\cr  
 j_\mu^{T} &= \bar q\sigma _{\mu \alpha } q^\alpha  b
\,,
&
j_\mu ^{T+A_T}  &= \bar q\sigma _{\mu \alpha } q^\alpha  (1 + \gamma ^5 )b
\,.
 \end{align}
Furthermore, we introduce longitudinal and transverse helicity projectors, 
 \begin{equation}
 P _L^{\mu \nu }(q^2)  = \frac{{q{}^\mu q^\nu  }}{{q^2}}
\,, \qquad
 P _T^{\mu \nu }(q^2)  = \frac{1}{D-1}\left( \frac{q^\mu  q^\nu}{q^2 }-g^{\mu \nu } \right)
\,,
 \end{equation}
which allow us to rewrite the correlation functions in terms of Lorentz scalars,
\begin{equation}
\Pi_{I}^X(q^2) \equiv P_{I}^{\mu \nu }(q^2) \, \Pi _{\mu \nu }^X(q^2)\,, \qquad
 \mbox{\small ($I=L,T$).}
\end{equation}
As $\Pi _{I}^X(q^2)$ is an analytic function, 
it satisfies the subtracted dispersion relation,
\begin{equation} \label{eq:chidef}
\chi^X_{I}(n) =\frac{1}{n!}\left. {\frac{d^n\Pi_{X}(q^2)}{{{dq^2}^n}}} \right|_{q^2  = 0}= \frac{1}{\pi } \, \int\limits_0^\infty  
dt \, \frac{\mathrm{Im} \, \Pi^X_I (t)}{\left( {t - q^2 } \right)^{n+1} }\bigg|_{q^2=0}\,,
\end{equation}
where the number of subtractions $n$ is chosen to render the resulting function 
$\chi^X_{I}(n)$ finite.

\subsection{Hadronic representation of the Correlator}

Unitarity allows us to express $\mathrm{Im}\, \Pi^X_{I}(q^2)$ as the positive
definite sum over all hadronic states $\Gamma$ with allowed quantum numbers:
 \begin{equation}
\mathrm{Im} \, \Pi^X_{I}(q^2) = \frac{1}{2} \,
\sum\limits_\Gamma\!
 \int d\rho_\Gamma \, (2\pi )^4 \,
\delta^4(q - p_\Gamma) \, P_I^{\mu \nu } \,
\left\langle 0 \right|j_\mu^X\left| \Gamma\right\rangle 
\left\langle \Gamma \right|j_\nu^{\dag X} \left| 0\right\rangle \,.
\end{equation}
where $p_\Gamma$ is the total momentum of the final state,
 and $d\rho_\Gamma$ contains the appropriate phase-space weighting. 
For a particular choice of intermediate state, $\Gamma = B L$, we define 
 \begin{equation}
\mathrm{Im} \, \Pi^X _{I,B L}(q^2) 
 =\eta \, \int d\rho_{BL} \, P_I^{\mu \nu } \,
\left\langle 0 \right|j_\mu^X  \left| B L \right\rangle
\left\langle  B L \right|j_\nu^{\dag X}  \left|0 \right\rangle\,,
\end{equation}
where $\eta$ is an isospin-degeneracy factor for a given channel, 
and we relegate the contribution from phase space to the function
\begin{equation}
d\rho_{BL}=\frac{1}{4\pi^2 } \,
\int {\frac{d^3 p_B}{2E_B} \, \frac{d^3 p_L}{2E_L} \, \delta^4(q - p_B  - p_L)} \,.
\end{equation}
Clearly, this results in the inequality
\begin{equation}\label{eq:unit}
{\mathop{\rm Im}\nolimits} \Pi^X_{I,B L}(t) \leq 
{\mathop{\rm Im}\nolimits} \Pi^X_{I}(t) \,.
\end{equation}
Now, by extending the FFs to analytic functions throughout the $t$\/--plane, 
except for along the branch cut at the threshold for production of real $BL$ pairs, 
one can use crossing symmetry to relate the matrix elements $\left\langle 0  \right|j^X  \left| BL \right\rangle$ to $\left\langle B \right|j^X  \left| L \right\rangle$. 
The latter can be rewritten in terms of FFs, as defined in Section~\ref{sec:FFdefs}. 
As stated earlier, we use helicity-based linear combinations of the traditional FFs, 
such that all production amplitudes `diagonalize':

\begin{eqnarray}
 \nonumber P _T^{\mu \nu } \,
\langle P|  j_\mu ^{V} | B\rangle 
\langle B| j_\nu ^{\dag V} | P\rangle&=&\frac{\lambda}{3 q^2} \left|\mathcal{A}_{V,0}\right|^2
\,, \\
 \nonumber P _L^{\mu \nu } \,
\langle P|  j_\mu ^{V} | B\rangle 
\langle B| j_\nu ^{\dag V} | P\rangle&=&\frac{\lambda}{q^2}\left|\mathcal{A}_{V,t}\right|^2
\,, \\
 P _T^{\mu \nu } \, 
\langle P|  j_\mu ^{T} | B\rangle 
\langle B| j_\nu ^{\dag T}| P\rangle&=&\frac{\lambda}{3}\left|\mathcal{A}_{T,0}\right|^2 \,,
\label{AFF2}
\end{eqnarray}
for $B$ decays into pseudoscalars, and
\begin{eqnarray}
\nonumber P _T^{\mu \nu } \,
 \langle V|   j_\mu ^{V - A} | B\rangle 
\langle B|j_\nu ^{\dag,V - A}| V\rangle
&=&\frac{\lambda}{3 q^2}\, \sum_{i=0}^2 \left|\mathcal{B}_{V,i}\right|^2
\,, \\
\nonumber P _L^{\mu \nu } \,
\langle V|j_\mu ^{V - A} |B\rangle 
\langle B|j_\nu ^{\dag,V - A} |V\rangle
&=&\frac{ \lambda}{q^2}\left|\mathcal{B}_{V,t}\right|^2
\,, \\
 P_T^{\mu \nu } \, 
\langle V | j_\mu ^{T+A_T} b|B\rangle 
\langle B | j_\nu ^{\dag,T+A_T} b| V\rangle
&=&\frac{\lambda}{3 }\,\sum_{i=0}^2 \left|\mathcal{B}_{T,i}\right|^2 \,.
\label{BFF2}
\end{eqnarray}
We can now express $\mathrm{Im} \, \Pi^X _{I,B L} $ in compact form, 
\begin{equation}
\mathrm{Im} \, \Pi^X _{I,BL}=
\eta \, \int d\rho_{BL} \, \frac{\lambda}{3 t} \left|A_I^X\right|^2
=\frac{\eta}{48\pi} \,
\frac{\lambda^{3/2}}{t^2}
\left|A_I^X\right|^2 \,,\label{eq:egBM}
\end{equation}
where the $\left|A_I^X\right|^2$ can be read off (\ref{AFF2},\ref{BFF2}),
\begin{align}
 \left|A_T^{V}\right|^2 &=\left|\mathcal{A}_{V,0}\right|^2 \,,
\qquad 
 \left|A_L^{V}\right|^2 = 3  \left|\mathcal{A}_{V,t}\right|^2 \,,
\qquad  
 \left|A_T^{T}\right|^2 =
  q^2  \left|\mathcal{A}_{T,0}\right|^2 \,,
\label{eq:Afunc1}
\end{align}
for decays into pseudoscalars, and
\begin{align}
 \left|A_T^{V-A}\right|^2 &=\sum_{i=0}^2 \left|\mathcal{B}_{V,i}\right|^2
\,, \qquad 
\left|A_L^{V-A}\right|^2 = 3 \left|\mathcal{B}_{V,t}\right|^2
\,, \qquad 
\left|A_T^{T+A_T}\right|^2 = q^2 \sum_{i=0}^2 \left|\mathcal{B}_{T,i}\right|^2 \,,
\label{eq:Afunc2}
\end{align}
for decays into vector mesons.

\subsection{OPE for the Correlator}

Alternatively, we can examine the correlator (\ref{eq:correlator}),
using an OPE for the T-ordered product of currents in the limit $q^2=0 \ll t_+$. 
The standard expansion takes the form \cite{Shifman:1978bx,Shifman:1978by,Novikov:1980uj}
 \begin{equation}
i \, \int dx \, e^{i \, q \cdot x} \, 
P^{\mu\nu}_I \, {\rm T} \left\{ j_\mu^X(x) \, j_\nu^{\dag\,X}(0) \right\}
=\sum\limits_{k = 1}^\infty  \, C^X_{I,k}(q) \, {\cal O}_k \,,
\end{equation}
where $C^{X}_{I,n}(q)$ are Wilson coefficients for a given current $X$ and projector $I$, and ${\cal O}_n$ are local gauge-invariant operators,
consisting of quark and gluon fields. 
Here, the operators are ordered by increasing dimension $k$.
We can use the above, to express the correlator,
 \begin{equation}\label{eq:OPE}
 \Pi^X_{I,\rm OPE}(q^2)=\sum\limits_{k = 1}^\infty 
 C^X_{I,k}(q^2) \left\langle O_k  \right\rangle \,.
 \end{equation}
Besides the identity operator, whose Wilson coefficient contains the purely
perturbative contribution to the correlator, we will specifically consider the 
first few operators related to the non-perturbative contribution from the quark condensate $\displaystyle \langle m_q \, \bar{q} q\rangle$, the gluon condensate  $\displaystyle\langle \frac{\alpha_s}{\pi} \, G^2\rangle$, and the mixed condensate 
$\displaystyle\left\langle g_s \, \bar q \left(\sigma \cdot G\right) q \right\rangle$.
We will elaborate on our calculation of the Wilson coefficients, $C^X_{I,k}(q^2)$, later. 
Specifically, we must calculate the Wilson coefficients entering the functions $\chi^X_I(n)$ in (\ref{eq:chidef}).

\subsection{Bounds on coefficients in the SE}

Using (\ref{eq:unit}), we find 
 \begin{equation} 
 \frac{1}{\pi } \,
 \int\limits_0^\infty dt \,
 \frac{\mathrm{Im} \, \Pi^X_{I,BL} (t)}{\left(t - q^2 \right)^{n+1} }\bigg|_{q^2=0}
=
 \frac{1}{\pi } \,
 \int\limits_{t_+}^\infty dt \,
 \frac{\eta \, \lambda^{3/2}(t)}{48\pi \, t^{n+3}} \left| A_I^X(t) \right|^2 \leq
\chi^{X}_{I}(n)\,,
\label{eq:unit2}
  \end{equation}
where $\chi^{X}_{I} \equiv \chi^{X}_{I,\rm OPE}$ is calculated from (\ref{eq:OPE}).
Mapping the pair-production region $t \geq t_+$ onto the unit circle $|z(t)|=1$,
this inequality could be written in the form
\begin{align}
\label{eq:phidet}
\frac{1}{2\pi i } \, 
\oint &\frac{dz}{z} \, |\phi_I^X \, A _I^X|^2(z) \leq 1
\quad \Leftrightarrow \quad 
\frac{1}{\pi} \, 
\int_{t_+}^\infty\frac{dt}{t-t_0} \, \sqrt{\frac{t_+-t_0}{t-t_+}}
\, |\phi_I^X \, A_I^X|^2(t) \leq 1
\,,
\end{align}
where the function $|\phi_I^X(t)|^2$ can be obtained by comparing
(\ref{eq:phidet}) and (\ref{eq:unit2}), and using $\lambda(t)=(t_+-t)(t_--t)$,
\begin{equation}\label{eq:bound2a}
 |\phi_I^X(t)|^2= \frac{\eta}{48 \pi \, \chi^{X}_I(n)}
\, \frac{(t-t_+)^2}{(t_+ -t_0)^{1/2}} \, \frac{(t-t_-)^{3/2}}{t^{n+2}}
 \, \frac{t-t_0}{t} \,.
\end{equation}
The isospin-degeneracy factor $\eta$ 
takes the values 3/2, 2 and 1 for $B\to\rho$,  $B\to  K^{(*)}$ and $B_s\to\phi$ respectively. 
We may now generically write the helicity-based FFs $A_I^X(t)$ as 
\begin{equation}
A_I^X(t) = \frac{(\sqrt{-z(t,0)})^m
(\sqrt{z(t,t_-)})^l }{B(t) \, \phi_I^X(t)} \, \sum_{k=0}^\infty \alpha_k \, z^k
\label{eq:ssepar}
\end{equation}
with real coefficients $\alpha_k$, and a Blaschke factor ${B(t) = \prod_i z(t,m_{R_i}^2)}$,
representing poles due to sub-threshold resonances of masses $m_{R_i}$, and 
satisfying $|B(t)| = 1$ in the pair-production region. The additional factors
${(\sqrt{-z(t,0)})^m}$ and 
${(\sqrt{z(t,t_-)})^l}$ have been added to take into account the unconventional
normalisation of our FF functions through factors of ${\sqrt{q^2}}$ and
$\sqrt{\lambda}$ (e.g.\ $m=1$ for ${\cal A}_{T,0}$, and
$l=-1$ for ${\cal A}_{V,t}$, cf.\ above).\footnote{These factors could also be
considered as part of the Blaschke factor. Note that under a change of
normalisation convention for the FFs, both, the so-constructed Blaschke factor
as well as the function $\phi(t)$ have to be modified, while the coefficients
$\alpha_k$ of the SE remain the same.}
The function $\phi_I^X(t)$ has to be constructed in such a way that its absolute value
satisfies Eq.~(\ref{eq:bound2a}), while (\ref{eq:ssepar}) 
retains the analytical properties of the FF. 
This can easily be achieved by replacing potential poles and cuts in 
$\sqrt{|\phi_I^X(t)|^2}$, by making replacements of the form 
\begin{equation}
 \frac{1}{t-X}\to \frac{-z(t,X)}{t-X}\, ,
\end{equation}
which is allowed as $\left| z(t,X) \right|=1$ in the pair-production region. 
This results in (see also \cite{Hill:2006ub})
\begin{equation}
\phi_I^X(t)=\sqrt{\frac{\eta}{48\pi\chi^{X}_{I}(n)}}
\, \frac{(t-t_+)}{(t_+-t_0)^{1/4}}
\left(\frac{z(t,0)}{-t}\right)^{(3+n)/2}
\left(\frac{z(t,t_0)}{t_0-t}\right)^{-1/2}
\left(\frac{z(t,t_-)}{t_--t}\right)^{-3/4}\,.
\end{equation} 
Inserting the parametrisation (\ref{eq:ssepar}) into (\ref{eq:phidet}),
and using $|z(t,t_0)|=|z(t,m_R^2)|=|z(t,0)|=1$,
the integration $dz/z = d\varphi$ along the unit circle is trivial, yielding
the desired bound on the coefficients $\alpha_k$, 
\begin{equation}
\sum_{k=0}^\infty \alpha_k^2<1 \,.
\end{equation}

For decays into vector mesons, using an analogous parametrisation 
as (\ref{eq:ssepar}) for each \emph{individual} FF contribution in (\ref{eq:Afunc2}),
one obtains a bound on the sum of the corresponding coefficients.
As an example, let us consider $A_T^{V-A}(t)$, where we parameterise
\begin{align}
\mathcal{B}_{V,0}(t) &= 
\frac{1}{B(t) \, \sqrt{z(t,t_-)} \, \phi_T^{V-A}(t)} \, \sum_{k=0}^{K-1} \beta_k^{(V,0)} \, z^k
\,,
\cr 
\mathcal{B}_{V,1}(t) &= 
\frac{\sqrt{-z(t,0)}}{B(t)\, \phi_T^{V-A}(t)} \,  \sum_{k=0}^{K-1} \beta_k^{(V,1)} \, z^k
\,,
\cr  
\mathcal{B}_{V,2}(t) &= 
\frac{\sqrt{-z(t,0)}}{B(t)\, \sqrt{z(t,t_-)} \, \phi_T^{V-A}(t)} \,  
\sum_{k=0}^{K-1} \beta_k^{(V,2)} \, z^k
\,,
\label{eq:BV012par}
\end{align}
resulting in the dispersive bound
\begin{align}
 \sum_{k=0}^{K-1}\left( (\beta_k^{(V,0)})^2 + (\beta_k^{(V,1)})^2 
 + (\beta_k^{(V,2)})^2 \right) <1 \,.
\end{align}


\subsection{The coefficients $\chi_I^X(n)$}

In Table~\ref{tab:chiIX} we summarize the numerical result of our calculation of
the various coefficients $\chi_I^X(n)$, which enter the functions $\phi_I^X(t)$
in the SE. 
We quote individual numbers for the perturbative
LO and NLO results, as well as from the condensate contributions,
for two different values of light-quark masses, $m_q=m_d$
and $m_q=m_s$. Also the number of subtractions is indicated.  Details of the calculation as 
well as analytical formulas can be found in Appendix~\ref{app:chiIX}.
As can be observed from Table~\ref{tab:chiIX}, the NLO perturbative corrections
are essential for a reliable estimate for the coefficients $\chi_I^X(n)$, while
the quark condensate gives only small contributions, and the gluon condensate
and the mixed quark-gluon condensate are negligible.

\begin{table}[tpbh]
\caption{\label{tab:chiIX} Summary of OPE results for the coefficients
$\chi_I^X(n)$. The following parameter values have been used
\cite{Bernard:2007ps,Williams:2007ey,Shifman:1978bx,Polyakov:1996kh}:
$\mu=m_b=4.2$~GeV,
 $m_d=4.8$~MeV,
 $m_s=104$~MeV,
 $\alpha_s=0.2185$, $\left\langle {\bar dd} \right\rangle=(278~\rm{MeV})^3 $,
  $\left\langle {\bar ss} \right\rangle=0.8 \left\langle {\bar dd} \right\rangle$,
 $\left\langle {\frac{\alpha_s}{\pi} \, G^2} \right\rangle  =0.038~\rm{GeV}^4$,
 $\left\langle {\bar qGq} \right\rangle=(1.4~\rm{GeV})^2 \left\langle {\bar qq} \right\rangle $.}

\begin{center}
\begin{tabular}{c|c|c|ccccc|cr}
\hline \hline \T \B
$q$
& Correlator
& Subtractions
& LO 
& NLO 
& $\left\langle {\bar qq} \right\rangle $ 
& $\left\langle {\frac{\alpha}{\pi} G^2} \right\rangle $ 
& $\left\langle {\bar qGq} \right\rangle$ & $\Sigma$  
\\[0.1em] \hline\hline 
 \T& $100 \times m_b^2 \chi^S$ & 2 & $1.265$
 & $0.589$
 & $0.029$
 & $0.001$
 & $-0.003$
 & $1.88$
 \\[0.1em] 
 &  $100 \times m_b^2\chi^P$ & 2 & $1.268$
 & $0.590$
 & $0.029$
 & $0.001$
 & $-0.003$
 & $1.88$
 \\[0.1em] 
 & $100 \times  \chi^V_L$ & 1 & $1.262$
 & $0.211$
 & $0.029$
 & $0.001$
 & $-0.003$
 & $1.50$
 \\[0.1em] 
 $d$ & $100 \times \chi^A_L$ & 1 & $1.271$
 & $0.205$
 & $0.029$
 & $0.001$
 & $-0.003$
 & $1.50$
 \\[0.1em] 
  & $100 \times m_b^2 \chi^V_T$ & 2 & $0.951$
 & $0.236$
 & $-0.029$
 & $-0.001$
 & $0.007$
 & $1.16$
 \\[0.1em] 
 & $100 \times m_b^2 \chi^A_T$ & 2 & $0.948$
 & $0.237$
 & $-0.029$
 & $-0.001$
 & $0.007$
 & $1.16$
 \\[0.1em] 
 & $100 \times m_b^2 \chi^T_T$ & 3 & $2.539$
 & $0.579$
 & $-0.029$
 & $-0.000$
 & $0.008$
 & $3.10$
 \\[0.1em] 
 \B & $100 \times m_b^2 \chi^{AT}_T$ & 3 & $2.527$
 & $0.586$
 & $-0.029$
 & $-0.001$
 & $0.008$
 & $3.09$
 \\ \hline
\hline
 \T & $100 \times m_b^2 \chi^S$ & 2 & $1.233$
 & $0.571$
 & $0.024$
 & $0.001$
 & $-0.003$
 & $1.83$
 \\[0.1em] 
 &  $100 \times m_b^2\chi^P$ & 2 & $1.296$
 & $0.608$
 & $0.022$
 & $0.001$
 & $-0.003$
 & $1.93$
 \\[0.1em] 
 & $100 \times \chi^V_L$ & 1 & $1.172$
 & $0.229$
 & $0.023$
 & $0.000$
 & $-0.003$
 & $1.42$
 \\[0.1em] 
 $s$ & $100 \times \chi^A_L$ & 1 & $1.361$
 & $0.187$
 & $0.023$
 & $0.002$
 & $-0.003$
 & $1.57$
 \\[0.1em] 
  & $100 \times m_b^2 \chi^V_T$ & 2 & $0.980$
 & $0.237$
 & $-0.022$
 & $0.000$
 & $0.005$
 & $1.20$
 \\[0.1em] 
 & $100 \times m_b^2 \chi^A_T$ & 2 & $0.916$
 & $0.238$
 & $-0.024$
 & $-0.002$
 & $0.006$
 & $1.13$
 \\[0.1em] 
 & $100 \times m_b^2 \chi^T_T$ & 3 & $2.652$
 & $0.569$
 & $-0.023$
 & $0.001$
 & $0.006$
 & $3.21$
 \\[0.1em]
 \B & $100 \times m_b^2 \chi^{AT}_T$ & 3 & $2.404$
 & $0.603$
 & $-0.024$
 & $-0.002$
 & $0.007$
 & $2.99$
\\
\hline
 \hline
\end{tabular}
\end{center}

\end{table}

\section{Form Factor Fits to Theoretical Data}

\subsection{Theory Input from Lattice and LCSR}
\label{sec:LatticeLCSR}

There are two main methods to obtain theoretical predictions on 
$B$-meson decay FFs: Light-cone sum rules (LCSRs) and Lattice QCD. 
As mentioned in the introduction, these techniques are largely complementary, as they perform best in different regimes of momentum transfer $q^2$. It is worth mentioning that certain decays, e.g.\ decays to unstable hadrons, are more challenging in Lattice QCD, and in some cases only quenched results exist for a subset of the FFs. 
On the other hand, LCSRs provide results for all decay channels considered 
in this work, including the complete set of seven FFs for $B \to K^*$ and $B_s\to\phi$, which
so far have not been fully addressed by Lattice QCD. However, as LCSR results are only valid in the low--$q^2$ regime, in these cases further theoretical information is needed to extrapolate
to large values of $q^2$, as will be discussed in the following subsections.

In our analysis, we will use the LCSR predictions from  Refs.~\cite{Ball:2004ye} and \cite{Ball:2004rg},
taking 3(4) points at low values of $q^2$ as input, see Table~\ref{tab:LCSRdata}. 
The quoted errors are extrapolated from the value quoted for $q^2=0$ in the references specified in the Table. Lattice data is available for $B\to \rho$ and $B\to K$ decays, and is as shown in Table~\ref{tab:Lattice}.\footnote{ We are very grateful to Sara Collins of the QCDSF collaboration for providing us with specific values for $B\to K$.} For those data points which have an asymmetric statistical or systematic error, in order to perform the fit, we take the FF to be the central value in this statistical or systematic range, and take half the range to be the statistical or systematic error \cite{D'Agostini:2004yu}.

For $B\to \rho$ and $B\to K$ decays, we use LCSR and Lattice data to interpolate between the low and high--$q^2$ region. The result can be compared to the case where we extrapolate to the high--$q^2$ region only on the basis of LCSR predictions. 
This procedure will
gives us an idea about the confidence in the extrapolations for those cases 
where Lattice data is lacking. 

\begin{table}[tpbh]

\caption{\label{tab:LCSRdata}
Overview of LCSR points used, transformed to the helicity amplitude basis.}

\begin{center}
\begin{tabular}{l|l|cccc|l}
\hline\hline 
\T \B Decay & FF & \multicolumn{4}{c|}{LCSR/$q^2$ (GeV${}^2$)} & Ref.\\
\hline
\hline
\T\B $B \to K$ &$q^2$ &3 & 6 & 9 &12 & Table 3, \cite{Ball:2004ye}\\
\hline 
\T&$\mathcal{A}_{V,0}$ & $0.40\pm 0.05$ & $0.48\pm 0.06$ & $0.59\pm 0.07$ &- &\\
&$\mathcal{A}_{V,t}$ & $0.40\pm 0.05$ & $0.51\pm 0.06$ & $0.65\pm 0.08$ & - &\\
\B&$\mathcal{A}_{T,0}$ & $0.13\pm 0.01$ & $0.22\pm 0.02$ & $0.34\pm 0.03$ & - &\\
\hline 
\T\B $B \to \rho$& $q^2$ &3 & 6 & 9 & 12 & Table 8, \cite{Ball:2004rg}\\
\hline 
\T&$\mathcal{B}_{V,0}$& $0.37\pm 0.12$ & $0.46\pm 0.13$ & $0.60\pm 0.14$ &-&\\
&$\mathcal{B}_{V,1}$& $0.16\pm 0.01$ & $0.27\pm 0.02$ & $0.41\pm 0.04$ &-&\\
&$\mathcal{B}_{V,2}$& $0.16\pm 0.02$ & $0.29\pm 0.03$ & $0.46\pm 0.04$ &-&\\
&$\mathcal{B}_{V,t}$& $0.37\pm 0.04$ & $0.46\pm 0.04$ & $0.58\pm 0.06$ &-&\\
&$\mathcal{B}_{T,0}$&$0.17\pm 0.35$&$0.3\pm 0.26$&$0.47\pm 0.23$&$0.71\pm 0.22$&\\
&$\mathcal{B}_{T,1}$&$0.45\pm 0.04$&$0.55\pm 0.05$&$0.69\pm 0.06$&$0.9\pm 0.08$&\\
\B&$\mathcal{B}_{T,2}$& $0.46\pm 0.04$&$0.58\pm 0.05$&$0.76\pm 0.07$&$1.0\pm 0.1$&\\

\hline
\T\B $B \to K^*$& $q^2$ & 3 & 6 & 9 &12 &  Table 8, \cite{Ball:2004rg}\\
\hline 
\T&$\mathcal{B}_{V,0}$& $0.45\pm 0.13$ & $0.56\pm 0.13$ & $0.73\pm 0.15$ &-&\\
&$\mathcal{B}_{V,1}$ & $0.19\pm 0.02$ & $0.32\pm 0.03$ & $0.49\pm 0.04$ &-&\\
&$\mathcal{B}_{V,2}$ & $0.20\pm 0.02$ & $0.35\pm 0.03$ & $0.57\pm 0.06$ & -&\\
&$\mathcal{B}_{V,t}$ & $0.44\pm 0.04$ & $0.54\pm 0.05$ & $0.67\pm 0.06$ & -&\\
&$\mathcal{B}_{T,0}$& $0.23\pm 0.36$ & $0.39\pm 0.27$ & $0.60\pm 0.24$ & $0.90\pm 0.22$&\\
&$\mathcal{B}_{T,1}$&$0 .59\pm 0.06$&$0 .72\pm 0.07$&$0 .89\pm 0.08$&$1 .2\pm 0.1$&\\
\B&$\mathcal{B}_{T,2}$&$0.61\pm 0.06$&$0.77\pm 0.07$&$1.0\pm 0.1$&$1.4\pm 0.1$&\\
\hline 
\T\B $B_s \to \phi$& $q^2$ & 3 & 6 & 9 &12 &  Table 8, \cite{Ball:2004rg}\\
\hline
\T&$\mathcal{B}_{V,0}$&$0.55\pm 0.12$&$0.68\pm 0.13$&$0.85\pm 0.14$& -&\\
&$\mathcal{B}_{V,1}$&$0.2\pm 0.02$&$0.34\pm 0.03$&$0.52\pm 0.04$& -&\\
&$\mathcal{B}_{V,2}$&$0.21\pm 0.02$&$0.38\pm 0.04$&$0.62\pm 0.06$&-&\\
&$\mathcal{B}_{V,t}$&$0.56\pm 0.04$&$0.68\pm 0.05$&$0.85\pm 0.06$& -&\\
&$\mathcal{B}_{T,0}$&$0.26\pm 0.39$&$0.44\pm 0.29$&$0.67\pm 0.26$&$1.0\pm 0.3$&\\
&$\mathcal{B}_{T,1}$& $0.59\pm 0.06$ & $0.72\pm 0.07$ & $0.89\pm 0.08$ & $1.2\pm 0.1$&\\
\B&$\mathcal{B}_{T,2}$& $0.61\pm 0.06$ & $0.77\pm 0.07$ & $1.0\pm 0.1$ & $1.4\pm 0.1$&\\
\hline \hline 
\end{tabular}
\end{center}
\end{table}

\begin{table}[tpbh]

\caption{\label{tab:Lattice}
Overview of Lattice points used, transformed to the helicity amplitude basis. Note that specific values for $B\to\rho$ are as in Table 2 of Ref.~\cite{Flynn:2007ii}.}

\begin{center}
\begin{tabular}{c|c|ccc|c}
\hline\hline 
\T \B Decay &  $q^2$ (GeV${}^2$)& \multicolumn{3}{c|}{FF}& Ref. \\
\hline\hline
\T \B  $B\to K$ && $\mathcal{A}_{V,0}$ & $\mathcal{A}_{V,t}$ & $\mathcal{A}_{T,0}$ & QCDSF \cite{AlHaydari:2009zr}\\
 \hline 
\T&14.5&$0.94\pm 0.19$&$1.1\pm 0.2$&-&\\
&15.6&$1.1\pm 0.2$&$1.3\pm 0.3$&-&\\
&16.7&$1.2\pm 0.2$&$1.5\pm 0.3$&-&\\
&17.9&$1.4\pm 0.3$&$1.8\pm 0.3$&-&\\
&19.&$1.6\pm 0.3$&$2.3\pm 0.4$&-&\\
&20.1&$1.9\pm 0.4$&$3.\pm 0.6$&-&\\
&21.3&$2.3\pm 0.4$&$4.4\pm 0.8$&-&\\
\B &22.4&$2.9\pm 0.6$&$8.7\pm 1.7$&-&\\
\hline
\T \B  $B\to \rho$ & & $\mathcal{B}_{V,0}$ & $\mathcal{B}_{V,1}$ & $\mathcal{B}_{T,2}$ &UKQCD \cite{Bowler:2004zb}\\
 \hline 
\T &12.7&$0.64\pm 0.78$&$0.34\pm 0.27$&$0.9\pm 0.18$&\\
&13.&$0.71\pm 0.72$&$0.39\pm 0.25$&$0.96\pm 0.18$&\\
&13.5&$0.8\pm 0.66$&$0.48\pm 0.22$&$1.1\pm 0.2$&\\
&14.&$0.9\pm 0.62$&$0.58\pm 0.19$&$1.2\pm 0.2$&\\
&14.5&$1.0\pm 0.6$&$0.68\pm 0.16$&$1.3\pm 0.2$&\\
&15.&$1.1\pm 0.6$&$0.78\pm 0.15$&$1.4\pm 0.2$&\\
&15.5&$1.3\pm 0.7$&$0.89\pm 0.15$&$1.6\pm 0.2$&\\
&16.&$1.4\pm 0.8$&$1.0\pm 0.2$&$1.8\pm 0.2$&\\
&16.5&$1.6\pm 0.9$&$1.2\pm 0.3$&$2.1\pm 0.2$&\\
&17.1&$1.8\pm 1.2$&$1.4\pm 0.4$&$2.4\pm 0.2$&\\
&17.6&$2.1\pm 1.5$&$1.7\pm 0.6$&$2.7\pm 0.3$&\\
 \B &18.2&$2.5\pm 2.$&$2.1\pm 0.9$&$3.3\pm 0.3$&\\
\hline\hline
\end{tabular}
\end{center}
\end{table}

\subsection{Parameterisation of FFs as Series Expansion}

For those channels where where Lattice data is not available, it is essential
to employ a FF parameterisation that takes into account the characteristic
features of the FF shape as determined from the analyticity and unitarity
consideration above. For every considered FF, we will therefore define a
parameterisation based on the SE,
\begin{align}
{\cal A}_{V,0}(t) &= \frac{1}{B(t) \, \phi_T^V(t)} \, \sum_{k=0}^{K-1} \alpha^{(V,0)}_k \, z^k \,,
\cr 
{\cal A}_{V,t}(t) &= \frac{1}{B(t) \, \sqrt{z(t,t_-)} \, \phi_L^V(t)} \, \sum_{k=0}^{K-1} \alpha^{(V,t)}_k \, z^k \,,
\cr 
 {\cal A}_{T,0}(t) &= \frac{\sqrt{-z(t,0)}}{B(t) \, \phi_T^T(t)} \, \sum_{k=0}^{K-1} \alpha^{(T,0)}_k \, z^k \,,
\end{align}
and
\begin{align}
{\cal B}_{V,t}(t) &= \frac{1}{B(t) \, \phi_L^{V-A}(t)} \, \sum_{k=0}^{K-1} 
  \beta^{(V,t)}_k \, z^k \,,
\cr 
{\cal B}_{T,0}(t) &= \frac{\sqrt{-z(t,0)}}{B(t) \,\sqrt{z(t,t_-)} \, \phi_T^{T+A_T}(t)} \, \sum_{k=0}^{K-1} 
  \beta^{(T,0)}_k \, z^k \,,
\cr 
{\cal B}_{T,1}(t) &= \frac{1}{B(t) \, \phi_T^{T+A_T}(t)} \, \sum_{k=0}^{K-1} 
  \beta^{(T,1)}_k \, z^k \,,
\cr 
{\cal B}_{T,2}(t) &= \frac{1}{B(t) \,\sqrt{z(t,t_-)} \, \phi_T^{T+A_T}(t)} \, \sum_{k=0}^{K-1} 
  \beta^{(T,2)}_k \, z^k \,,
\end{align}
and ${\cal B}_{V,0-2}$ already given in (\ref{eq:BV012par}).
Here we have used our FF convention defined in (\ref{eq:AVdef},\ref{eq:ATdef},\ref{eq:BVdef},\ref{eq:BTdef}) and 
explicitly quoted the pre-factors, necessary to obtain the 
correct analytical behaviour of our FFs.\footnote{In Refs.~\cite{Becher:2005bg} and \cite{Caprini:1997mu,Bourrely:2008za}, the predictions from perturbative QCD 
for the scaling of the FFs at 
large values of $q^2$ have been used as an additional 
constraint on the shape of the FFs. We have found that these constraints do not 
influence the FF fits in the decay region significantly. As the asymptotic behaviour
of exclusive observables in QCD is still a matter of controversy, we therefore
find it safer and simpler not to include these constraints in our parameterisation.} 
In our fits below, we will
find that in general the SE can be truncated after the first two terms,
i.e.\ the parameter $K$ can be set to 2. We should point out, however,
that this does not necessarily imply that higher-order terms in the SE are
negligible: Although $|z|^2 \ll 1$ in the semi-leptonic region, one may still
have $ |\alpha_2 \, z^2| \sim |\alpha_1 z|$ if the coefficients satisfy $\alpha_1 \ll \alpha_2 \lesssim 1$. 
From the theoretical point of view, this 
reflects an irreducible source of uncertainty, which we will discuss in some detail for some specific examples
in Sec.~\ref{sec:BK}. 
From the practical point of view, we consider the truncated SE 
as a reasonable parameterisation which is easy to implement (and easy to refine) in 
phenomenological studies. 

For simplicity, we will not explicitly implement the theoretical relations
(\ref{eq:AVat0},\ref{eq:BTat0}), that some of the FFs fulfil at $q^2=0$,
into the fit, because they are automatically satisfied by the rather precise
input from LCSR at this point. However, the helicity-based FF definition 
further implies a relation between the FFs ${\cal B}_{V,0}$ and 
${\cal B}_{V,2}$, and similarly between ${\cal B}_{T,0}$ and 
${\cal B}_{T,2}$, see (\ref{eq:addrel}) in the appendix, which we will
implement as an additional constraints on the corresponding coefficients
in the SE.
From the above parameterisations, the SSE is obtained by the replacements
\begin{align}
 \phi_I^X(t) & \to 1 \,,
\quad 
 B(t)  \to P(t) \,,
\quad
 \sqrt{-z(t,0)}  \to \sqrt{q^2}/m_B \,,
\quad  
 \sqrt{z(t,t_-)}  \to \sqrt{\lambda}/m_B^2 \,,
\end{align}
with new coefficients $\tilde \alpha_k$ and $\tilde \beta_k$.

\subsubsection{Unitarity constraints}

For the SE parameterisation, the unitarity  constraints take the form
\begin{align}
& \sum_{k=0}^{K-1}\,(\alpha_k)^2 \leq 1 \quad \mbox{for ${\cal A}_{V,0}$ and ${\cal A}_{T,0}$,}
\qquad 
 3 \sum_{k=0}^{K-1}\,(\alpha_k)^2 \leq 1 \quad \mbox{for  ${\cal A}_{V,t}$,}
\label{eq:aconstrSE}
\end{align}
and 
\begin{align}
3 &\sum_{k=0}^{K-1}\,(\beta_k^{(V,t)})^2 \leq 1 \quad \mbox{for ${\cal B}_{V,t}$,}
\cr 
& \sum_{k=0}^{K-1}\left\{ (\beta_k^{(V,0)})^2 +(\beta_k^{(V,1)})^2+(\beta_k^{(V,2)})^2
 \right\}
  \leq 1 \quad \mbox{for ${\cal B}_{V,0}$, ${\cal B}_{V,1}$, and ${\cal B}_{V,2}$,}
\cr 
& \sum_{k=0}^{K-1}\left\{ (\beta_k^{(T,0)})^2 +(\beta_k^{(T,1)})^2+(\beta_k^{(T,2)})^2
 \right\}
  \leq 1 \quad \mbox{for ${\cal B}_{T,0}$, ${\cal B}_{T,1}$, and ${\cal B}_{T,2}$.}
\label{eq:bconstrSE}
\end{align}

For the SSE parameterisation, imposing the unitarity bound is more complicated, as shown  in Ref.~\cite{Bourrely:2008za}. We repeat the derivation of this bound in order to define notation used later. One first compares the SE and SSE parameterisations:
\begin{equation}
 \sum_{k=0}^{K-1}\,\alpha_k\,z^k= \Lambda(z)\sum_{k=0}^{K-1} \tilde\alpha_k\, z^k
\end{equation}
One can simply obtain $\Lambda(z)$ by combining the prefactors from the SE expansion
with the prefactors from the SSE expansion, and expressing the result as a function of $z(t,t_0)$. Since $z$ is a small parameter, we can expand $\Lambda(z)$ in powers of $z$:
\begin{equation}
\Lambda(z)=\sum_k \,\zeta_k\,z^k.
\end{equation}
We can therefore obtain a relation between the coefficients $\alpha_k$ and $\tilde\alpha_k$,
\begin{equation}
 \alpha_i=\sum_{k=0}^{\mathrm{min}[K-1,i]} \zeta_{i-k} \, \tilde\alpha_k \,,
\qquad 0\leq i\leq K-1 \,,
\end{equation}
which results in bounds of the type
\begin{equation}
 \sum_{j,k=0}^{K-1} C_{jk} \, \tilde\alpha_j \, \tilde\alpha_k \leq 1\,,
\end{equation}
where
\begin{equation}
\label{eq:Cjk}
 C_{jk}=\sum_{i=0}^{K-1-\mathrm{max}[j,k]} \zeta_i\, \zeta_{i+|j-k|}
\end{equation}
is a positive definite matrix.

\subsection{Fitting prescription}
\label{sec:fit}
We perform a fit to the LCSR data, as well as, where possible, a combined fit to the LCSR and Lattice data, by minimising a $\chi^2$ function defined by
\begin{equation}
\chi ^2 (\vec \theta ) = 
\left( F_i  - F( t_i ,\vec \theta \,) \right)
\left[ V^{ - 1}  \right]_{ij} \!
\left( F_j  - F( t_j ,\vec \theta \,) \right) \,,
\end{equation}
where $\vec \theta$ contains the parameters of a given FF parameterisation,
$F_i$ are the FF values from LCSR/Lattice at given points $t_i$, 
and $V_{ij}$ are elements of the covariance matrix as defined below.

As explained above, we are going to investigate parameterisations based on
 two variants of the SE, where the parameters will be subject to additional constraints
derived from dispersive bounds on the FFs.
\begin{itemize}
 \item In the conventional series expansion (SE), we use (\ref{par:SE}), and 
  truncate the series after the first 2 terms, such that
$$
   \vec\theta= \{\alpha_0, \alpha_1 \} \,, \qquad \sum \alpha_i^2 \stackrel{!}{<} 1 \,.
$$
 \item The simplified series expansion (SSE) uses (\ref{par:SSE}), with
$$
   \vec\theta= \{\tilde\alpha_0, \tilde\alpha_1 \} \,, \qquad \sum_{i,j=0}^{1}
 C_{ij} \, \tilde\alpha_i \tilde\alpha_j  \stackrel{!}{<} 1 \,,
$$
where the matrix $C_{ij}$ is defined in (\ref{eq:Cjk}).
\end{itemize}

In constructing the covariance matrix, when we do a combined fit to LCSR and Lattice data, we assume the matrix to be block diagonal with independent blocks for Lattice and LCSR, equivalent to $\chi^2=\chi_{\rm LCSR}^2+\chi_{\rm Lat}^2$, where 
\begin{equation}
\chi_{\rm LCSR} ^2 (\vec \theta ) = 
\left( F_i  - F( t_i ,\vec \theta \,) \right)
\left[ V_{\rm LCSR}^{ - 1}  \right]_{ij} \!
\left( F_j  - F( t_j ,\vec \theta \,) \right) \,,
\end{equation}
and
\begin{equation}
\chi_{\rm Lat} ^2 (\vec \theta ) = 
\left( F_i  - F( t_i ,\vec \theta \,) \right)
\left[ V_{\rm Lat}^{ - 1}  \right]_{ij} \!
\left( F_j  - F( t_j ,\vec \theta \,) \right) \,,
\end{equation}

 We consider the statistical and systematic contributions to the Lattice errors separately. Where results were not available in the literature, we received the breakdown by private communication with the authors. 
In obtaining the covariance matrix, we make the following conservative assumptions:
\begin{itemize}
	\item Statistical errors of Lattice data are $50\%$ correlated \cite{AlHaydari:2009zr,Bowler:2004zb}.
	\item Systematic errors of Lattice data are $100\%$ correlated \cite{AlHaydari:2009zr,Bowler:2004zb}.
	\item Errors of LCSR data are due to parametric as well as to systematic uncertainties
              from different sources. In order to provide a concrete number for the $\chi^2$ value characeterizing the quality of
              the fit, we have estimated  the errors of LCSR data at different values $t_i$ to be $75\%$ correlated.\footnote{We have
 checked that using $90\%$ or $50\%$ correlation, instead --- of course --- globally changes the very value of $\chi^2$, but does not
  influence the optimal parameter values. A similar comment applies to the number of individual LCSR points used in the fit.}
\end{itemize}
This prescription leads to a covariance matrix $V^{ij}=\rm{cov}[t^i,t^j]$, containing
\begin{eqnarray}
V_{\rm LCSR}^{ij} & =& \frac{1}{4}\kappa^i \kappa^j \delta _{ij}+\frac{3}{4}\kappa^i\kappa^j \qquad \mathrm{and}\\
V_{\rm Lat}^{ij}  &=& \frac{1}{2}\sigma^i \sigma^j \delta _{ij}+\frac{1}{2}\sigma^i\sigma^j  + \varepsilon^i \varepsilon^j
\end{eqnarray}
where $\sigma_i$ are the statistical errors, $\varepsilon_i$ are the systematic errors for the Lattice data, and $\kappa_i$ are the errors for the LCSR predictions.
 
Minimising $\chi^2(\vec \theta)$ then yields the best fit parameters $\vec\theta^*$, 
as well as the covariance matrix of the fit, $U_{ij}=\rm{cov}[\theta_i,\theta_j]$,
\begin{equation}
\left( {U^{ - 1} } \right)_{ij}  
= \frac{1}{2}\left. {\frac{{\partial ^2 \chi ^2 (\vec \theta )}}{{\partial \theta _i \, \partial \theta _j }}} \right|_{\vec \theta  = \vec\theta^* } \,,
\label{eq:U}
\end{equation}
from which we calculate the error associated to the fitted FF function:
\begin{equation}
\Delta F(t,\vec\theta^* ) ={\left. {\frac{{\partial F(t,\vec \theta )}}{{\partial \theta _i }}} \right|_{\vec \theta  = \vec\theta^* } } \, U_{ij} \,  {\left. {\frac{{\partial F(t,\vec \theta )}}{{\partial \theta _j }}} \right|_{\vec \theta  = \vec\theta^* } } 
\end{equation}

\subsection{Results}\label{sec:Heavy-Light}

Having established the fitting procedure, we consider FFs for the decays $B\to \rho$, $B\to K$, $B\to K^*$ and $B_s\to \phi$. We concentrate on radiative and rare semi-leptonic decays, as previously the dispersive bounds had not been calculated for the tensor current, so could not be applied to these decays. The phenomenological motivations for studying the chosen decays are as follows. First, they involve flavour changing neutral currents via e.g.\ electroweak penguins, so they 
are particularly sensitive to new physics. Secondly, the di-lepton signature can easily be detected at the LHC, and the three-body or four-body final state (for 
subsequent decays  $K^*\to K\pi$ and $\phi\to KK$) involves many promising observables
related to various angular distributions  \cite{Altmannshofer:2008dz,Bobeth:2008ij,Bobeth:2007dw, Bharucha:2010bb, Egede:2008uy}. 

From the theoretical point of view, the $B\to V\gamma$ decay as well
as the low-$q^2$ region of $B \to L \,\ell^+\ell^-$ transitions allow for
a systematic inclusion of radiative corrections within the QCD factorization
approach at leading order in the $1/m_b$ expansion \cite{Beneke:2001at,Beneke:2004dp}.
In this region, the transition FFs (which still determine a major part  
of non-perturbative input) can be obtained from LCSR estimates alone.
As it has been discussed, for instance, in \cite{Grinstein:2004vb}, the
high-$q^2$ region may also be interesting in order to constrain NP contributions
(notably to the short-distance Wilson coefficients $C_9$ and $C_{10}$), and therefore
our extrapolations of LCSR results for the tensor FFs in that region
will be particularly relevant for this purpose.
In the following subsection, we present the results of fitting the specific FFs to both,
the SE and SSE parameterisations, using LCSR and Lattice data where appropriate 
as discussed in Section \ref{sec:LatticeLCSR}.
For the light meson masses, we use $m_K=494$~MeV, $m_{K^*}=892$~MeV, $m_\rho=776$~MeV,
$m_\phi=1.02$~GeV.

\FloatBarrier

\paragraph{\boldmath $B \to K$ form factors:\unboldmath}

\label{sec:BK}

\begin{figure}[tpbh]
\centering
\includegraphics[width=0.48\textwidth]{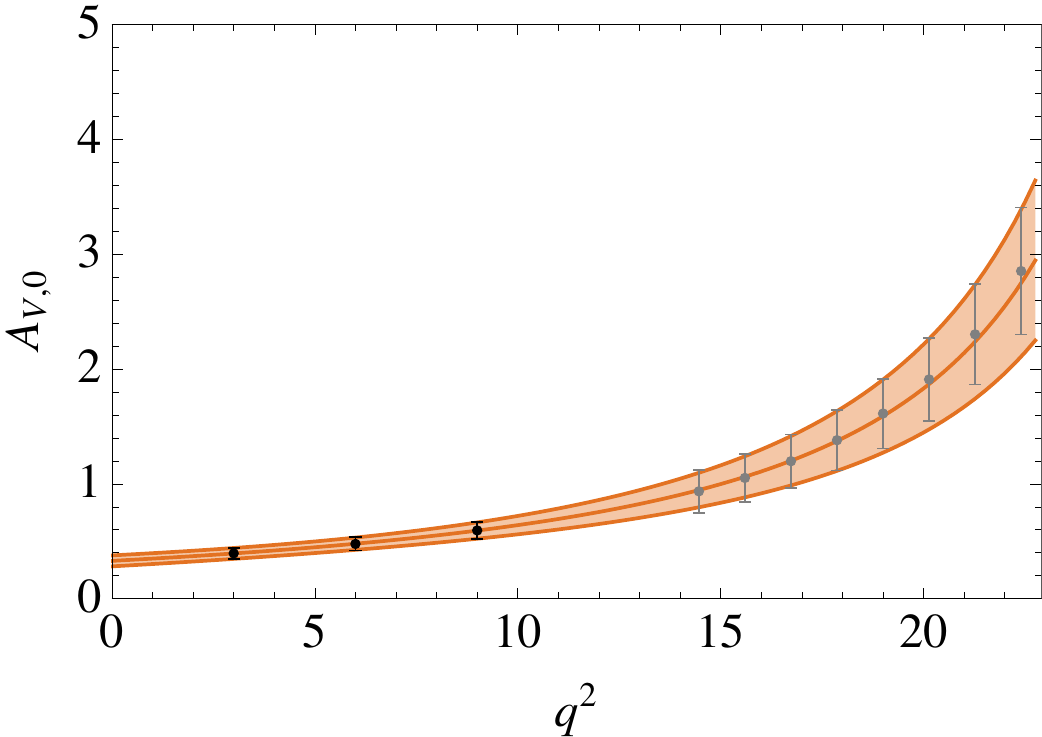}\hfill
\includegraphics[width=0.48\textwidth]{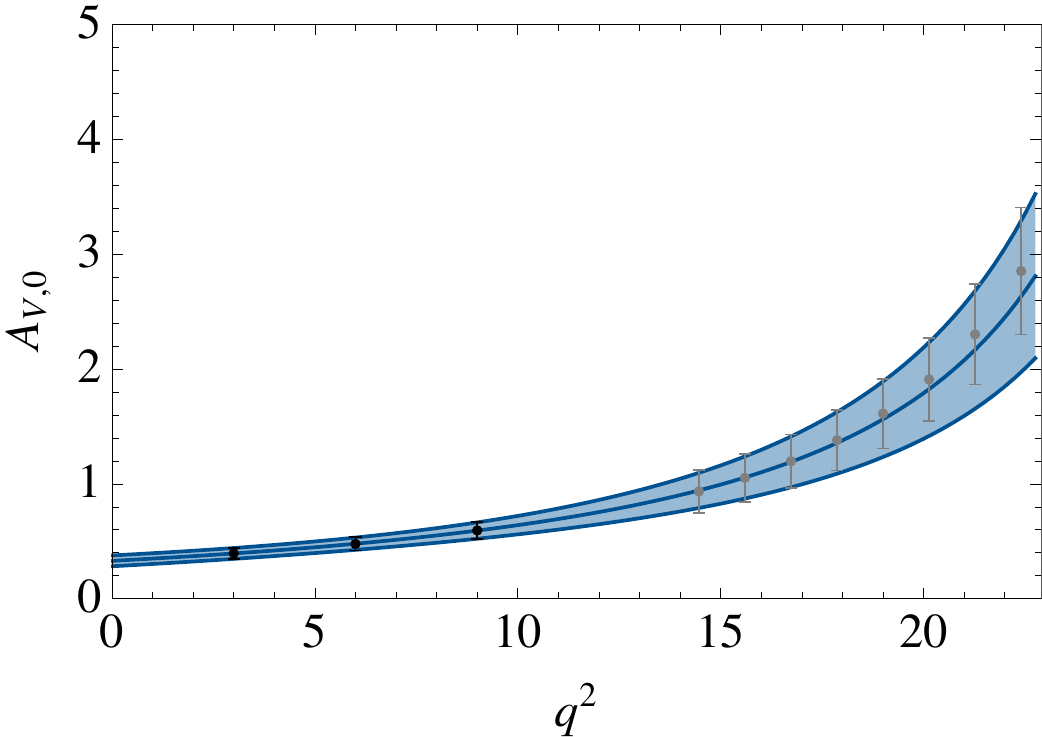}
 \includegraphics[width=0.48\textwidth]{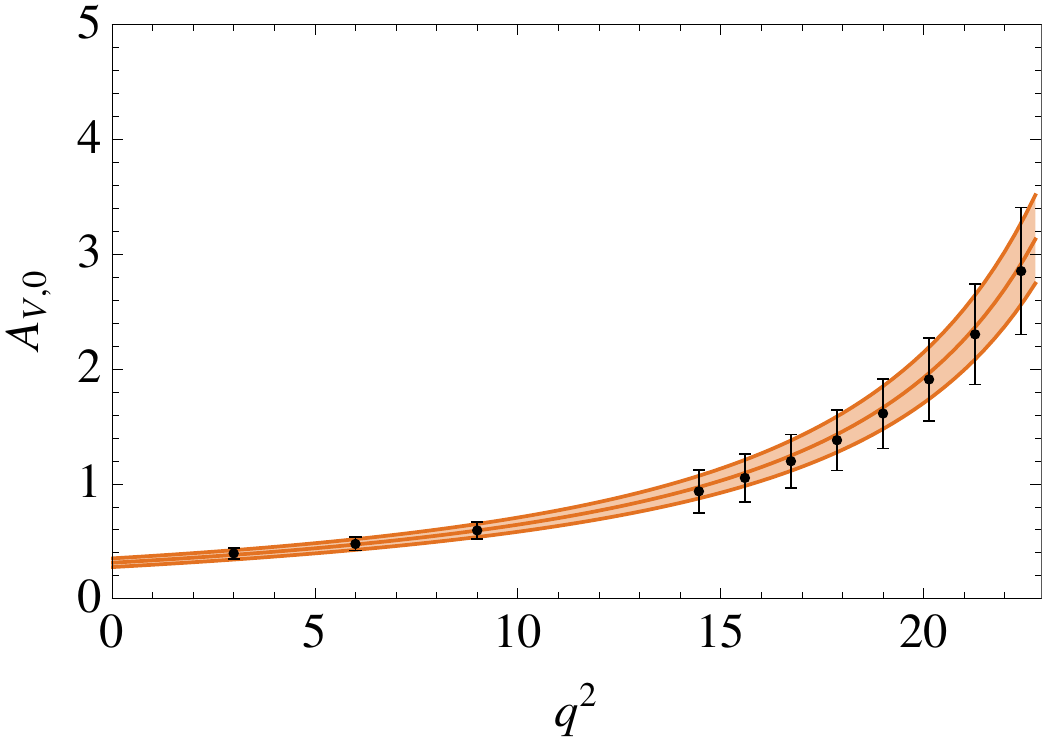}\hfill \includegraphics[width=0.48\textwidth]{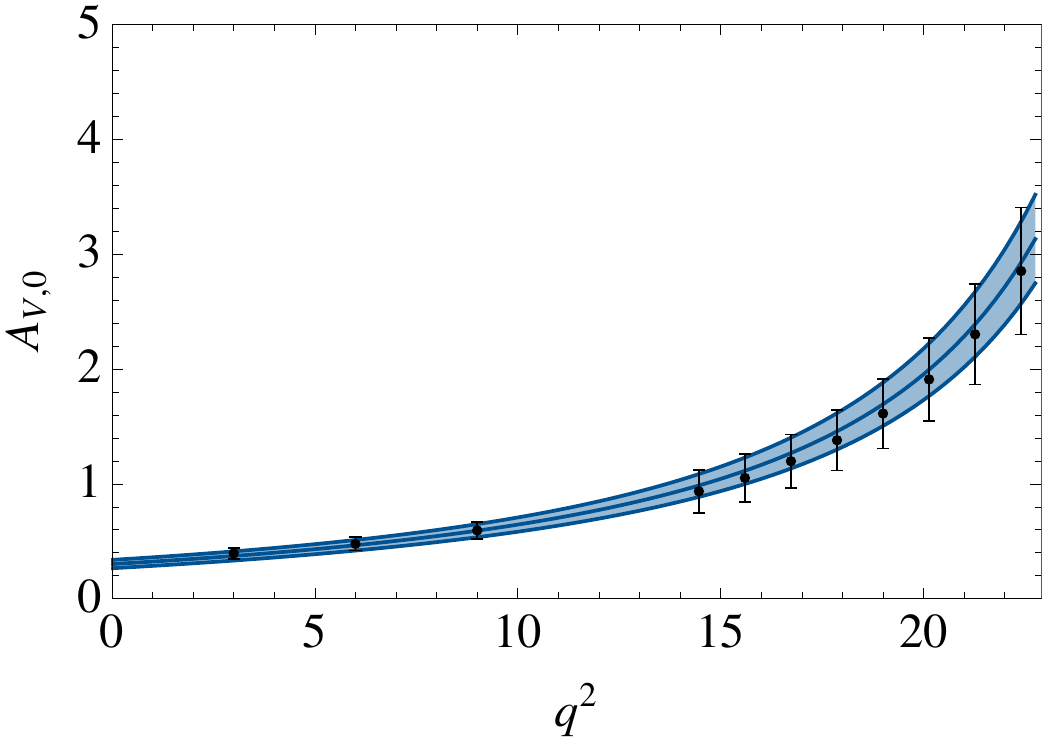}
\caption{$B \to K$: Fit of SE (left) and SSE (right) parameterisation to LCSR (top) and to LCSR and Lattice (bottom) for $\mathcal{A}_{V,0}$. The LCSR and Lattice data are shown by black points with error bars in the appropriate $q^2$ range.}
\label{fig:AV0K}
\end{figure}

\begin{figure}[tpbh]
\centering
\includegraphics[width=.48\textwidth]{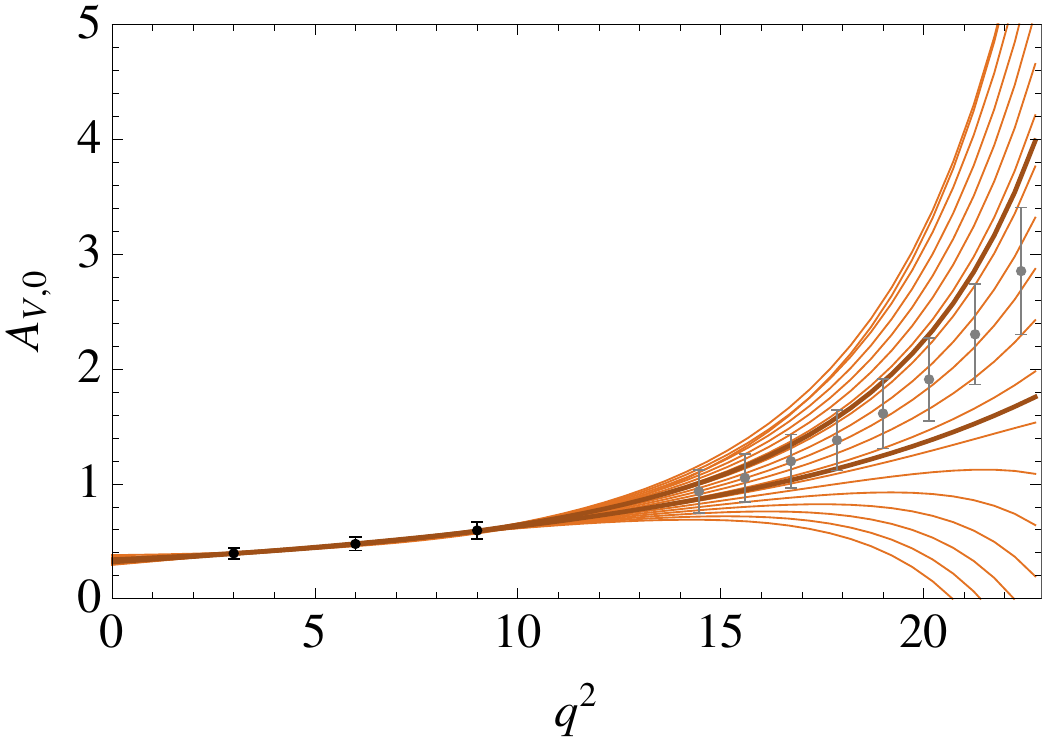} \hfill 
\includegraphics[width=.48\textwidth]{AV0KSE.pdf} 
\\
\includegraphics[width=.48\textwidth]{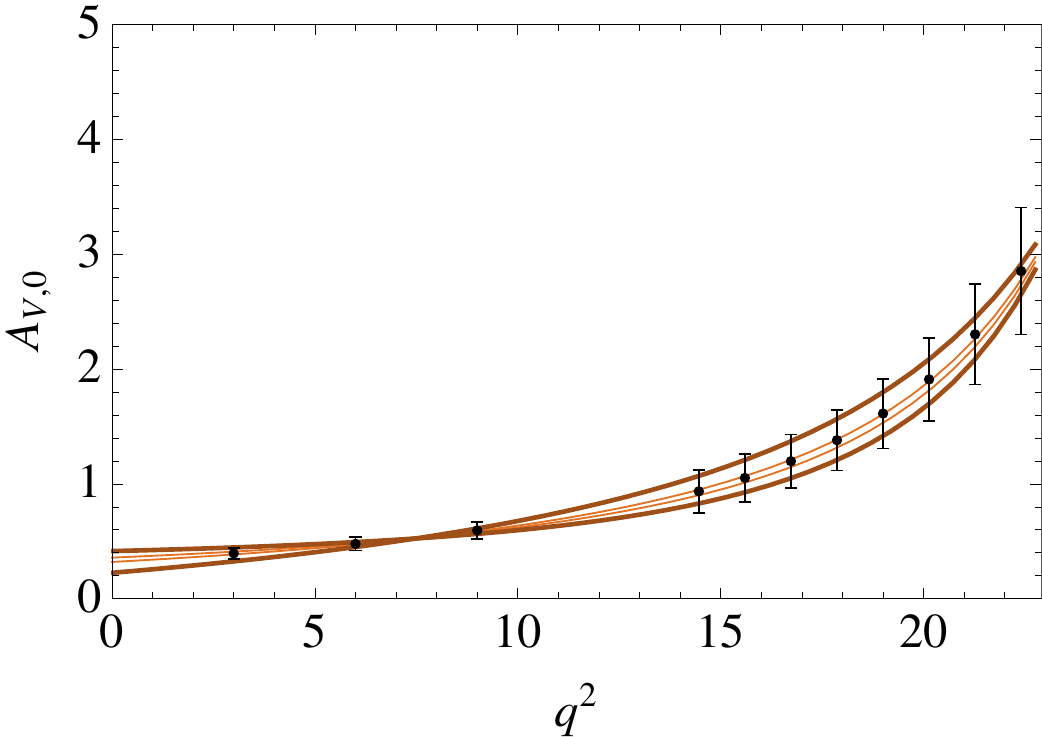} \hfill 
\includegraphics[width=.48\textwidth]{AV0KSELat.pdf} 
\caption{ 
$B \to K$: 
Fit of SE  parameterisation to LCSR or LCSR/Lattice for $\mathcal{A}_{V,0}$ 
with the parameter $a_2$ varied between $[-0.9,+0.9]$ (thin lines).
The thick, dark lines show $a_2 = \pm 0.25$.
For comparison, on the right we show again the corresponding fit and error 
estimate for a SE truncated after $a_1$.
}
\label{fig:a2comp}
\end{figure}

\begin{figure}[tpbh]
\centering
\includegraphics[width=0.48\textwidth]{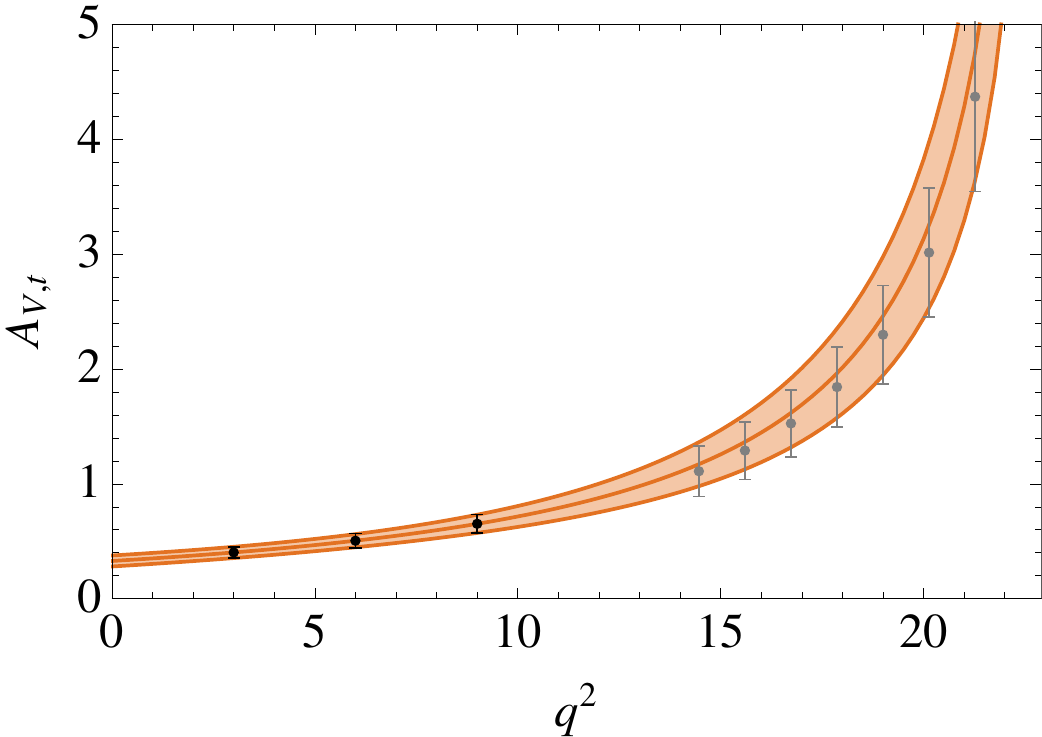}\hfill
\includegraphics[width=0.48\textwidth]{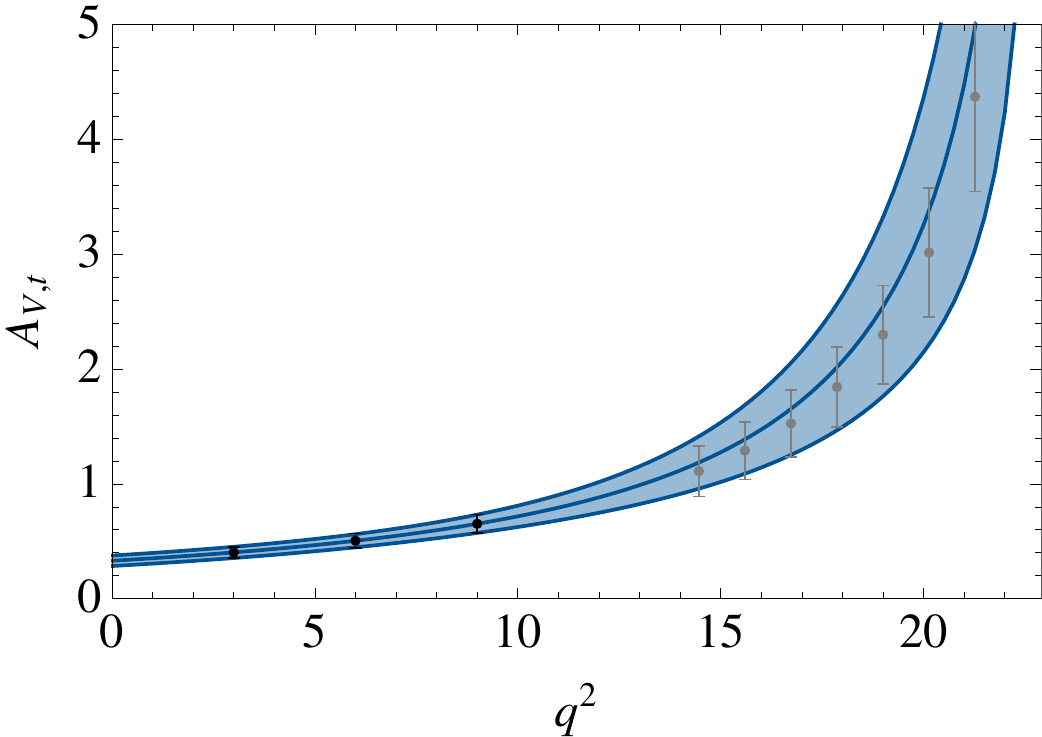} \includegraphics[width=0.48\textwidth]{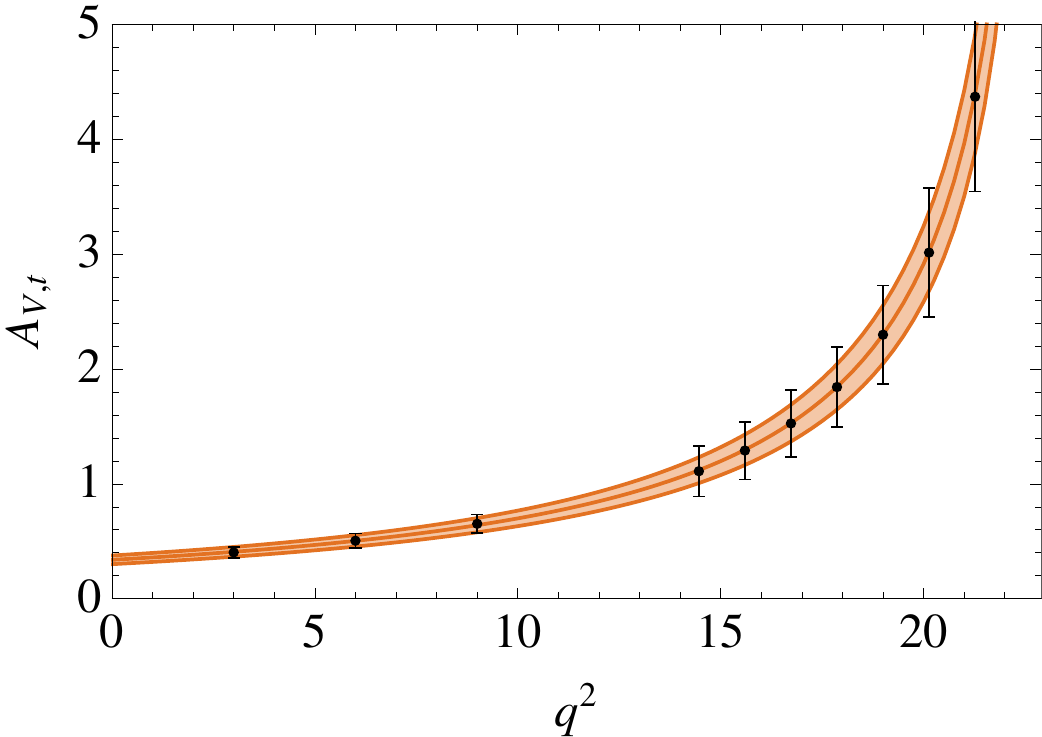}\hfill \includegraphics[width=0.48\textwidth]{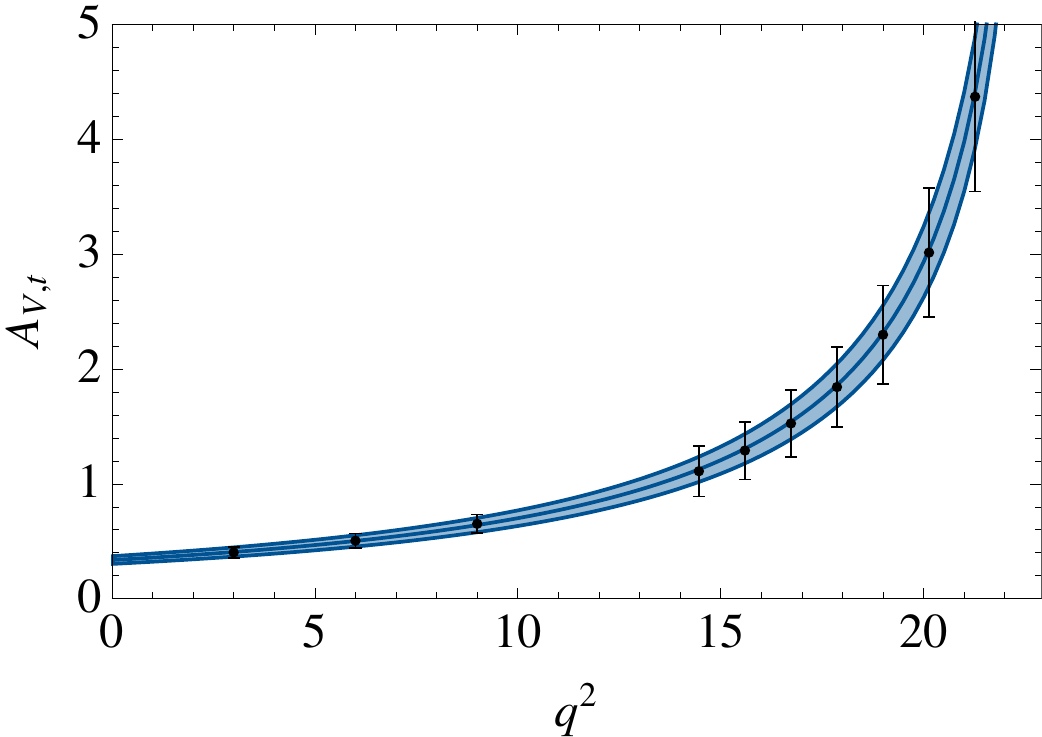}
\caption{$B \to K$: Fit of SE (left) and SSE (right) parameterisations to LCSR (top) and to LCSR and Lattice (bottom) for $\mathcal{A}_{V,t}$. The LCSR and Lattice data are shown by black points with error bars in the appropriate $q^2$ range.}
\label{fig:AVtK}
\end{figure}

\begin{figure}[tpbh]
\centering
\includegraphics[width=0.48\textwidth]{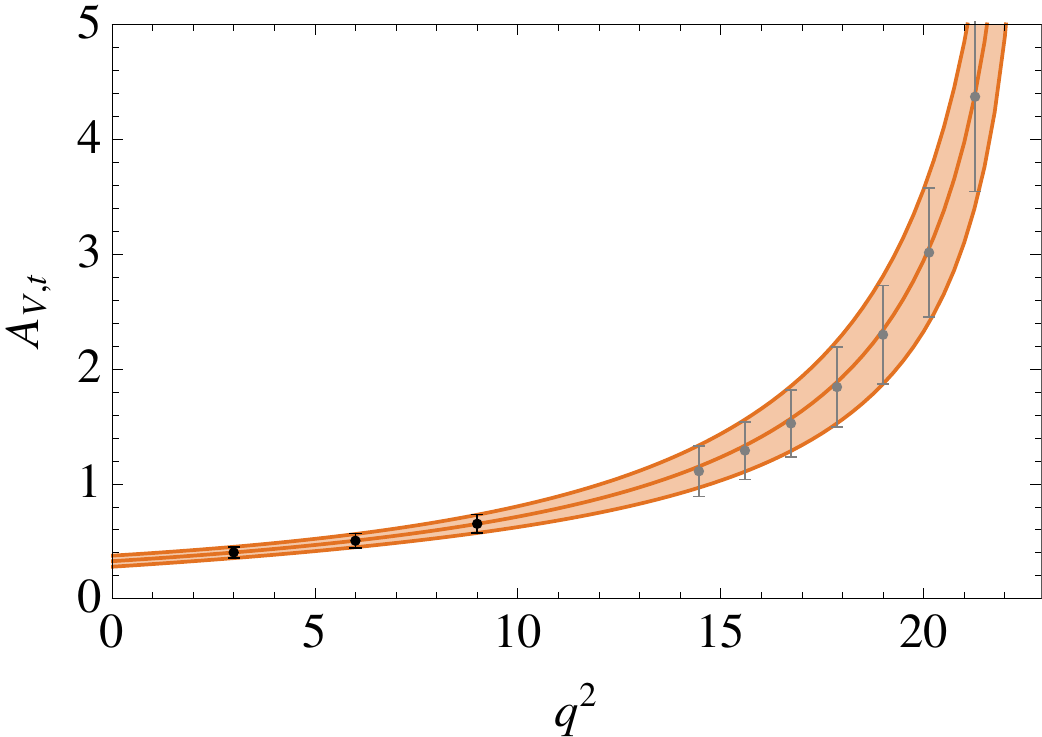}\hfill 
\includegraphics[width=0.48\textwidth]{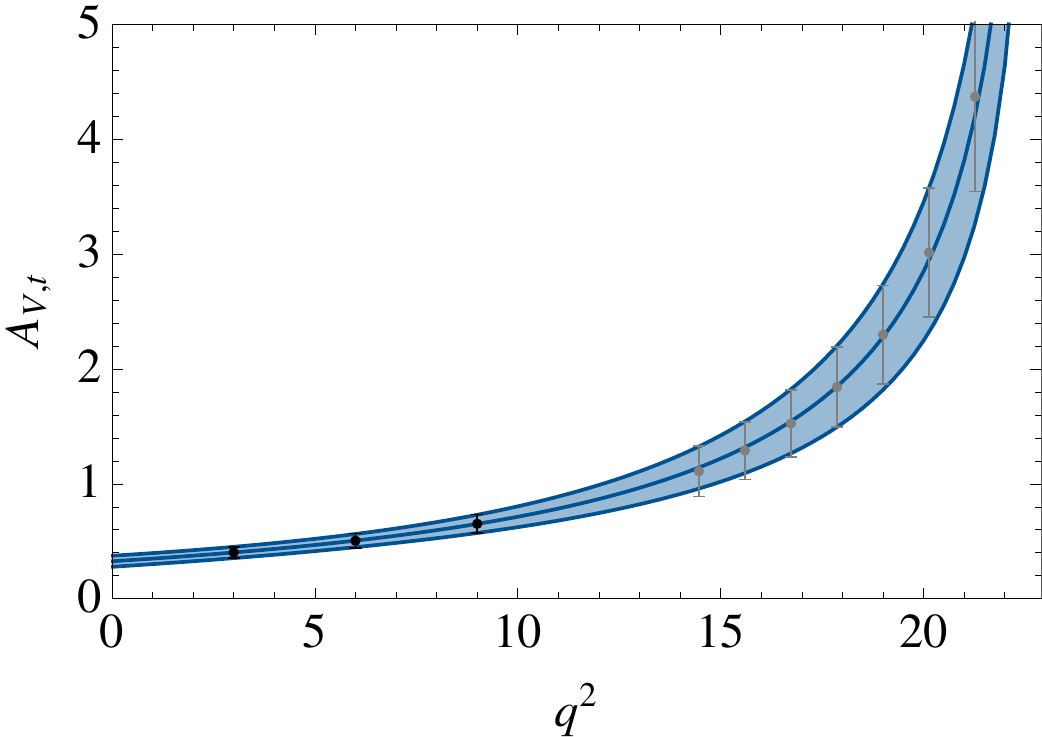}
\includegraphics[width=0.48\textwidth]{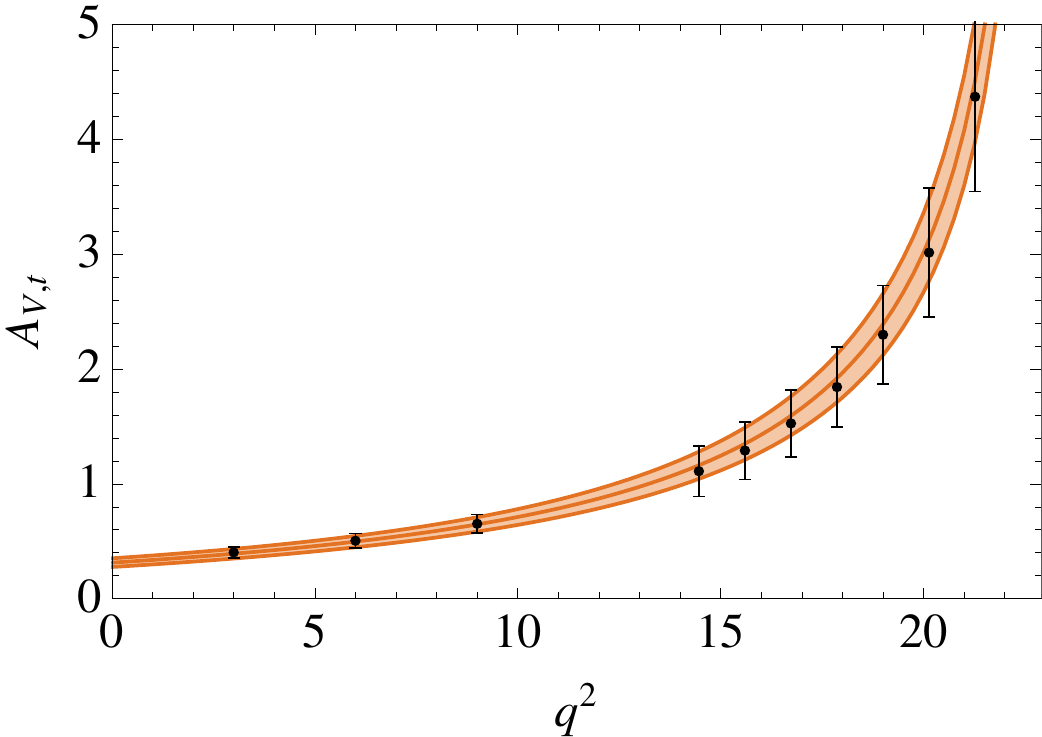}\hfill \includegraphics[width=0.48\textwidth]{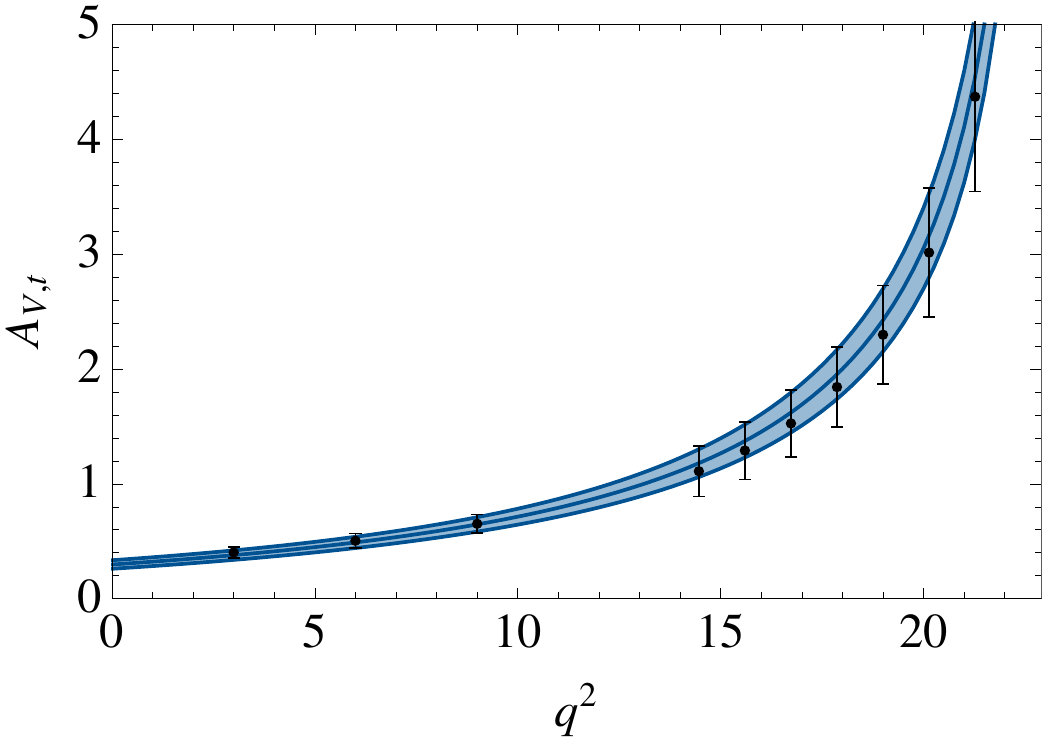}
\caption{$B \to K$: The same as Fig.~\ref{fig:AVtK} but without using
the scalar $B_s$ resonance in the fit ansatz.}
\label{fig:AVtKnores}
\end{figure}

\begin{figure}[tpbh]
\centering
\includegraphics[width=.48\textwidth]{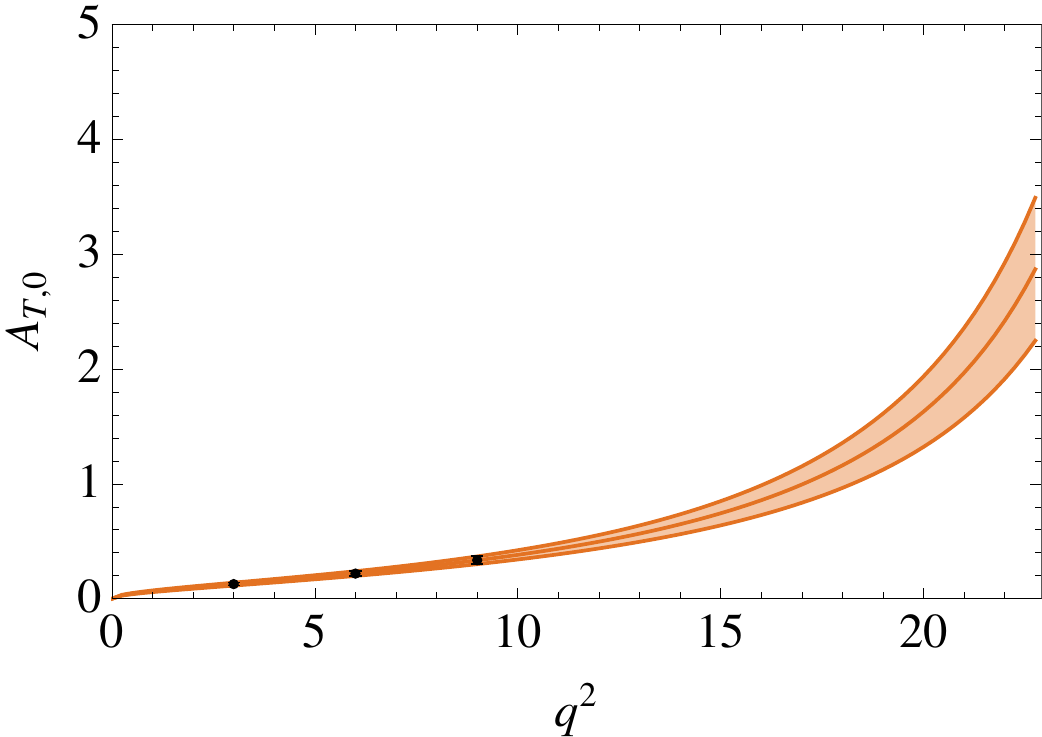} \hfill \includegraphics[width=.48\textwidth]{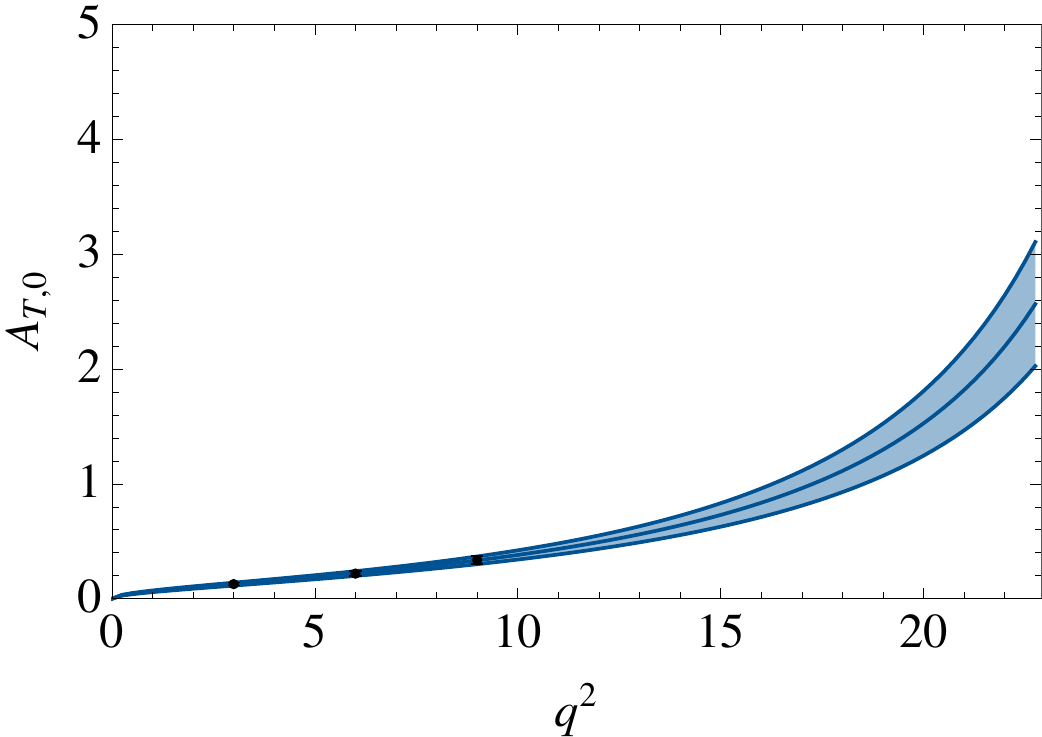} 
\caption{$B \to K$: Fit of SE (left) and SSE (right) parameterisations to LCSR for $\mathcal{A}_{T,0}$. The LCSR data is shown by black points with error bars.}
\label{fig:AT0K}
\end{figure}

In Figs.~\ref{fig:AV0K}--\ref{fig:AT0K}, we show the fit for the various $B\to K$ FFs,
which enter, for instance, the radiative $B\to K \ell^+\ell^-$ and $B\to K\nu\bar \nu$ decays.
We compare the result of the SE and SSE parameterisations using LCSR data, and investigate the changes when the Lattice data is included. 
The numerical results for the best-fit parameters of the SE and SSE fit are found in 
corresponding Tables~\ref{tab:AKSE}--\ref{tab:AKSSE}. The covariance matrices for these fits can also be found in Appendix~\ref{app:covmat}. 

Generally, both parameterisations are seen to fit the data well, and importantly, 
we find agreement with the Lattice predictions for $\mathcal{A}_{V,0}$ and  $\mathcal{A}_{V,t}$,
even when they are not included in the fit. We therefore consider our extrapolation of 
LCSR data for the tensor FF $\mathcal{A}_{T,0}$ to the high-$q^2$ region,
where Lattice data does not exist, as sufficiently reliable.
The quality of the fits is astonishingly good, considering
the $\chi^2$ values for only two free parameters in the expansion. The differences between
the SE and SSE are only marginal, which can be traced back to the usage of the optimised
value for the auxiliary parameter $t_0$ in (\ref{eq:t0opt}).  The dispersive bounds 
turn out to be far from being separated, and therefore they have only little impact on the
FF fit. This observation is in line with other studies of heavy-to-light FFs in the literature,
see e.g.\ \cite{Becher:2005bg,Arnesen:2005ez,DescotesGenon:2008hh}.

\begin{figure}[tpbh]
\centering
\includegraphics[width=.48\textwidth]{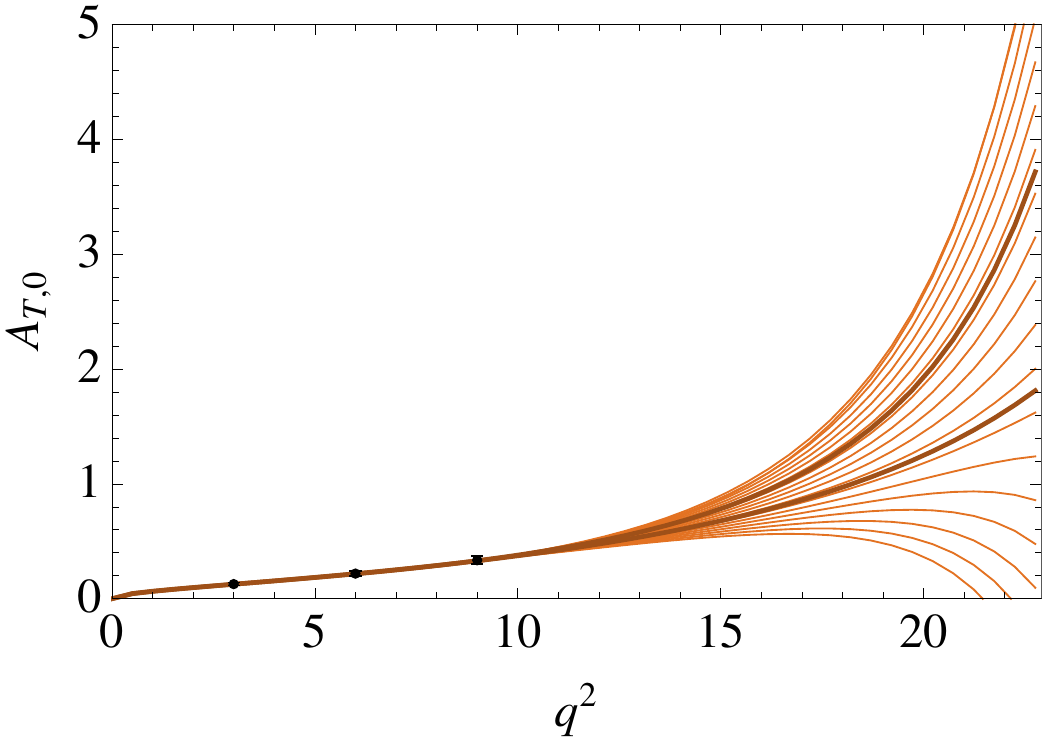} \hfill 
\includegraphics[width=.48\textwidth]{AVT0KSE.pdf} 
\caption{
Same as Fig.~\ref{fig:a2comp}, but for $\mathcal{A}_{T,0}$ (no lattice data).
}
\label{fig:a2comp2}
\end{figure}

In order to address the question of potential contributions from higher-order
terms in the SE, we consider the LCSR prediction for the vector and tensor 
form factors
${\cal A}_{V,0}$, ${\cal A}_{T,0}$ and fit to a SE with $K=3$, where the coefficient $a_2$
of the $z^2$ term in the expansion has been fixed to values between $[-0.9,+0.9]$,
representing almost the
maximal range allowed by the unitarity
constraints. The results of these fits are shown in Figs.~\ref{fig:a2comp},\ref{fig:a2comp2}. 
Let us first consider the case, where only LCSR data is used in the fit.
We observe that --- as expected --- 
the constraints from the LCSR points at low values of $q^2$ are not sufficient 
to determine the behaviour at large $q^2$, if higher-order terms in the SE 
(with no further phenomenological constraints) are allowed for. However, the
behaviour of the form factor corresponding to the extreme values of $a_2$ does
not appear very realistic (even a rough numerical estimate of
coupling constants for the first low-lying resonances
with the considered hadronic transition would be sufficient to exclude the
curves at the margin).
Therefore the associated error estimate appears
too pessimistic to us. Moreover,
typically, exclusive semi-leptonic branching fractions are suppressed relative to
the inclusive ones by a factor of about 20. This suggests that the right-hand side
of the dispersive bounds should not exceed a value of 5\% (instead of 100\%). 
Correspondingly, a realistic range of allowed $a_2$--values should rather be taken
as $[-0.25,+0.25]$, which is indicated by the thick central lines in  Fig.~\ref{fig:a2comp}.
The associated uncertainty is only slightly larger than the one obtained 
from the variation of $(a_0,a_1)$ for the SE fit with $K=2$.
If, on the other hand, the lattice information (in the case of ${\cal A}_{V,0}$) is 
taken into account, the cases with $|a_2| \gtrsim 0.25$ actually do not yield a
satisfactoring fit anymore, and again the error estimate of the linear fit seems
to be sufficient to estimate the fit errors.
 In view of the generic
difficulties in estimating systematic theoretical uncertainties, we thus consider the 
SE/SSE parameterisation with $K=2$ and the associated error estimates as sufficiently reliable for
practical applications.

Another comment applies to the scalar FFs ${\cal A}_{V,t}$: As 
shown in Table~\ref{tab:reson}, the combined heavy-quark/chiral-symmetry limit 
considered in \cite{Bardeen:2003kt} predicts a scalar $B_s$ resonance \emph{below}
$BK$-production threshold (such a state is also favoured by a Lattice computation
in \cite{Green:2003zza}).
 On the other hand, the PDG only finds resonances at masses near/above
the production threshold. We have therefore chosen to compare two variants of the fit,
with/without a scalar resonance.\footnote{Notice that BZ \cite{Ball:2004ye} use an effective
resonance mass above production threshold to parameterise the scalar FFs.}
 As can be seen, the fit \emph{with} a scalar resonance
from \cite{Bardeen:2003kt} describes the combined Lattice/LCSR data significantly better than
the fit without a low-lying resonance (where in the latter case again the dispersive
bounds constrain the FF to lie systematically below the Lattice data). 
However, within the present uncertainties of Lattice and LCSR data, this
could only be taken as a very indirect argument in favour of a scalar resonance in
the anticipated mass region.

\begin{table}[tpbh]

\caption{$B \to K$: Fit of SE parameterisation to LCSR or LCSR/Lattice
results, for $\mathcal{A}_{V,0}$ ($X=1$), $\mathcal{A}_{V,t}$
($X=3$) and $\mathcal{A}_{T,0}$ ($X=1$). \label{tab:AKSE}}

\begin{center}
\begin{tabular}{c|c|c|c|c|c|c}
\hline \hline \T \B
$A_X$ &$m_R$ &$\alpha_0$ &$\alpha_1$ &Fit to& $\chi^2_{\rm{fit}}$ & $X
\sum\limits_i{\alpha _i^2} $  \\
\hline \hline\T $\mathcal{A}_ {V,0}$ &$5 .41$&$-2.5\times
10^{{-2}}$&$7 .2\times 10^{{-2}}$&LCSR and Lattice&$0 .329$ &$5
.77\times 10^{{-3}}$ \\
 $\mathcal{A}_ {V,t}$ &-&$-6.8\times 10^{{-2}}$&$0 .20$&LCSR and
Lattice&$0 .200$ &$0 .129$ \\
 $\mathcal{A}_ {V,t}$ &$5 .72$&$-4.5\times 10^{{-2}}$&$8 .9\times
10^{{-2}}$&LCSR and Lattice&$0 .234$ &$2 .99\times 10^{{-2}}$ \\
\hline \hline\T $\mathcal{A}_ {V,0}$ &$5 .41$&$-2.4\times
10^{{-2}}$&$6 .2\times 10^{{-2}}$&LCSR&$5 .07\times 10^{{-3}}$&$4
.43\times 10^{{-3}}$ \\
 $\mathcal{A}_ {V,t}$ &-&$-6.7\times 10^{{-2}}$&$0 .18$&LCSR&$1
.44\times 10^{{-3}}$ &$0 .111$ \\
 $\mathcal{A}_ {V,t}$ &$5 .72$&$-4.8\times 10^{{-2}}$&$0 .11$&LCSR&$1
.54\times 10^{{-4}}$ &$4 .34\times 10^{{-2}}$ \\
\B $\mathcal{A}_ {T,0}$ &$5 .41$&$-2.8\times 10^{{-2}}$&$6.0\times
10^{{-2}}$&LCSR&$3 .94\times 10^{{-3}}$&$4 .40\times 10^{{-3}}$ \\
\hline \hline
\end{tabular}
\end{center}
\end{table}


\begin{table}[tpbh]

\caption{$B \to K$: Fit of SSE parameterisation to LCSR or
LCSR/Lattice results, for $\mathcal{A}_{V,0}$ ($X=1$),
$\mathcal{A}_{V,t}$
($X=3$) and $\mathcal{A}_{T,0}$ ($X=1$).\label{tab:AKSSE}}

\begin{center}
\begin{tabular}{c|c|c|c|c|c|c}
\hline \hline \T \B
$A_X$ &$m_R$ &$\tilde\alpha_0$ &$\tilde\alpha_1$ &Fit to&
$\chi^2_{\rm{fit}}$ &$X \sum\limits_{i,j}{C_{i,j}\tilde\alpha _i
\tilde\alpha_j} $\\
\hline \hline\T $\mathcal{A}_ {V,0}$ &$5 .41$&$0 .50$&$-1.4$&LCSR and
Lattice&$0 .940$ &$6 .51\times 10^{{-3}}$ \\
 $\mathcal{A}_ {V,t}$ &-&$0 .54$&$-1.7$&LCSR and Lattice&$0 .904$ &$0 .142$ \\
 $\mathcal{A}_ {V,t}$ &$5 .72$&$0 .28$&$0 .35$&LCSR and Lattice&$0
.128$ &$3 .15\times 10^{{-2}}$ \\
\hline \hline\T $\mathcal{A}_ {V,0}$ &$5 .41$&$0 .48$&$-1.0$&LCSR&$5
.15\times 10^{{-3}}$&$4 .04\times 10^{{-3}}$ \\
 $\mathcal{A}_ {V,t}$ &-&$0 .52$&$-1.4$&LCSR&$2 .27\times 10^{{-3}}$
&$9 .55\times 10^{{-2}}$ \\
 $\mathcal{A}_ {V,t}$ &$5 .72$&$0 .30$&$0 .20$&LCSR&$7 .17\times
10^{{-5}}$ &$5 .32\times 10^{{-2}}$ \\
\B $\mathcal{A}_ {T,0}$ &$5 .41$&$0 .48$&$-1.1$&LCSR&$8 .15\times
10^{{-3}}$&$3 .06\times 10^{{-3}}$ \\
\hline \hline
\end{tabular}
\end{center}
\end{table}

\FloatBarrier

\paragraph{\boldmath $B \to \rho$ form factors: \unboldmath}

\begin{figure}[tpbh]
\centering
\includegraphics[width=0.48\textwidth]{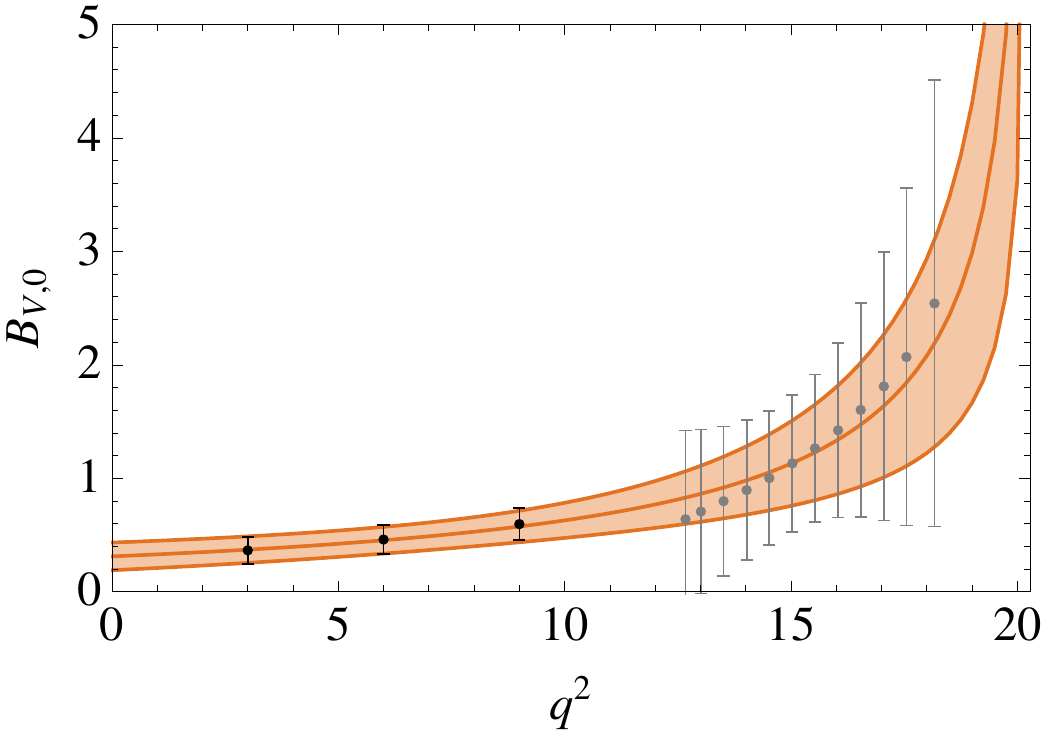}\hfill 
\includegraphics[width=0.48\textwidth]{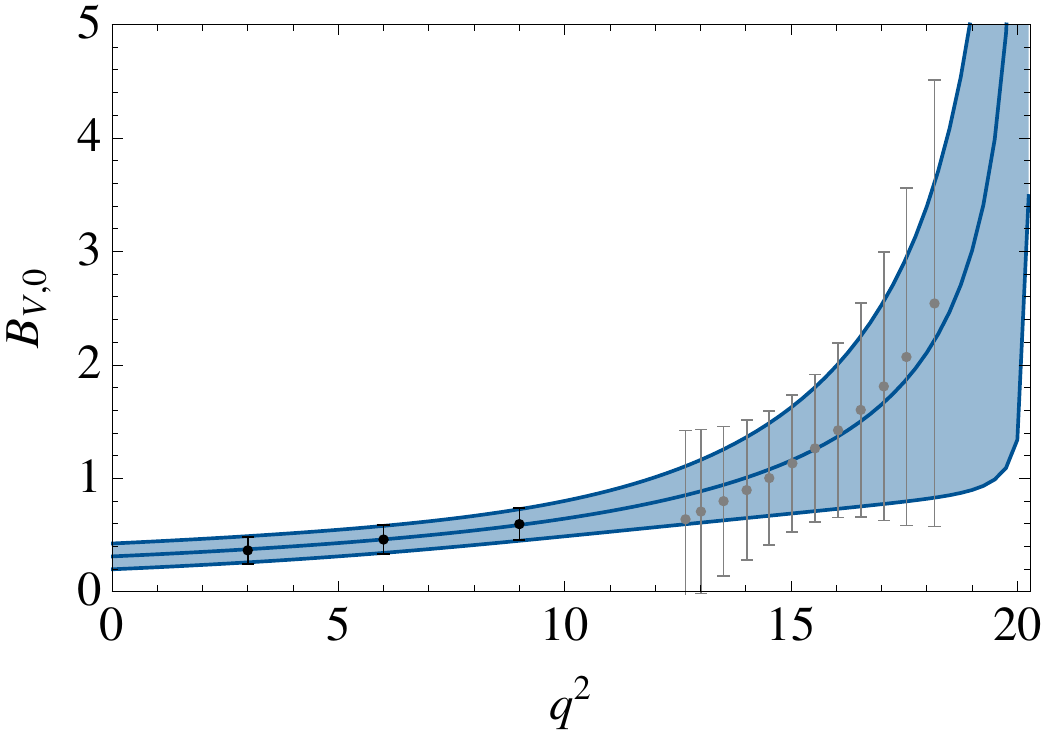}
\includegraphics[width=0.48\textwidth]{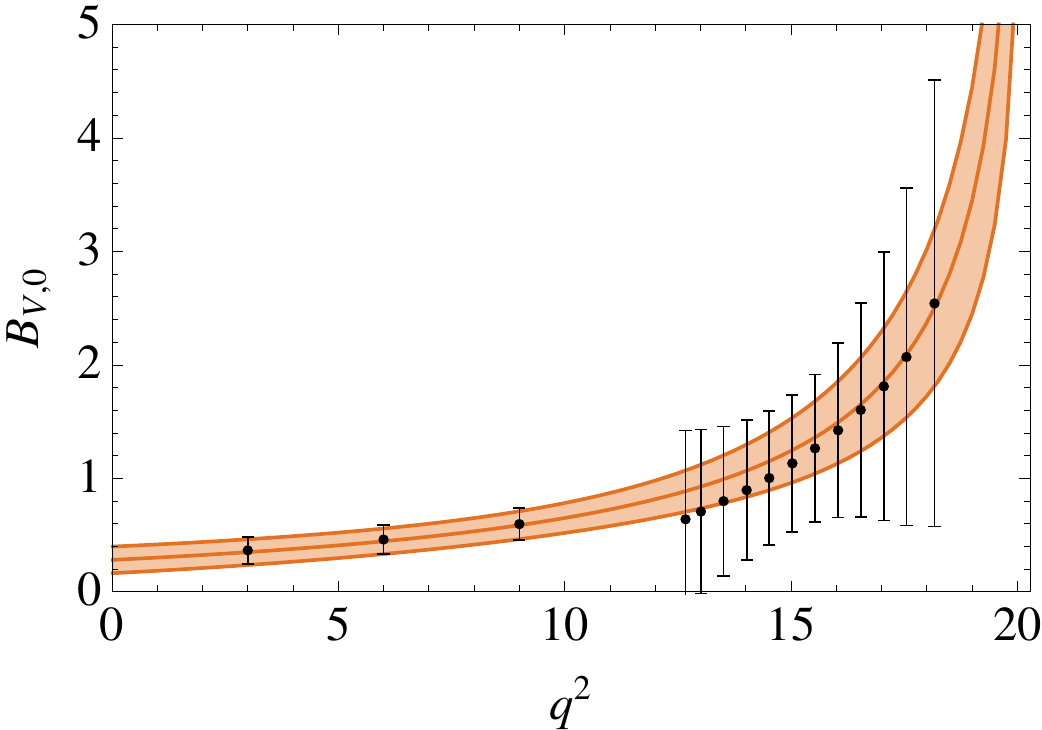}\hfill \includegraphics[width=0.48\textwidth]{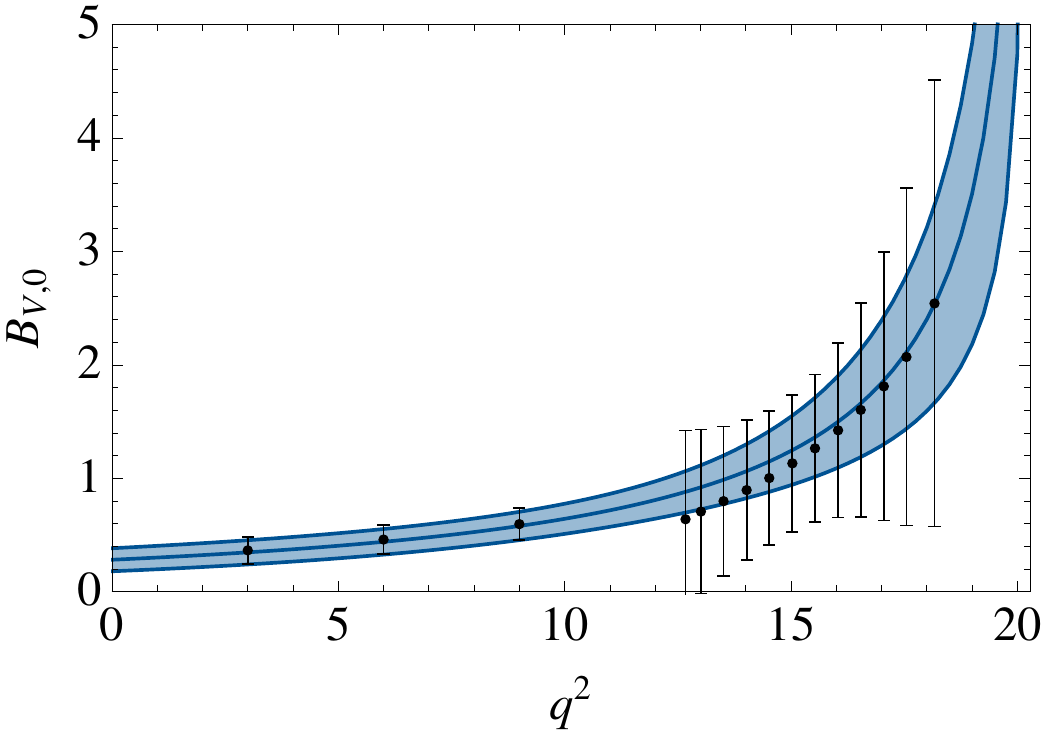}
\caption{$B \to \rho$: Fit of SE (left) and SSE (right) parameterisations to LCSR (top) and to LCSR and Lattice (bottom) for $\mathcal{B}_{V,0}$. The LCSR and Lattice data are shown by black points with error bars in the appropriate $q^2$ range.}
\label{fig:BV0rho}
\end{figure}
\begin{figure}[tpbh]
\centering
\includegraphics[width=0.48\textwidth]{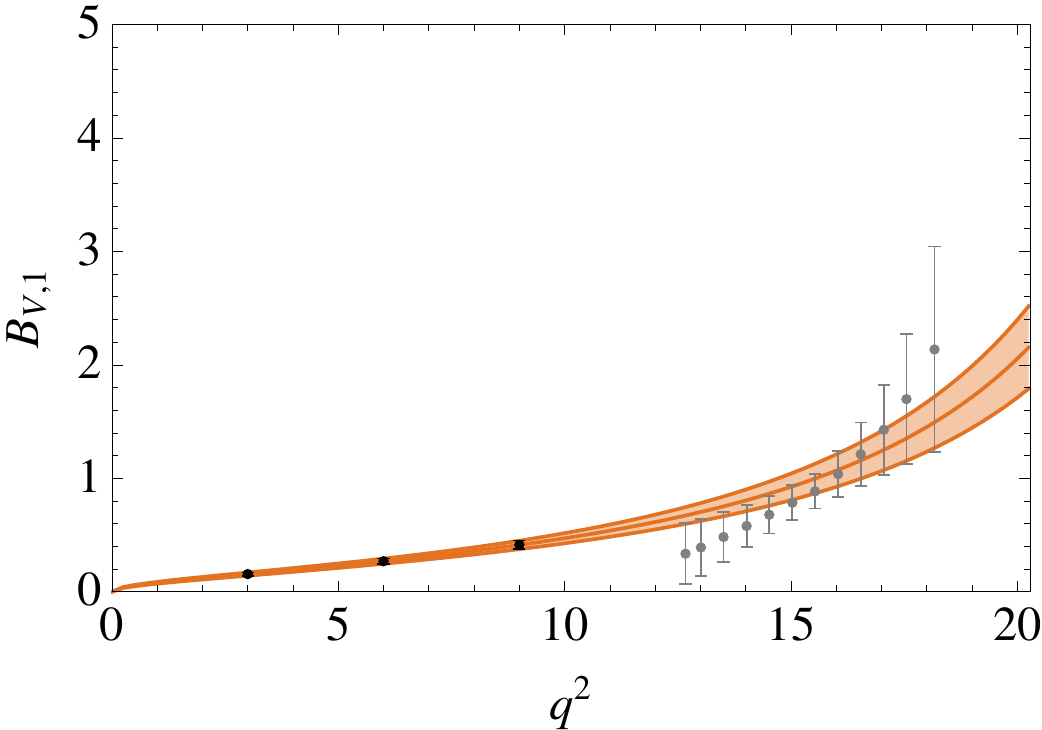}\hfill 
\includegraphics[width=0.48\textwidth]{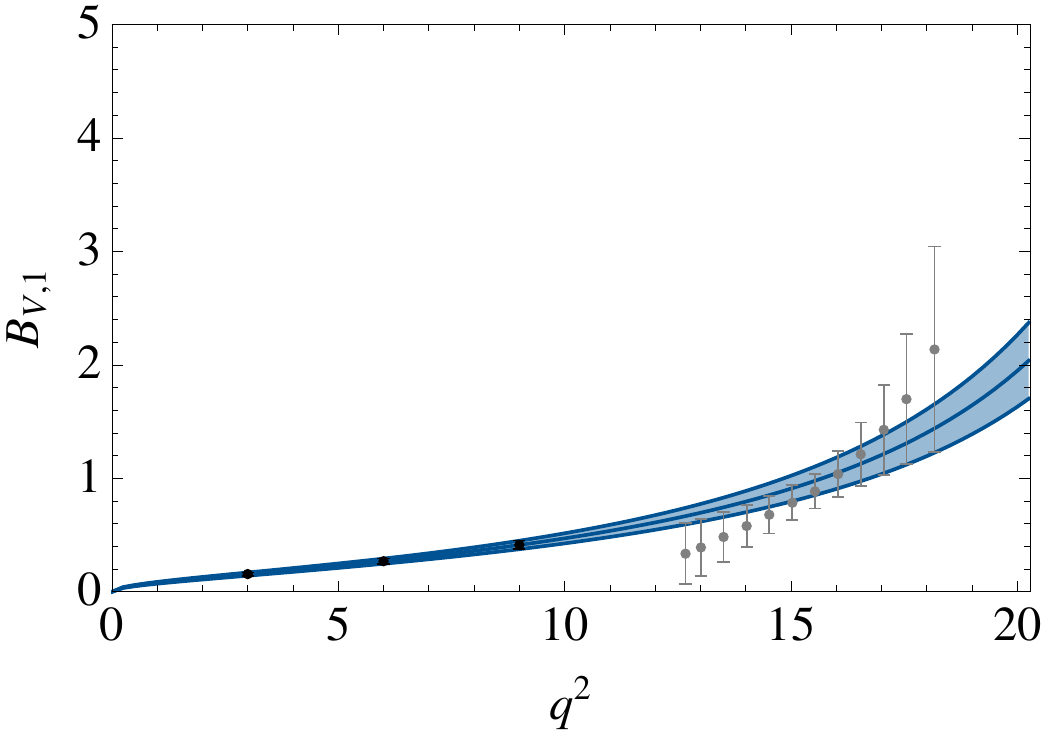}
\includegraphics[width=0.48\textwidth]{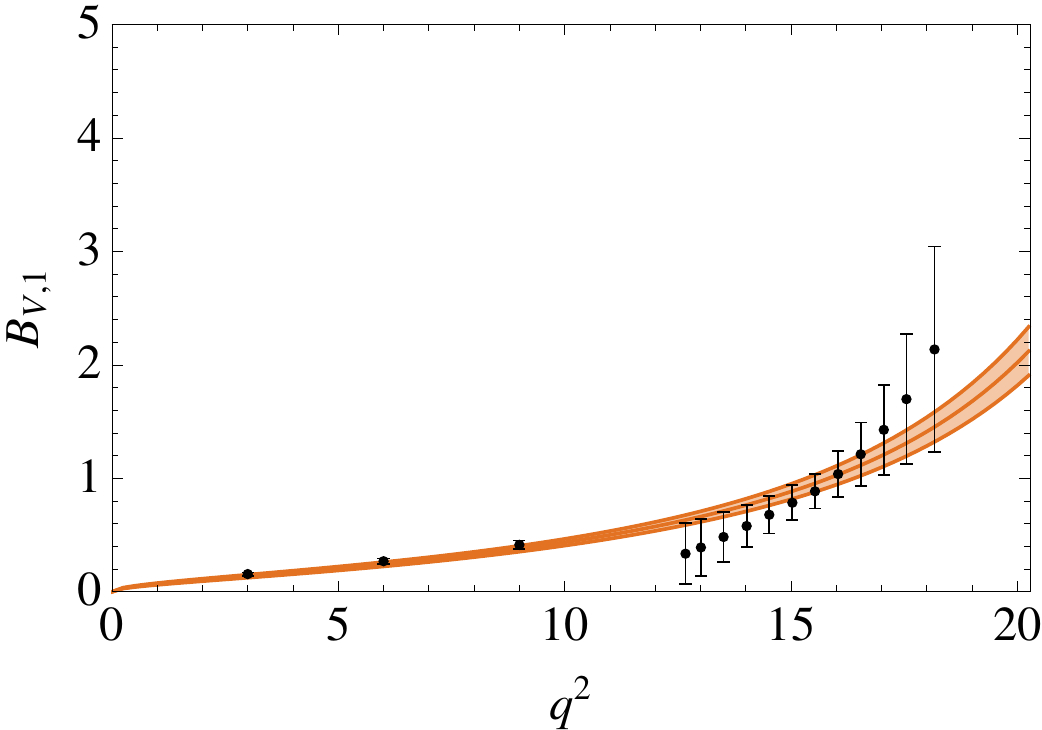}\hfill \includegraphics[width=0.48\textwidth]{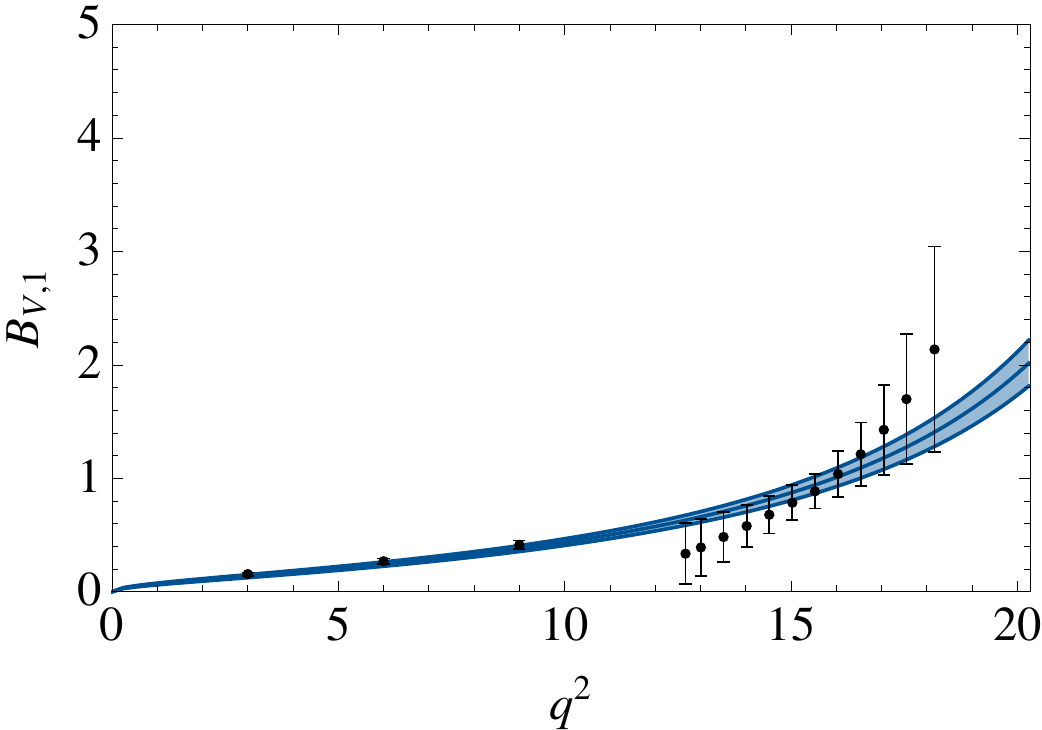}
\caption{$B \to \rho$: Fit of SE (left) and SSE (right) parameterisations to LCSR (top) and to LCSR and Lattice (bottom) for $\mathcal{B}_{V,1}$. The LCSR and Lattice data are shown by black points with error bars in the appropriate $q^2$ range.}
\label{fig:BV1rho}
\end{figure}
\begin{figure}[tpbh]
\centering
\includegraphics[width=0.48\textwidth]{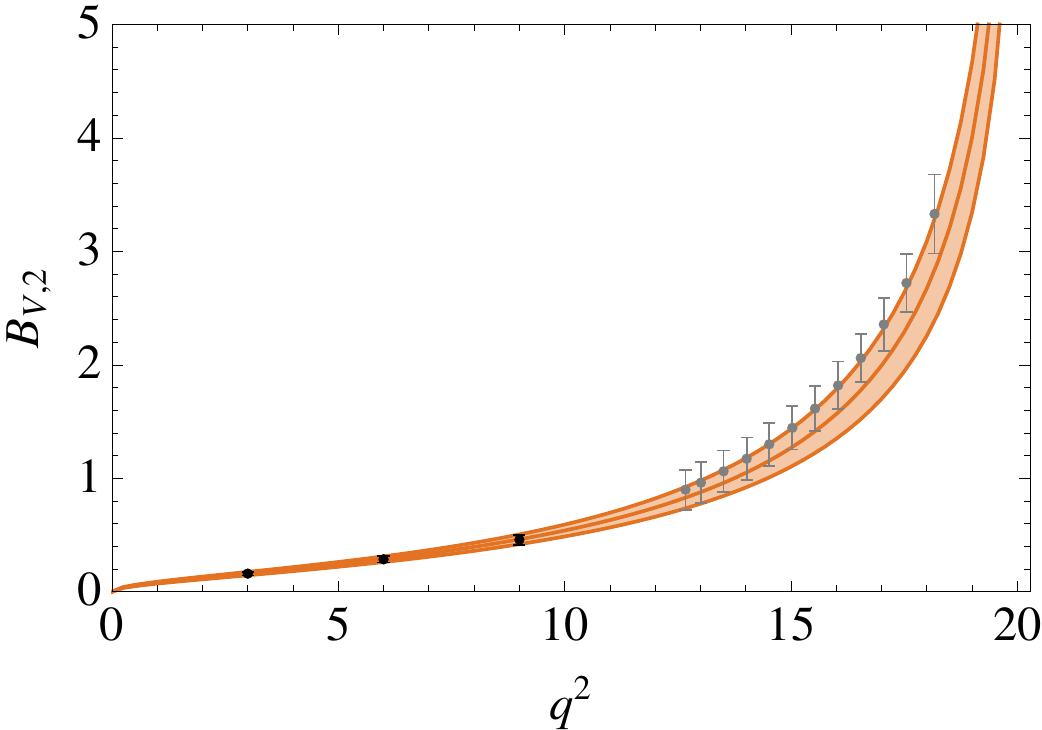}\hfill \includegraphics[width=0.48\textwidth]{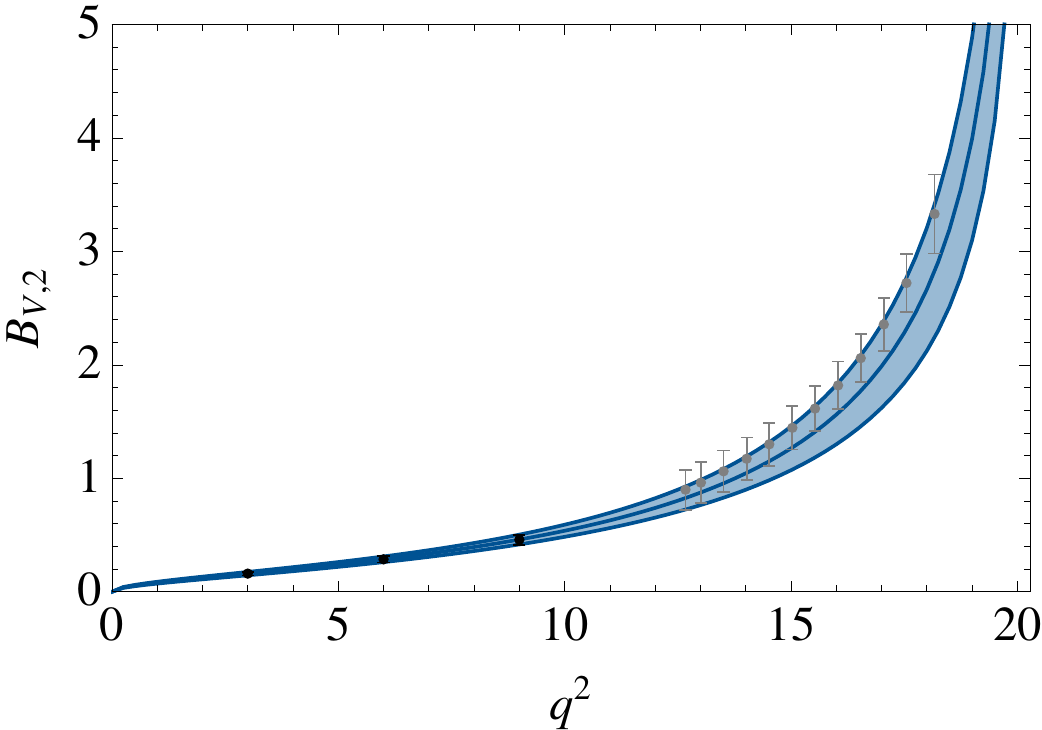}
\includegraphics[width=0.48\textwidth]{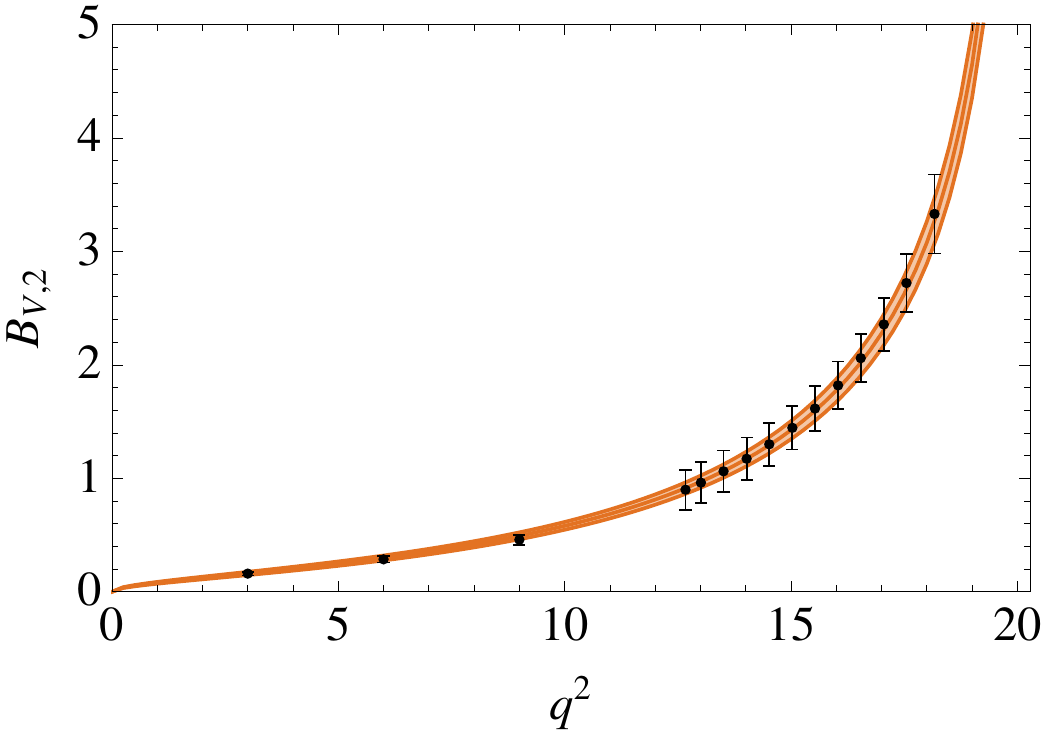}
\hfill \includegraphics[width=0.48\textwidth]{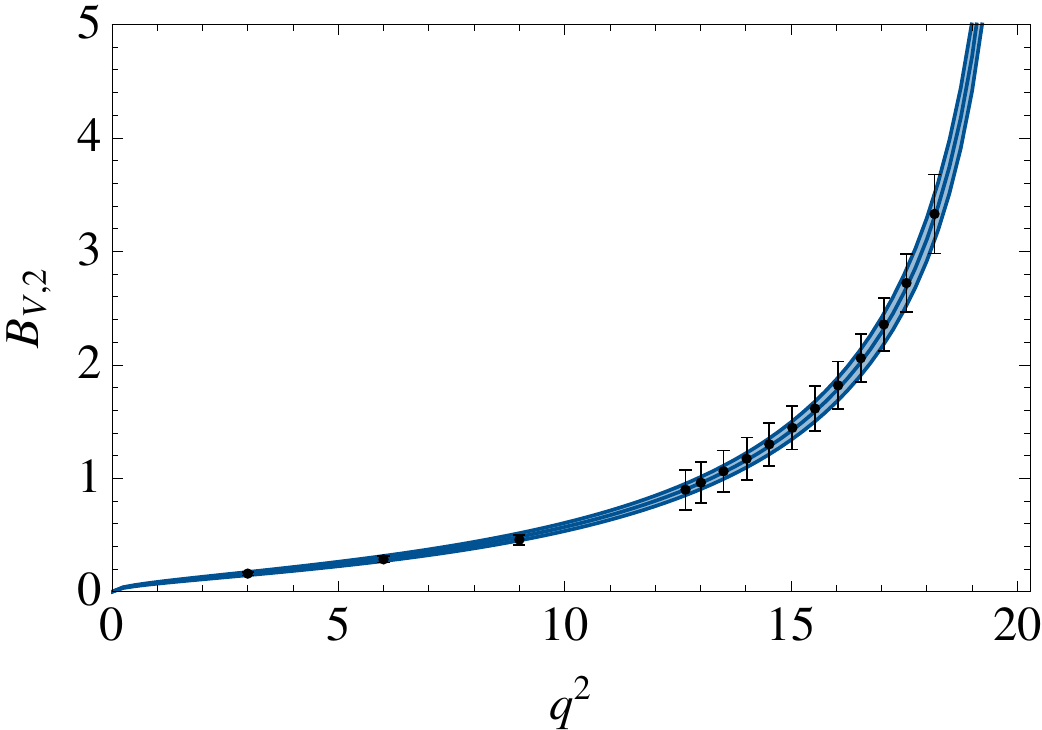}
\caption{$B \to \rho$: Fit of SE (left) and SSE (right) parameterisations to LCSR (top) and to LCSR and Lattice (bottom) for $\mathcal{B}_{V,2}$. The LCSR and Lattice data are shown by black points with error bars in the appropriate $q^2$ range.}
\label{fig:BV2rho}
\end{figure}
\begin{figure}[tpbh]
\centering
\includegraphics[width=.48\textwidth]{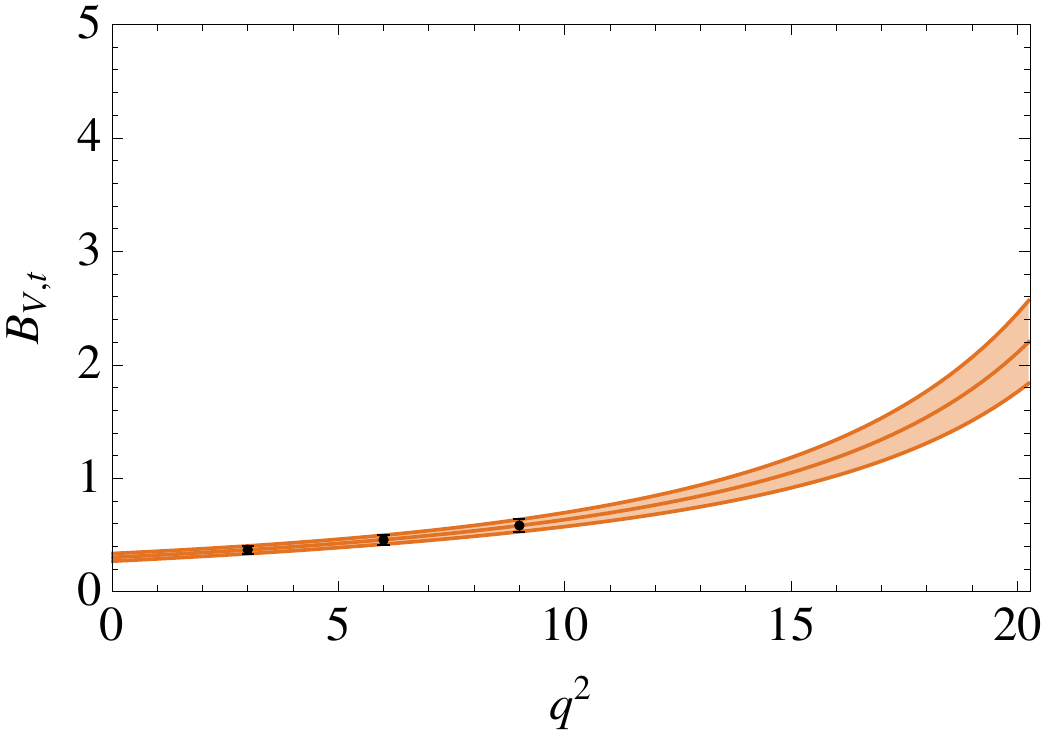} \hfill \includegraphics[width=.48\textwidth]{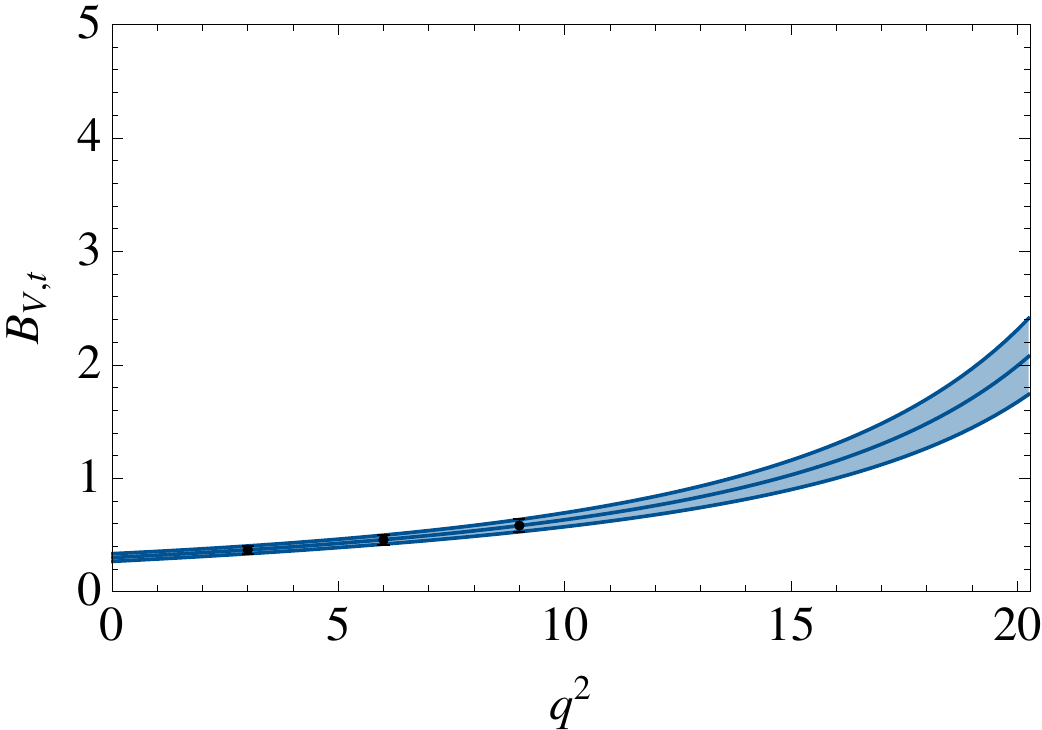} 
\caption{$B \to \rho$: Fit of SE (left) and SSE (right) parameterisations to LCSR for $\mathcal{B}_{V,t}$. The LCSR data is shown by black points with error bars.}
\label{fig:BVtrho}
\end{figure}
\begin{figure}[tpbh]
\centering
\includegraphics[width=.48\textwidth]{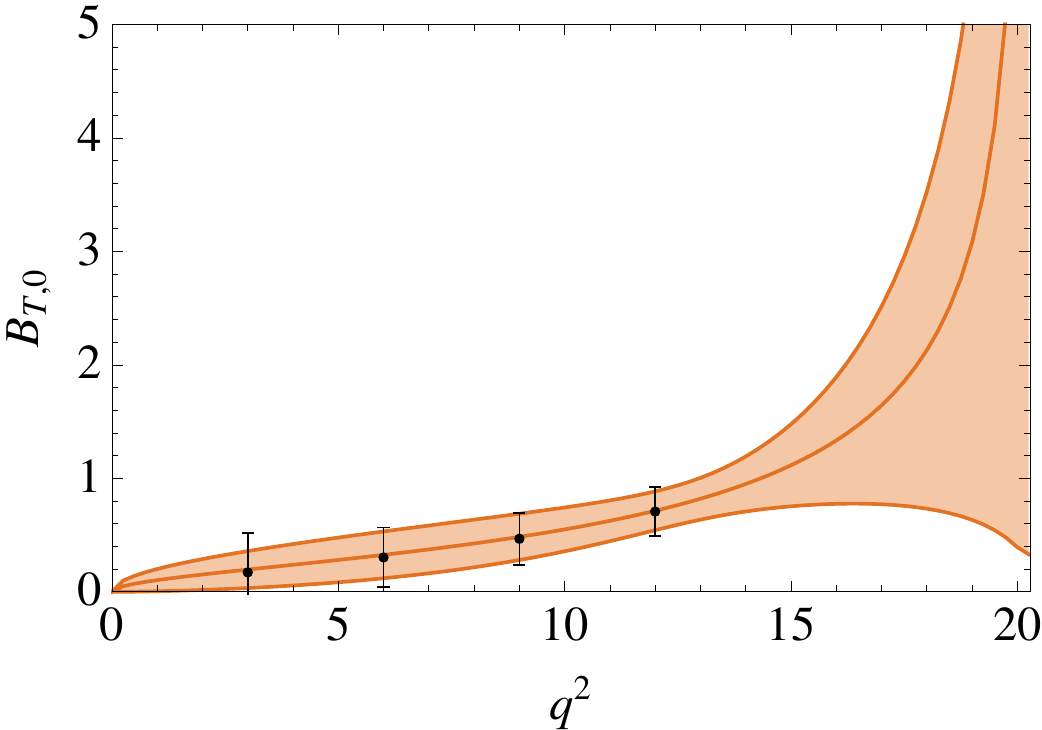} \hfill \includegraphics[width=.48\textwidth]{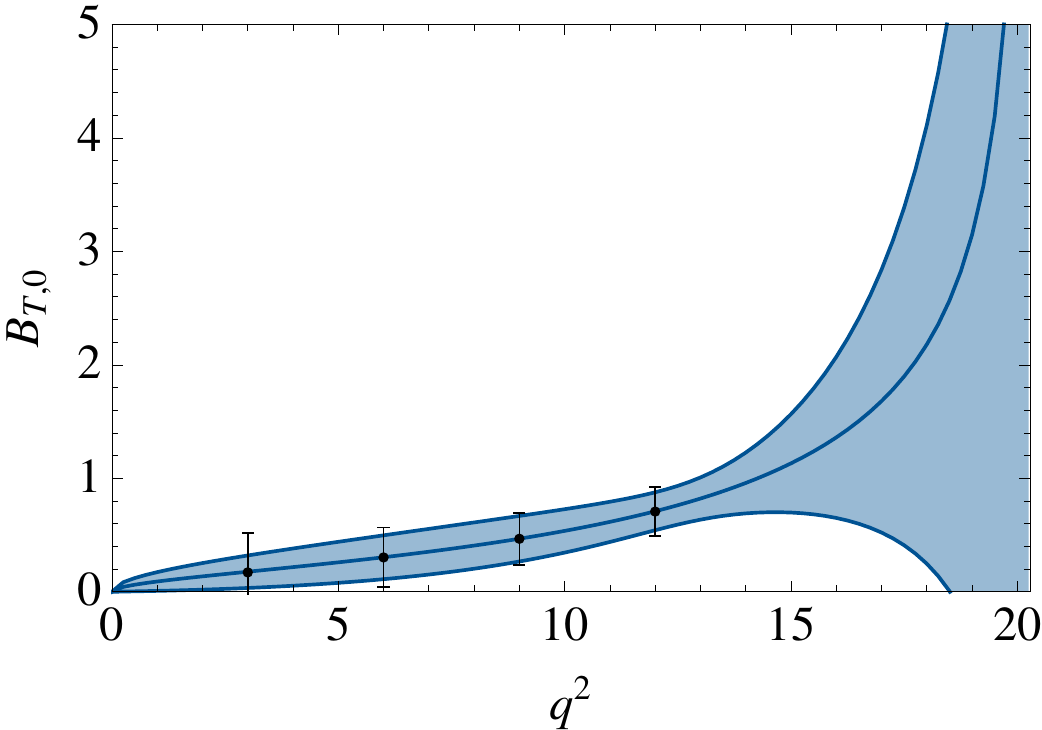} 
\caption{$B \to \rho$: Fit of SE (left) and SSE (right) parameterisations to LCSR for $\mathcal{B}_{T,0}$. The LCSR data is shown by black points with error bars.}
\label{fig:BT0rho}
\end{figure}
\begin{figure}[tpbh]
\centering
\includegraphics[width=.48\textwidth]{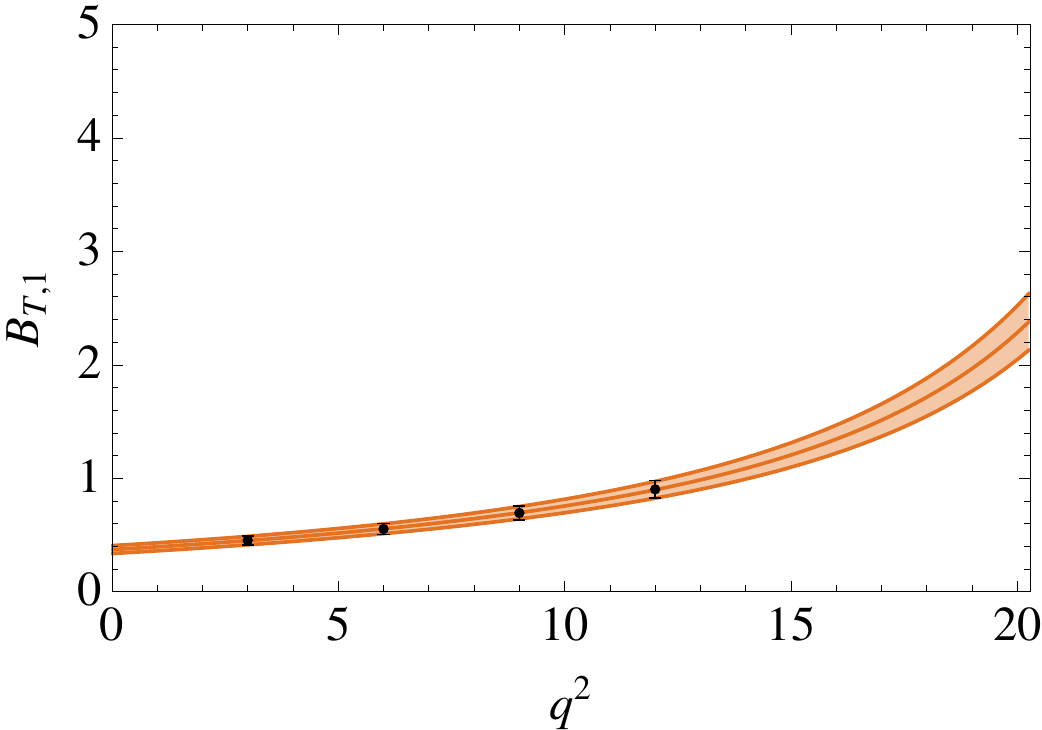} \hfill \includegraphics[width=.48\textwidth]{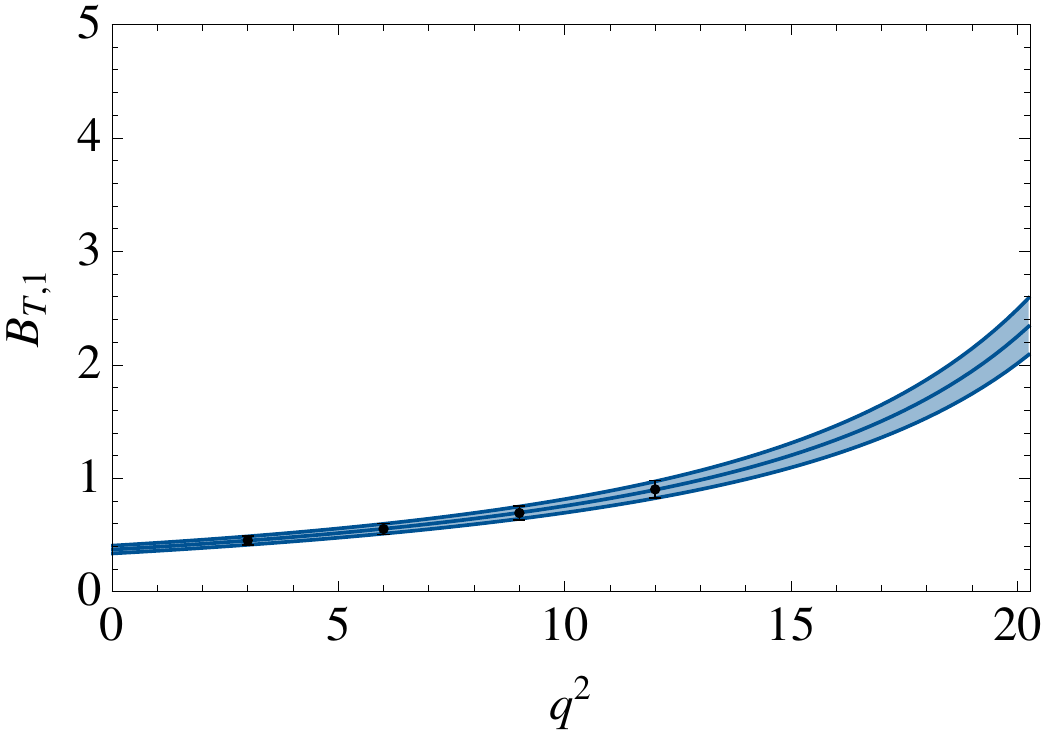} 
\caption{$B \to \rho$: Fit of SE (left) and SSE (right) parameterisations to LCSR for $\mathcal{B}_{T,1}$. The LCSR data is shown by black points with error bars.}
\label{fig:BT1rho}
\end{figure}
\begin{figure}[tpbh]
\centering
\includegraphics[width=.48\textwidth]{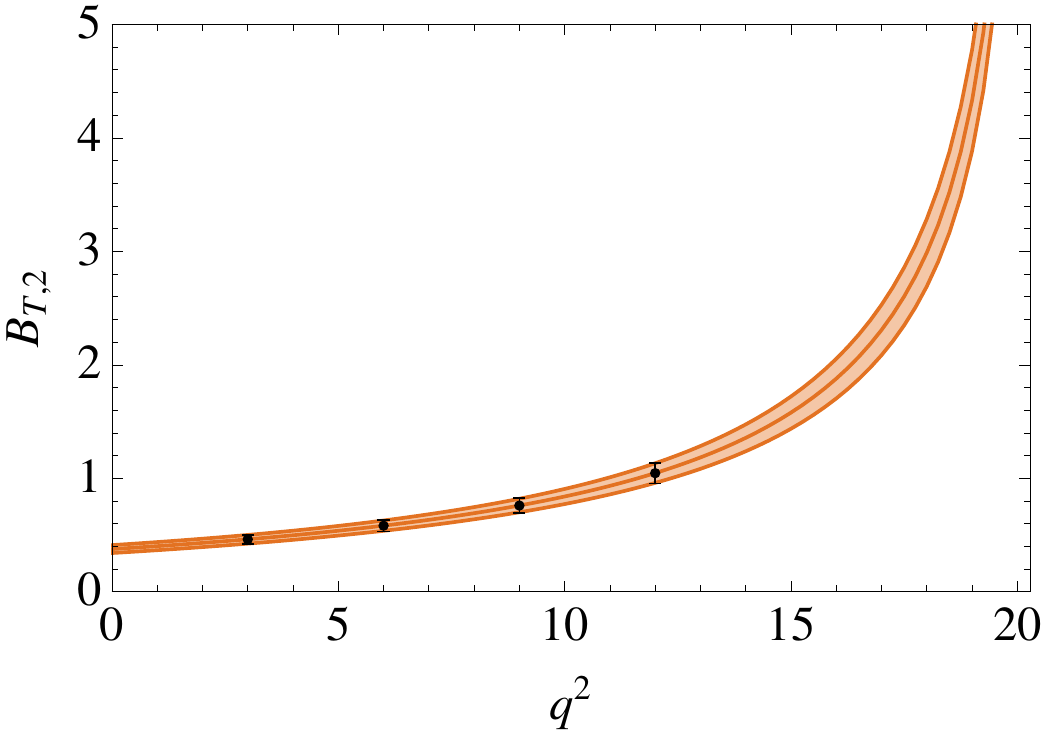} \hfill \includegraphics[width=.48\textwidth]{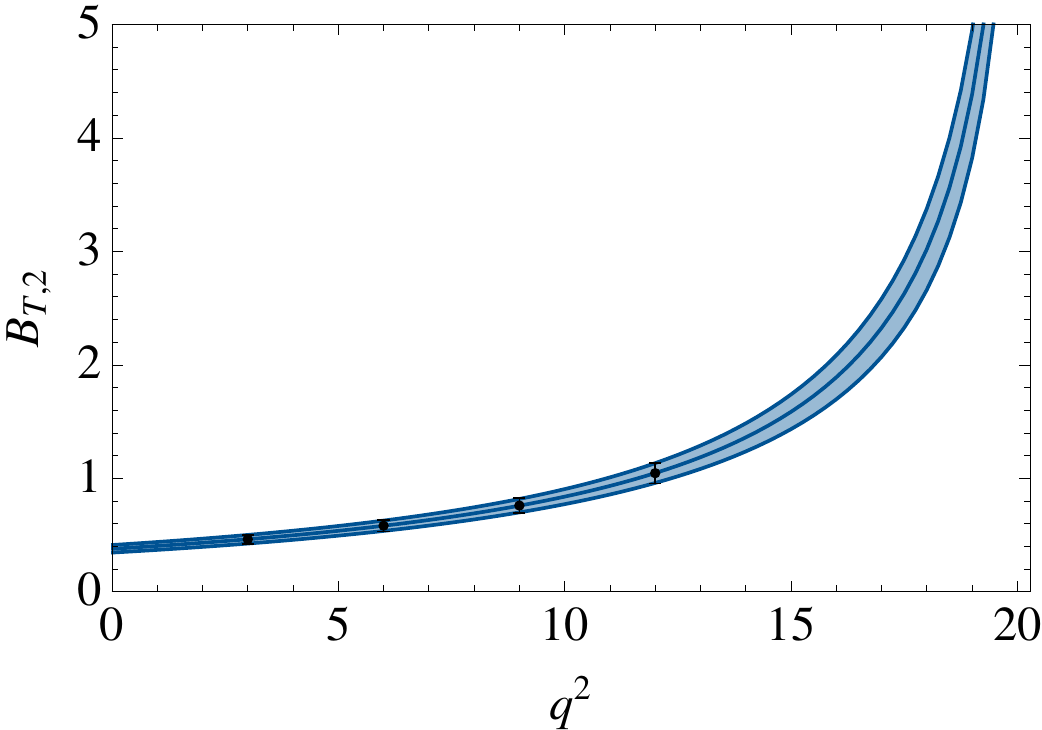} 
\caption{$B \to \rho$: Fit of SE (left) and SSE (right) parameterisations to LCSR for $\mathcal{B}_{T,2}$. The LCSR data is shown by black points with error bars.}
\label{fig:BT2rho}
\end{figure}

Our FF fits for $B \to \rho$ transitions, relevant for the radiative
$B \to \rho\gamma$ and $B \to \rho \ell^+\ell^-$ decays, are summarized
in Figs.~\ref{fig:BV0rho}--\ref{fig:BT2rho} and Tables~\ref{tab:BrhoSE} 
and \ref{tab:BrhoSSE}, where we again compare the fit to SE and SSE 
parameterisations. As in the case of $B\to K$ FFs, we generally
observe similar good results for SE and SSE fits,  with the dispersive
bounds again playing only a minor role in restricting the coefficients of the SE/SSE.
The covariance matrices for the fits can again be found in 
Appendix~\ref{app:covmat}.

Lattice results are restricted to the (axial--)vector FFs, and we again study how
the fits change when the Lattice data is included:
In case of the FF ${\cal B}_{V,0}$, the uncertainties on the Lattice data
are rather large, and the fit is in any case dominated by the LCSR points
at low values of $q^2$. Still, we find that the best-fit curve also well
describes the central values of the Lattice estimates.
The situation is somewhat different for ${\cal B}_{V,1}$, where
the central values of the Lattice points do not quite agree with the extrapolation
of the LCSR prediction. The fit is consistent within Lattice uncertainties, but
a rather large value of $\chi^2$, dominated by the deviations from the Lattice
points, is generated.
On the other hand, for ${\cal B}_{V,1}$ the Lattice data are competitive with the
LCSR input, and we can again observe that the extrapolation of the LCSR predictions
describes the Lattice points very well, while inclusion of the Lattice data in this
case leads to a very precise FF description.

In the remaining cases, we again provide the extrapolations for the pseudoscalar
and tensor FFs from LCSR input, where Lattice data have not been available.
Here, it is to be mentioned that the uncertainties for the FF ${\cal B}_{T,0}$ 
are quite large, because we had to determine the LCSR input values from the
\emph{difference} of two FFs in (\ref{eq:BTdef}). Of course, it would be 
desirable to directly calculate the FF ${\cal B}_{T,0}$ in the LCSR approach
which should lead to significantly smaller uncertainties for the input data
and the extrapolation to large values of $q^2$. A similar comment applies to
the FF ${\cal B}_{V,0}$.

\begin{table}[tpbh]

\caption{$B \to \rho$: Fit of SE parameterisation to LCSR or LCSR/Lattice results for $\mathcal{B}_{V,0-2}$ ($X=1$), $\mathcal{B}_{V,t}$ ($X=3$) and $\mathcal{B}_{T,0-2}$ ($X=1$).\label{tab:BrhoSE}}

\begin{center}
\begin{tabular}{c|c|c|c|c|c|c}
\hline \hline
\T \B $B_X$ &$m_R$ &$\beta_0$ &$\beta_1$ &Fit to& $\chi^2_{\rm{fit}}$ & $X \sum\limits_i{\beta _i^2} $\\
\hline \hline\T $\mathcal{B}_ {V,0}$ &$5 .72$&$-8.0\times 10^{{-3}}$&$2 .5\times 10^{{-2}}$&&&\\
$\mathcal{B}_ {V,1}$ &$5 .33$&$-3.5\times 10^{{-2}}$&$0 .11$&LCSR and Lattice&$32 .1$ &$1 .98\times 10^{{-2}}$ \\
$\mathcal{B}_ {V,2}$ &$5 .72$&$-2.5\times 10^{{-2}}$&$7 .8\times 10^{{-2}}$&&&\B \\
\hline \hline\T $\mathcal{B}_ {V,0}$ &$5 .72$&$-7.5\times 10^{{-3}}$&$1 .4\times 10^{{-2}}$&&&\\
$\mathcal{B}_ {V,1}$ &$5 .33$&$-3.7\times 10^{{-2}}$&$8 .9\times 10^{{-2}}$&LCSR&$9 .56\times 10^{{-2}}$ &$1 .28\times 10^{{-2}}$ \\
$\mathcal{B}_ {V,2}$ &$5 .72$&$-2.3\times 10^{{-2}}$&$5 .2\times 10^{{-2}}$&&&\B \\
\hline \hline\T \B $\mathcal{B}_ {V,t}$ &$5 .28$&$-3.2\times 10^{{-2}}$&$8 .9\times 10^{{-2}}$&LCSR&$3 .81\times 10^{{-3}}$&$2 .66\times 10^{{-2}}$ \\
\hline \hline\T $\mathcal{B}_ {T,0}$ &$5 .72$&$-1.4\times 10^{{-2}}$&$-8.3\times 10^{{-3}}$&&&\\
$\mathcal{B}_ {T,1}$ &$5 .33$&$-1.0\times 10^{{-2}}$&$3 .4\times 10^{{-2}}$&LCSR&$4 .18\times 10^{{-2}}$ &$1 .86\times 10^{{-3}}$ \\
$\mathcal{B}_ {T,2}$ &$5 .72$&$-6.3\times 10^{{-3}}$&$1 .7\times 10^{{-2}}$&&&\B \\
\hline \hline
\end{tabular}
\end{center}
\end{table}


\begin{table}[tpbh]

\caption{$B \to \rho$: Fit of SSE parameterisation to LCSR or  LCSR/Lattice results for $\mathcal{B}_{V,0-2}$ ($X=1$), $\mathcal{B}_{V,t}$ ($X=3$) and $\mathcal{B}_{T,0-2}$ ($X=1$).\label{tab:BrhoSSE}}

\begin{center}
\begin{tabular}{c|c|c|c|c|c|c}
\hline \hline
\TT \BB $B_X$ &$m_R$ &$\tilde\beta_0$ &$\tilde\beta_1$ &Fit to& $\chi^2_{\rm{fit}}$ & $X \sum\limits_{i,j}{C_{i,j}\tilde\beta _i \tilde\beta_j} $ \\
\hline \hline\T $\mathcal{B}_ {V,0}$ &$5 .72$&$0 .26$&$0 .14$&&&\\
$\mathcal{B}_ {V,1}$ &$5 .33$&$0 .51$&$-1.7$&LCSR and Lattice&$33.0$ &$1 .85\times 10^{{-2}}$ \\
$\mathcal{B}_ {V,2}$ &$5 .72$&$0 .40$&$-0.15$&&&\B \\
\hline \hline\T $\mathcal{B}_ {V,0}$ &$5 .72$&$0 .26$&$0 .50$&&&\\
$\mathcal{B}_ {V,1}$ &$5 .33$&$0 .54$&$-1.4$&LCSR&$4 .34\times 10^{{-2}}$ &$1 .10\times 10^{{-2}}$ \\
$\mathcal{B}_ {V,2}$ &$5 .72$&$0 .37$&$0 .24$&&&\B \\
\hline \hline\T \B $\mathcal{B}_ {V,t}$ &$5 .28$&$0 .43$&$-1.3$&LCSR&$8 .49\times 10^{{-3}}$&$2 .16\times 10^{{-2}}$ \\
 \hline \hline\T $\mathcal{B}_ {T,0}$ &$5 .72$&$0 .35$&$0 .94$&&&\\
$\mathcal{B}_ {T,1}$ &$5 .33$&$0 .52$&$-1.5$&LCSR&$3 .57\times 10^{{-2}}$ &$1 .79\times 10^{{-3}}$ \\
$\mathcal{B}_ {T,2}$ &$5 .72$&$0 .34$&$0 .31$&&&\B \\
\hline \hline
\end{tabular}
\end{center}
\end{table}

\FloatBarrier 

\paragraph{\boldmath $B \to K^*$ and $B_s \to \phi$ form factors: \unboldmath}

The analysis of $B\to K^*$ transitions is more difficult on the Lattice, as the $K^*$--meson is unstable. Quenched calculations on the Lattice have been attempted for the tensor FFs needed in $B\to K^*\gamma$, at $q^2=0$, but we do not include these results in our analysis as the other FFs for this decay have not so far been calculated. Therefore, we can only fit to the LCSR data, and our numerical results for the best-fit parameters of the SE and SSE fit are found in Tables~\ref{tab:BKsSE} and \ref{tab:BKsSSE}. The covariance matrices for the fits can also be found in Appendix~\ref{app:covmat}.

\begin{table}[tpbh]

\caption{$B \to K^*$: Fit of SE parameterisation to LCSR results for $\mathcal{B}_{V,0-2}$ ($X=1$), $\mathcal{B}_{V,t}$ ($X=3$) and $\mathcal{B}_{T,0-2}$ ($X=1$). \label{tab:BKsSE}}
\begin{center}
\begin{tabular}{c|c|c|c|c|c|c}
\hline \hline
\T \B $B_X$ &$m_R$ &$\beta_0$ &$\beta_1$ &Fit to& $\chi^2_{\rm{fit}}$ & $X \sum
\limits_i{\beta _i^2} $\\
\hline \hline\T $\mathcal{B}_ {V,0}$ &$5 .83$&$-9.4\times 10^{{-3}}$&$1 .4\times 10^{{-2}}$&&&\\
$\mathcal{B}_ {V,1}$ &$5 .41$&$-5.0\times 10^{{-2}}$&$0 .10$&LCSR&$0 .149$ &$1 .86\times 10^{{-2}}$ \\
$\mathcal{B}_ {V,2}$ &$5 .83$&$-3.0\times 10^{{-2}}$&$6 .8\times 10^{{-2}}$&&&\B \\
\hline \hline\T \B $\mathcal{B}_ {V,t}$ &$5 .37$&$-4.4\times 10^{{-2}}$&$0 .11$&LCSR&$1 .72\times 10^{{-3}}$&$4 .25\times 10^{{-2}}$ \\
\hline \hline\T $\mathcal{B}_ {T,0}$ &$5 .83$&$-1.9\times 10^{{-2}}$&$-1.9\times 10^{{-2}}$&&&\\
$\mathcal{B}_ {T,1}$ &$5 .41$&$-1.4\times 10^{{-2}}$&$4 .1\times 10^{{-2}}$&LCSR&$1 .88\times 10^{{-2}}$ &$3 .14\times 10^{{-3}}$ \\
$\mathcal{B}_ {T,2}$ &$5 .83$&$-8.0\times 10^{{-3}}$&$2 .2\times 10^{{-2}}$&&&\B \\
\hline \hline
\end{tabular}
\end{center}
\end{table}


\begin{table}

\caption{$B \to K^*$: Fit of SSE parameterisation to LCSR results for $\mathcal{B}_{V,0-2}$ ($X=1$), $\mathcal{B}_{V,t}$ ($X=3$) and $\mathcal{B}_{T,0-2}$ ($X=1$).\label{tab:BKsSSE}}

\begin{center}
\begin{tabular}{c|c|c|c|c|c|c}
\hline\hline
\TT \BB $B_X$ &$m_R$ &$\tilde\beta_0$ &$\tilde\beta_1$ &Fit to& $\chi^2_{\rm 
{fit}}$ & $X \sum\limits_{i,j}{C_{i,j}\tilde\beta _i \tilde\beta_j} $ \\
\hline \hline\T $\mathcal{B}_ {V,0}$ &$5 .83$&$0 .31$&$0 .74$&&&\\
$\mathcal{B}_ {V,1}$ &$5 .41$&$0 .62$&$-1.4$&LCSR&$5 .84\times 10^{{-2}}$ &$1 .63\times 10^{{-2}}$ \\
$\mathcal{B}_ {V,2}$ &$5 .83$&$0 .45$&$0 .35$&&&\B \\
\hline \hline\T \B $\mathcal{B}_ {V,t}$ &$5 .37$&$0 .49$&$-1.4$&LCSR&$4 .63\times 10^{{-3}}$&$3 .62\times 10^{{-2}}$ \\\hline \hline \hline\T $\mathcal{B}_ {T,0}$ &$5 .83$&$0 .45$&$1 .4$&&&\\
$\mathcal{B}_ {T,1}$ &$5 .41$&$0 .60$&$-1.5$&LCSR&$9 .65\times 10^{{-3}}$ &$2 .75\times 10^{{-3}}$ \\
$\mathcal{B}_ {T,2}$ &$5 .83$&$0 .42$&$0 .45$&&&\B \\
\hline \hline
\end{tabular}
\end{center}
\end{table}

As in the case of $B \to K^*$, Lattice QCD predictions for $B_s \to \phi$ FFs
are lacking, and we fit to the LCSR data, only. Our numerical results for the best-fit parameters of the SE and SSE parameterisations are found in Tables~\ref{tab:BsphiSE} and \ref{tab:BsphiSSE}. The covariance matrices for the fits can also be found in Appendix~\ref{app:covmat}.
In all cases, we find a good description of the LCSR input at low $q^2$, and from the
experience in $B\to K$ and $B\to \rho$ transitions we expect the extrapolation to
high $q^2$ to be sufficiently reliable. Still, input from Lattice computations 
-- if feasible -- for
$B \to K^*$ and $B_s \to \phi$ transitions at intermediate values of $q^2$ would
be highly welcome.

\begin{table}[tpbh]

\caption{$B_s \to \phi$: Fit of SE parameterisation to LCSR results for $\mathcal{B}_{V,0-2}$ ($X=1$), $\mathcal{B}_{V,t}$ ($X=3$) and $\mathcal{B}_{T,0-2}$ ($X=1$).\label{tab:BsphiSE}}

\begin{center}
\begin{tabular}{c|c|c|c|c|c|c}
\hline \hline
\T \B $B_X$ &$m_R$ &$\beta_0$ &$\beta_1$ &Fit to& $\chi^2_{\rm{fit}}$ & $X \sum\limits_i{\beta _i^2} $ \\
\hline \hline\T $\mathcal{B}_ {V,0}$ &$5 .83$&$-5.8\times 10^{{-3}}$&$3 .5\times 10^{{-3}}$&&&\\
$\mathcal{B}_ {V,1}$ &$5 .41$&$-3.4\times 10^{{-2}}$&$9 .6\times 10^{{-2}}$&LCSR&$0 .124$ &$1 .29\times 10^{{-2}}$ \\
$\mathcal{B}_ {V,2}$ &$5 .83$&$-1.8\times 10^{{-2}}$&$4 .7\times 10^{{-2}}$&&&\B \\
\hline \hline\T \B $\mathcal{B}_ {V,t}$ &$5 .37$&$-3.4\times 10^{{-2}}$&$9 .3\times 10^{{-2}}$&LCSR&$8 .93\times 10^{{-3}}$&$2 .96\times 10^{{-2}}$ \\
\hline \hline\T $\mathcal{B}_ {T,0}$ &$5 .83$&$-1.1\times 10^{{-2}}$&$-1.5\times 10^{{-2}}$&&&\\
$\mathcal{B}_ {T,1}$ &$5 .41$&$-9.0\times 10^{{-3}}$&$3 .5\times 10^{{-2}}$&LCSR&$4 .45\times 10^{{-2}}$ &$1 .86\times 10^{{-3}}$ \\
$\mathcal{B}_ {T,2}$ &$5 .83$&$-4.6\times 10^{{-3}}$&$1 .4\times 10^{{-2}}$&&&\B \\
\hline \hline
\end{tabular}
\end{center}
\end{table}


\begin{table}[tpbh]

\caption{$B_s \to \phi$: Fit of SSE parameterisation to LCSR results for $\mathcal{B}_{V,0-2}$ ($X=1$), $\mathcal{B}_{V,t}$ ($X=3$) and $\mathcal{B}_{T,0-2}$ ($X=1$).\label{tab:BsphiSSE}}

\begin{center}
\begin{tabular}{c|c|c|c|c|c|c}
\hline \hline
\TT \BB $B_X$ &$m_R$ &$\tilde\beta_0$ &$\tilde\beta_1$ &Fit to& 
$\chi^2_{\rm{fit}}$ & $X \sum\limits_{i,j}{C_{i,j}\tilde\beta _i \tilde\beta_j} $	
\\
\hline \hline\T $\mathcal{B}_ {V,0}$ &$5 .83$&$0 .37$&$1 .1$&&&\\
$\mathcal{B}_ {V,1}$ &$5 .41$&$0 .67$&$-2.1$&LCSR&$1 .44\times 10^{{-2}}$ &$1 .15\times 10^{{-2}}$ \\
$\mathcal{B}_ {V,2}$ &$5 .83$&$0 .50$&$0 .19$&&&\B \\
\hline \hline\T \B $\mathcal{B}_ {V,t}$ &$5 .37$&$0 .61$&$-1.8$&LCSR&$1 .57\times 10^{{-2}}$&$2 .58\times 10^{{-2}}$ \\
\hline \hline\T $\mathcal{B}_ {T,0}$ &$5 .83$&$0 .51$&$1 .7$&&&\\
$\mathcal{B}_ {T,1}$ &$5 .41$&$0 .66$&$-2.2$&LCSR&$3 .49\times 10^{{-2}}$ &$1 .66\times 10^{{-3}}$ \\
$\mathcal{B}_ {T,2}$ &$5 .83$&$0 .46$&$0 .26$&&&\B \\
\hline \hline
\end{tabular}
\end{center}
\end{table}

\FloatBarrier

\section{Discussion and Conclusions}

\label{sec:concl}

We have shown that the form factors (FFs) relevant for
radiative and semi-leptonic decays of $B$ and $B_s$ mesons into
light pseudoscalar or vector mesons can be conveniently
parameterised as a series expansion (SE) in the variable $z(t)$
(see the definition in (\ref{eq:zdef})).
With the current accuracy of theoretical estimates from 
light-cone sum rules (LCSRs) and (where available) Lattice QCD, 
we found that keeping only two terms in the expansion
and correctly implementing the analytical behaviour due
to below-threshold resonances, 
results in a very good description of the FFs 
over the whole range of momentum transfer in
the physical decay region.

The coefficients of the SE are further
constrained by dispersive bounds, exploiting the 
crossing symmetry between the physical $B$-meson decay
and the pair-production of heavy and light mesons by
the considered decay current. In order to put the discussion
for the various FFs on a
common footing, we found it convenient to use a FF basis
where the decay/production currents are projected by 
transverse, longitudinal and time-like polarisation
vectors with respect to momentum transfer $t$. Considering the
corresponding projections for the current correlators, the
constraints take the simple form as indicated in
(\ref{eq:aconstrSE},\ref{eq:bconstrSE}).
We stress that for decays  into vector mesons the dispersive
bounds constrain the \emph{sum} of (squared) coefficients for
the three axial-vector FFs, as well as for the three
tensor FFs. In a simultaneous fit of all FFs,
these constraints are thus stronger than those for the
individual FFs in that sum.

In order to determine the correct normalization of the SE,
given by the profile functions $\phi(z(t))$, we calculate the current correlators using an OPE, including NLO perturbative corrections and 
the leading non-perturbative contributions from quark, gluon and
mixed condensates. In particular, we provide the NLO results for the
tensor-current correlation functions, which are relevant for the 
FFs appearing in radiative and rare semi-leptonic
$B$ decays.

With these theoretical tools at hand, we have performed numerical fits
to LCSR (Lattice) predictions at low (medium) momentum transfer 
for all the FFs appearing in 
$B \to K,\rho,K^*$ and $B_s \to \phi$
transitions. We have also investigated a simplified form of the 
SE, where the profile functions $\phi(z(t))$ 
are re-expanded in powers of $z(t)$, while the dispersive bounds take
a somewhat more complicated form. We find that
both the standard and the simplified SE give a
similarly good description of the FF functions.
In those cases, where Lattice estimates of the FFs is lacking, 
the SE is used to extrapolate the LCSR predictions to the high-$q^2$
region. Comparing fits with/without using the available 
Lattice data for $B \to K$ and $B\to \rho$ transitions, we judge these
extrapolations to be rather reliable. Some of our results could be
further improved in the future, by addressing some of the following 
issues: The experimental 
confirmation of a scalar $B_s$ resonance below $B$--$K$ threshold,
contributing to the scalar $B \to K$ FF. Decreasing the
uncertainties in Lattice predictions for $B \to \rho$ axial-vector
FFs. The calculation of LCSRs directly in the helicity basis.
The reliable computation of $B\to K^*$ and $B_s \to \phi$ FFs on
the Lattice.

In conclusion, we have shown that the parameterisation of heavy-to-light
FFs as a (truncated) SE in $z(t)$ in combination
with theoretical estimates from LCSRs and Lattice QCD is very useful, not
only for the determination of the CKM element $|V_{ub}|$ from charged
semi-leptonic $B\to\pi$ or $B\to \rho$ decays, but also for the description
of FFs for radiative and semi-leptonic $b\to s$ and $b\to d$ transitions,
which will continue to play a major role for the indirect search of new physics
effects from rare flavour decays.

\section*{Acknowledgements}

AKMB is very grateful to her supervisor Patricia Ball for many helpful
discussions, and also acknowledges an STFC studentship.

\appendix

\section{Kinematics and Polarization Vectors}
 
In the following, we consider the 
rest frame of the decaying $B$-meson, 
with the 3-momentum of the final-state meson
pointing in the $z$-direction.
The polarisation vectors for a (virtual) vector state,
with 4-momentum $q^\mu=(q^0,0,0,-|\vec q\,|)$, are defined as
 \begin{eqnarray}
 \varepsilon _ \pm ^\mu(q) &=&  \mp \frac{1}{{\sqrt 2 }} \, (0,1, \mp i,0) \,, 
\qquad 
 \varepsilon _0^\mu  (q) =   \frac{1}{{\sqrt {q^2 } }} \, (|\vec q\,| ,0,0, - q^0 ) \,, 
\cr 
 \varepsilon _t^\mu  (q) &=& \frac{1}{{\sqrt {q^2 } }} \, q^\mu \,.
\label{eq:defeps}
 \end{eqnarray}
For the decay of a $B$-meson at rest into a light meson with mass $m_L$ 
and momentum $\vec k$, we have in particular 
\begin{eqnarray}
q^0 &=&m_B-E=\frac{m_B^2  - m_L^2  + q^2 }{2m_B } \,, 
\qquad |\vec q\,|=|\vec k\,| = \frac{\sqrt \lambda  }{2m_B } \,,
\end{eqnarray} 
with $\lambda$ defined in (\ref{eq:lambdadef}). 
We also define the linear combinations
 \begin{eqnarray}
 \varepsilon _1^\mu  (q) &=& 
\frac{{\varepsilon _ - ^\mu  (q) - \varepsilon _ + ^\mu  (q)}}{{\sqrt 2 }} = (0,1,0,0) 
\,, \qquad 
 \varepsilon _2^\mu  (q) = 
\frac{{\varepsilon _ - ^\mu  (q) + \varepsilon _ + ^\mu  (q)}}{{\sqrt 2 }} = (0,0,i,0) \,.
 \end{eqnarray}
In the same way, the polarisation vectors for an on-shell  
$K^*$ meson with momentum $k^\mu=(E,0,0,|\vec k\,|)$ are given as
 \begin{eqnarray}
 \varepsilon _ \pm ^\mu  (k) &=&  \mp \frac{1}{{\sqrt 2 }}(0,1, \pm i,0) 
\,, \qquad  
 \varepsilon _0^\mu  (k) =   \frac{1}{m_{K^*}}(|\vec k\,| ,0,0,E) \,.
  \end{eqnarray}

\section{FF Properties}

\label{sec:newFF}

In the following, we summarize a few useful properties of the helicity-based FFs, 
following from the definitions in 
(\ref{eq:AVdef},\ref{eq:ATdef},\ref{eq:BVdef},\ref{eq:BTdef}).
From the e.o.m.\ for vanishing momentum transfer, $q^2 \to 0$, one derives
\begin{align}
 \mathcal{A}_{V,0}(0)  =
 \mathcal{A}_{V,t}(0)  & = f_0(0) = f_+(0) \,,
\label{eq:AVat0}
\end{align}
and
\begin{align}
\mathcal{B}_{V,0}(0)=\mathcal{B}_{V,t}(0) & = A_0(0) = A_3(0) \,, 
\cr 
\label{eq:BTat0}
\mathcal{B}_{T,1}(0)=\mathcal{B}_{T,2}(0)& =\sqrt{2} \, T_1(0) = \sqrt{2} \, T_2(0)
\,,
\end{align}  
while the FFs  $\mathcal{A}_{T,0}$, 
$\mathcal{B}_{V,1}$, $\mathcal{B}_{V,2}$, and
$\mathcal{B}_{T,0}$ vanish like $\sqrt{q^2}$. Similarly, at the kinematic
endpoint $q^2=t_- =(m_B-m_L)^2$, we obtain the relations
\begin{align}
\label{eq:addrel}
\lim_{q^2\to t_-} \, \frac{{\cal B}_{V,2}(q^2)}{{\cal B}_{V,0}(q^2)} & =
\lim_{q^2\to t_-} \, \frac{{\cal B}_{T,2}(q^2)}{{\cal B}_{T,0}(q^2)} = \sqrt2 \,,
\end{align}
which has been implemented in our FF parameterisation in 
Section~\ref{sec:fit}.

In the infinite-mass limit $m_b \to \infty$, one can project the 
heavy quark field in the decay current onto its large component in HQET,
which results in the well-known HQET spin-symmetry relations. With our
FF conventions, they read
\begin{align}
2 m_B \sqrt{q^2} \, {\cal A}_{T,0} &= 
 (m_B^2+q^2) \, {\cal A}_{V,0} - (m_B^2-q^2) \, {\cal A}_{V,t} \,, 
\end{align}
and
\begin{align}
 2 m_B \sqrt{q^2} \, {\cal B}_{T,0} &= 
 (m_B^2+q^2) \, {\cal B}_{V,0} + (m_B^2-q^2) \, {\cal B}_{V,t} \,,
\cr 
2 m_B \sqrt{q^2} \, {\cal B}_{T,1} &= 
 (m_B^2+q^2) \, {\cal B}_{V,1} +  (m_B^2-q^2) \, {\cal B}_{V,2} \,,
\cr 
 2 m_B \sqrt{q^2} \, {\cal B}_{T,2} &= 
 (m_B^2+q^2) \, {\cal B}_{V,2} +  (m_B^2-q^2) \,{\cal B}_{V,1} \,.
\end{align}

Furthermore, in the limit of large recoil energy to the final state meson, one obtains additional
relations \cite{Charles:1998dr,Beneke:2000wa} that follow from the factorisation of
soft and collinear QCD dynamics  \cite{Beneke:2003xr,Lange:2003pk,Beneke:2003pa}. 
Excluding radiative corrections for $q^2, m_{L}^2 \ll m_B^2$,
and using the FFs ${\cal A}$ and ${\cal B}$, these relations read:
\begin{align}
& {\cal A}_{V,0} \simeq {\cal A}_{V,t} \simeq \frac{m_B}{\sqrt{q^2}} \, {\cal A}_{T,0} 
\,, 
\end{align}
and
\begin{align}
& {\cal B}_{V,0} \simeq {\cal B}_{V,t} \simeq \frac{m_B}{\sqrt{q^2}} \, {\cal B}_{T,0} 
\,, 
\qquad {\cal B}_{V,1} \simeq {\cal B}_{V,2} 
\simeq \frac{\sqrt{q^2}}{m_B} \, {\cal B}_{T,1}
\simeq   \frac{\sqrt{q^2}}{m_B} \, {\cal B}_{T,2} \,.
\end{align}

It is also useful to write the differential decay rates for $B \to P \ell^+\ell^-$
and $B \to V \ell^+\ell^-$ in the naive-factorization approximation in terms of
the new FFs. 
For the decays into pseudoscalar mesons, we
obtain the relatively simple expression
\begin{align}
 \frac{d\Gamma[B \to P \ell^+\ell^-]}{dq^2} &= 
2 \, {\cal N}^{\,2} \, \frac{ \left( 2 m_b \, C_7^{\rm eff}\right)^2}{q^2} 
                   \, ({\cal A}_{T,0})^2
 +
4\, {\cal N}^{\,2} \, \frac{ {\rm Re} \left[2 m_b \, C_7^{\rm eff} C_9^{\rm eff}{}^*\right]}{\sqrt{q^2}} 
                   \, {\cal A}_{T,0} \, {\cal A}_{V,0} 
\cr 
& \quad +
2 \, {\cal N}^{\,2} \left( |C_{9}^{\rm eff}|^2 + |C_{10}|^2 \right)
                             \, ({\cal A}_{V,0})^2 
\,,
\end{align}
where the overall normalization is given by
\begin{align}
{\cal N} &=  \left|V_{tb}^{\vphantom{*}} \, V_{ts}^*\right| 
 \left[\frac{G_F^2 \alpha^2}{3\cdot 2^{10}\pi^5 m_B^3}
 \lambda^{3/2}\nonumber\right]^{1/2} \,.
\end{align}
For decays into vector mesons, we decompose the doubly differential decay width as
\cite{Kruger:2005ep,Altmannshofer:2008dz}
\begin{align}
\frac{d^2\Gamma[B \to V \ell^+\ell^-]}{dq^2 \, d\cos\theta} 
&=& 
\frac38 \left[(1+\cos^2\theta) \, H_T(q^2) + 2 \cos\theta \, H_A(q^2) 
 + 2 \, (1-\cos^2 \theta) \, H_L(q^2) \right]
\,.
\end{align}
Here $\theta$ is the angle between the positively charged lepton and the
3-momentum of the $\bar B^0$ or $B^-$\/--meson, and the functions $H_T$, $H_L$, $H_T$ 
are related to the transverse rate, the longitudinal rate and the forward-backward
asymmetry, respectively. In terms of the new FFs, the naive-factorization approximation
for the SM contribution results in 
\begin{align}
 H_T(q^2) &= 
2 \, {\cal N}^{\,2} \, \frac{ \left( 2 m_b \, C_7^{\rm eff}\right)^2}{q^2} 
                   \left(  ({\cal B}_{T,1})^2 + ({\cal B}_{T,2})^2 \right)
\cr 
& \quad +
4 \, {\cal N}^{\,2} \, \frac{ {\rm Re} \left[2 m_b \, C_7^{\rm eff} C_9^{\rm eff}{}^*\right]}{\sqrt{q^2}} 
 \left(  {\cal B}_{T,1} \, {\cal B}_{V,1}  +   {\cal B}_{T,2} \, {\cal B}_{V,2}\right)
\cr 
& \quad +
2 \, {\cal N}^{\,2} \left( |C_{9}^{\rm eff}|^2 + |C_{10}|^2 \right)
                             \left( ({\cal B}_{V,1})^2 + ({\cal B}_{V,2})^2 \right) 
  \,,
\\[0.25em]
 H_L(q^2)&=
2 \, {\cal N}^{\,2} \, \frac{ \left( 2 m_b \, C_7^{\rm eff}\right)^2}{q^2} 
                   \, ({\cal B}_{T,0})^2
 +
4\, {\cal N}^{\,2} \, \frac{ {\rm Re} \left[2 m_b \, C_7^{\rm eff} C_9^{\rm eff}{}^*\right]}{\sqrt{q^2}} 
                   \, {\cal B}_{T,0} \, {\cal B}_{V,0} 
\cr 
& \quad +
2 \, {\cal N}^{\,2} \left( |C_{9}^{\rm eff}|^2 + |C_{10}|^2 \right)
                             \, ({\cal B}_{V,0})^2 
  \,,
\\[0.25em]
H_A(q^2) &=
 8 \, {\cal N}^{\,2} \, {\rm Re} \left\{ 
 C_9^{\rm eff}{}^* \, C_{10} \, {\cal B}_{V,1} \, {\cal B}_{V,2}
+
 \frac{m_b}{\sqrt{q^2}} \, C_7^{\rm eff} \, C_{10} 
 \left( {\cal B}_{T,1} \, {\cal B}_{V,2} + {\cal B}_{T,2} \, {\cal B}_{V,1} \right)
\right\} \,. 
\end{align}
Finally, we note the simple FF dependence of the differential decay width and the longitudinal polarization fraction $F_L$ of the vector meson in the decays $B \to V \nu \bar\nu$. 
In the absence of right-handed currents, we obtain \cite{Altmannshofer:2009ma}
\begin{equation}
\frac{d\Gamma[B\to V \nu \bar \nu]}{dq^2 } =  12 \, \mathcal{N}^2\, C_L^\nu \left( (\mathcal{B}_{V,0})^2+(\mathcal{B}_{V,1})^2+(\mathcal{B}_{V,2})^2\right)\,,
\end{equation}
and
\begin{equation}
F_L(q^2) =  \frac{(\mathcal{B}_{V,0})^2}{(\mathcal{B}_{V,0})^2+(\mathcal{B}_{V,1})^2+(\mathcal{B}_{V,2})^2}
\,.
\end{equation}
At the kinematic endpoint ${\cal B}_{V,1}(t_-)$ vanishes, and (\ref{eq:addrel}) implies
$F_L(t_-)=1/3$, as required on general grounds.

\section{Calculation of Wilson Coefficients $\chi_I^X$}

\label{app:chiIX}

\subsection{Perturbative Contribution}

\begin{figure}[t!!!pbht]
\begin{center}
\includegraphics[scale=0.9]{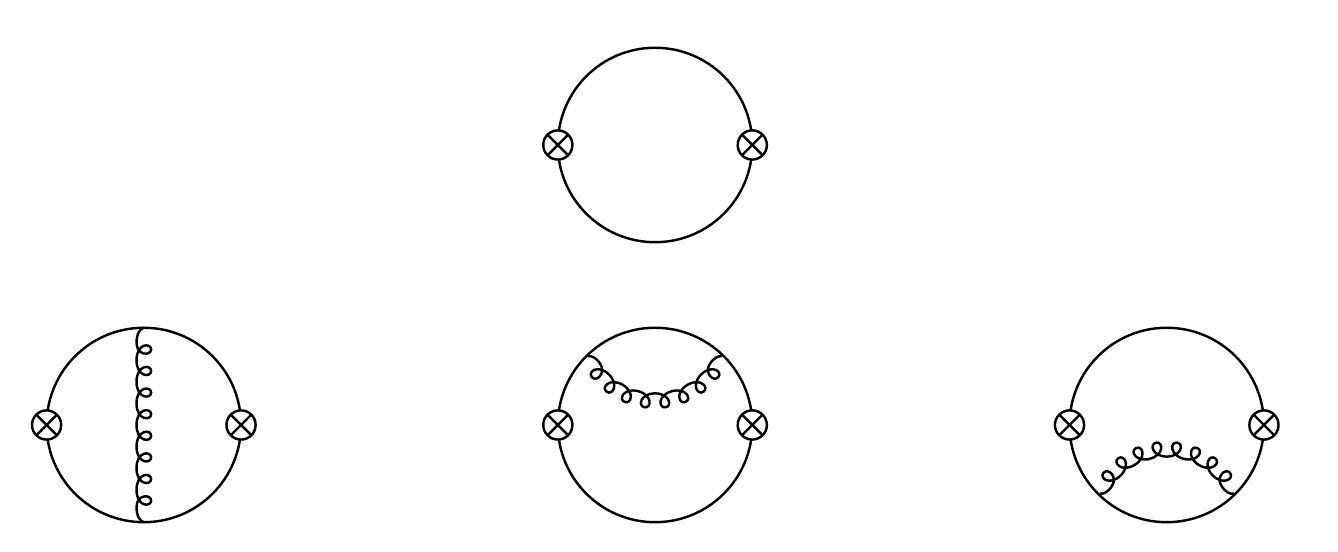} 
\end{center} 
\caption{One- and two-loop diagrams contributing to the correlation function. 
The crossed circle indicates the insertion of the corresponding scalar, vector
or tensor currents. The counter-term diagrams related to the fermion self-energies 
are not shown.}
\label{perturbative_diagrams}
\end{figure}

In this section, we will briefly sketch the evaluation of the one- and two-loop diagrams 
(see Fig.~\ref{perturbative_diagrams}) contributing to the perturbative part of the
 correlation functions.
We will specify the necessary number $n$ of subtractions for the scalar, 
vector and tensor correlators, and determine the corresponding values of $\chi_I^X(n)$
from the Taylor expansion of the Wilson coefficients at $q^2=0$. This leads to a 
major simplification in the calculation, which allows to
eliminate external momenta in propagator denominators and to use tensor reduction
and recursion relations to express the two-loop integrals in terms of two fundamental
master integrals.
Furthermore, we will follow the procedure
explained in \cite{Jamin:1992se} and absorb the IR-sensitive contributions
to the Feynman integrals (in the limit $m \to 0$) into the corresponding
condensate terms, such that our results have a finite limit when $m \to 0$.

We will find in useful to present the result in terms of the dimensionless variable
\begin{equation}
v \equiv \frac{M-m}{M+m} \,,
\end{equation}
where $M$ and $m$ are the masses of the heavy and light quark in the loop.
We further define the functions
\begin{eqnarray}
f_1(v)& \equiv & \frac{1-v^2}{v} \, \textrm{atanh}[v]
 \,, 
\nonumber \\[0.2em]
f_2(v) & \equiv & 
\frac{1}{v} \, \ln \left[\frac{1-v}{1+v}\right]
- \frac{2}{1-v} \, \ln \left[\frac{1+v}{2}\right]
- \frac{2}{1+v} \, \ln \left[\frac{1-v}{2}\right]
 \,,
\nonumber \\[0.2em]
f_3(v)& \equiv &
  \frac{1}{v} \, \textrm{Li}_2\left[\frac{4 v}{(1+v)^2}\right]
  -\frac{1}{v} \,\textrm{Li}_2\left[-\frac{4 v}{(1-v)^2}\right] 
  -\frac{4 \,(1 + v^2)}{v^2} \, \textrm{atanh}^2[v] 
 \,,
\end{eqnarray}
which are manifestly symmetric under exchange
of light and heavy quarks ($ v \to -v$),
and take finite values in the limits $v \to \{-1,0,1\}$.

We will quote our results for scalar, vector and tensor currents.
The expressions for currents with opposite parity can be simply
obtained by changing $v \to 1/v$. Our expressions for scalar and
vector currents coincide with \cite{Jamin:1992se}; the results for
the tensor currents are new.

\paragraph{Scalar Correlator:}

For the correlator of two scalar currents, we obtain
\begin{eqnarray}
 \chi^{S}(n=2) \Big|_{\rm LO}
& = &
\frac{(3 + v^2) (3v^2-1)}{64 \pi ^2 (M+m)^2 \, v^4}
 \ \stackrel{v \to 1}{\rightarrow} \ 
\frac{1}{8 \pi ^2  M^2} \,,
\\[0.3em]
 \chi^{S}(n=2) \Big|_{\rm NLO}
&  = &
\frac{\alpha_s C_F}{4 \pi}
\, \frac{1}{64 \pi ^2 (M+m)^2 \, v^4}
\Bigg\{ 
\cr 
&&  6 
\left( 3 f_1 \, (1-v^2)^2 + (3+v^2)(3 v^2-1) \right)
\left(  f_2 \, (1-v^2) - 4 \, \ln \left[\frac{m+M}{\mu }\right] \right)
\cr 
&&
- f_1^2 \left(11 v^4-50
   v^2+23\right)
+  f_1 \left(47 v^4-126
   v^2+103\right)
\cr
&& + 4 \, f_3 \, v^2 \, (5
   v^2-1)
   + 2 \left(29 v^4+65 v^2-40\right)
\Bigg\} 
\nonumber \\[0.2em]
&\stackrel{v \to 1}{\rightarrow}& 
\frac{1}{8 \pi ^2 M^2} 
\,
\frac{\alpha_s C_F}{4 \pi} 
\Bigg\{
 -24 \, \ln \left[\frac{M}{\mu }\right]
    +\frac{2\pi ^2}{3}+\frac{27}{2} 
\Bigg\}
\,.
\end{eqnarray}

\paragraph{Vector Correlator:}

For the different projections of the correlator of two vector currents, we 
obtain
\begin{eqnarray}
 \chi^V_L(n=1) \Big|_{\rm LO}
& = &
\frac{(3 + v^2) (3v^2-1)}{64 \pi^2 \, v^2}
\ \stackrel{v \to 1}{\rightarrow} \
\frac{1}{8\pi^2} \,, 
\\[0.3em]
 \chi^V_L(n=1)\Big|_{\rm NLO}
& = & \frac{\alpha_s C_F}{4 \pi}
\, \frac{1}{64 \pi ^2 \, v^2}
\Bigg\{ 
\cr 
&&
 f_1^2 \left( 25 v^4 + 14 v^2 - 23 \right)
+ 2 f_1 \left( 19v^4-6v^2+23 \right)
\cr 
&& + 4 f_3 \, v^2 \, (5 v^2 -1)
 - 23 + 14 v^2 + 13 v^4
\Bigg\} \,,
\cr 
& \stackrel{v \to 1}{\rightarrow} &\frac{\alpha_s C_F}{4 \pi}
\, \frac{1}{8 \pi ^2 } \left( \frac12 + \frac{2\pi^2}{3} \right) \,,
\end{eqnarray}
and 
\begin{eqnarray}
 \chi^V_T(n=2) \Big|_{\rm LO}
& = &
\frac{-21 v^6 + 53 v^4 + 13 v^2 + 3}{512 \pi^2 \, (M+m)^2 \, v^4}
\ \stackrel{v \to 1}{\rightarrow} \
\frac{3}{32\pi^2 M^2} \,,
\\[0.3em]
 \chi^V_T(n=2)\Big|_{\rm NLO}
& = & \frac{\alpha_s C_F}{4 \pi}
\, \frac{1}{1536 \pi^2 \, (M+m)^2 \, v^4} 
\Bigg\{ 
\cr 
&&
-f_1^2 \left(803v^6-863 v^4 -155 v^2 - 73 \right)
- 2 f_1 \left(677v^6-741 v^4+279 v^2+73 \right)
\cr 
&& - 4 f_3 \, v^2 \left(19v^4-86v^2-5 \right)
 + 73 + 323 v^2 + 755 v^4 - 551 v^6
\Bigg\} \,,
\cr 
& \stackrel{v \to 1}{\rightarrow} &\frac{\alpha_s C_F}{4 \pi}
\, 
\frac{3 }
{32 \pi^2 M^2 } 
\left( \frac{25}{6} + \frac{2 \pi^2}{3}
\right)  \,.
\end{eqnarray}

\paragraph{Tensor Correlator:}

The relevant projection of the tensor current gives rise to
\begin{eqnarray}
\left. {\chi ^{T}_T (n = 3)} \right|_{LO}
& = &
\frac{-9 f_1 \left(v^2-1\right)^2 \left(3 v^2+1\right)+4 \left(-9 v^6+21 v^4+v^2+3\right)}{256 \pi ^2 \left(m+M\right)^2 v^4}
\cr  
	 & \mathop  \to \limits^{v \to 1}  &  \frac{1}{4 \pi ^2 M^2} 
\\[0.3em]
\left. {\chi ^{T}_T (n = 3)} \right|_{NLO}
 & = &
\frac{\alpha_s C_F }{4\pi}
\,
\frac{1}{384 \pi^2 (M+m)^2 v^4} 
 \Bigg\{
\cr 
&&
12
\left(
3 (v^2-1)^2 (3v^2+1) f_1 
 -3 - v^2 - 21 v^4 + 9 v^6
\right)
\cr 
&& \qquad \times 
\left(  f_2 \, (1-v^2) - 4 \, \ln \left[\frac{m+M}{\mu }\right] \right)
\cr 
&& - f_1^2 \left(
766 v^6-598v^4-142v^2-218
\right)
\cr 
&&
- f_1 \left( 1091 v^6 -1137 v^4 +297 v^2+325
\right)
\cr 
&&
- 8 f_3 \, v^2 \left( 7 v^4-26 v^2 -5 \right)
+ 107 + 69 v^2 + 469 v^4 - 325 v^6
\Bigg\}
\cr 
	 & \mathop  \to \limits^{v \to 1}  &
 \frac{\alpha_s C_F}{4\pi} 
\,
 \frac{1}{4\pi^2 M^2}
  \left(
\frac{10}{3} + \frac{2\pi^2}{3} 
+ 8 \ln \left[\frac{M}{\mu}\right]\right)
\,. 
\end{eqnarray}

\subsection{Condensate Contribution to the Correlation Functions}

In this section we provide the expressions for the contributions of the
gluon condensate, the quark condensate and the mixed quark-gluon condensate to 
the various current correlators.
The contributions to the coefficient of the scalar and vector correlators to all orders in the quark mass and lowest order in the coupling constant can already be found in \cite{Jamin:1992se}, and we reproduce the results given in that paper. 
We extend this analysis by determining the coefficient functions for the tensor correlators. 
For the quark and the quark-gluon condensate, we employ techniques analogous to that given in \cite{Jamin:1992se} and closely follow their notation. 
In case of the gluon condensate, we use the plane-wave technique.

\paragraph{Quark Condensate and Quark--Gluon Condensate:}

\begin{figure}[t!pbh]
\begin{center}
\includegraphics[scale=1]{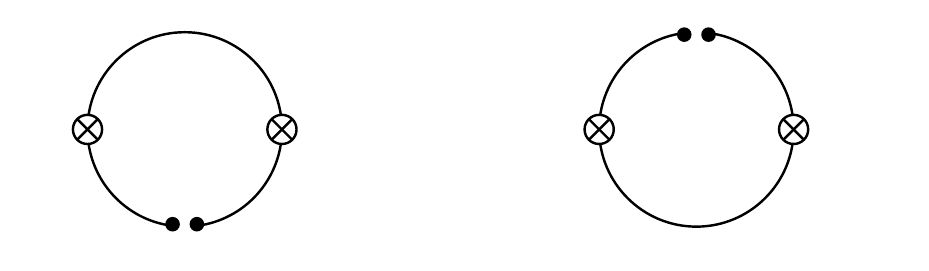}
\end{center} 
\caption{Diagrams involving the non-local quark condensate, indicated by the two solid dots. 
The crossed circle symbolises the insertion of the currents.}
\label{nonlocalquarkcondensate}
\end{figure}

\begin{figure}[t!pbh]
\begin{center} 
\includegraphics[scale=1]{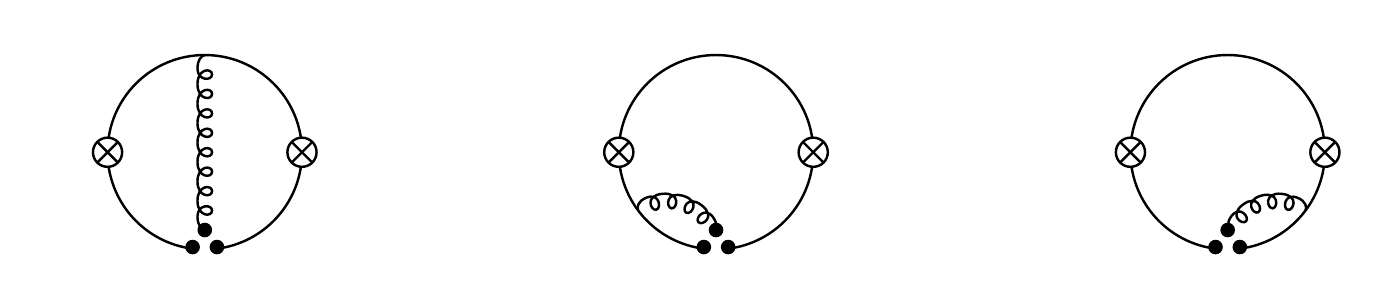} 
\end{center} 
\label{nonlocalquarkgluonmixdedcondensate}
\caption{Diagrams involving the non-local mixed quark-gluon condensate, indicated 
by the three solid dots.}
\end{figure}

The starting point for calculating the coefficient functions to all orders in 
the quark masses is a closed expression for the non-local quark condensate. 
The position-space expressions
for the projection of the non-local quark condensates on the local quark condensate 
$\qq$ and the local mixed quark-gluon condensate $\qFq$ read
\begin{eqnarray}
\lnp\bar q_{\alpha}(0)q_{\beta}(x)\rnp_{\bar qq} \;& = &\; \frac{1}{4m}\,
\qq\,\Gamma\left(\frac{D}{2}\right)(i\!\Slash\partial+m)_{\beta\alpha}
\sum_{n=0}^{\infty} \frac{(-m^{2}x^{2}/4)^{n}}{n!\,\Gamma(n+D/2)} \,, \nn
\\
\lnp\bar q_{\alpha}(0)q_{\beta}(x)\rnp_{\bar qFq} & = &
-\,\frac{1}{8m^{3}}\,\qFq\,\Gamma\left(\frac{D}{2}\right) \nn \\
\smvs
& & {} \times \sum_{n=0}^{\infty} \Big[\, (n-1)i\!\!\Slash\partial+n\,m\,\Big] 
_{\beta\alpha}\,\frac{(-m^{2}x^{2}/4)^{n}}{n!\,\Gamma(n+D/2)} \,.
\end{eqnarray}
Here $\alpha$ and $\beta$ indicate the spinor indices.
The corresponding projection of the non-local mixed quark-gluon condensate reads
\begin{eqnarray}
&& \lnp g_{s}\bar q_{\alpha}(0)F_{\mu\nu}(0)\,q_{\beta}(x)\rnp_{\bar qFq}
 =  \frac{1}{4(D-1)(D-2)\,m^{2}}\,\qFq\,\Gamma\left(\frac{D}{2}\right)
 \nn \\
&&  \qquad {}
\times  \left[ \Big(\big(\gamma_{\mu}\partial_{\nu}-\gamma_{\nu}\partial_{\mu}
\big)+m \sigma_{\mu\nu}\Big) \big(i\!\!\Slash\partial+m\big) \right]_{\beta\alpha} \,
\sum_{n=0}^{\infty} \frac{(-m^{2}x^{2}/4)^{n}}{n!\Gamma(n+D/2)} \,. 
\end{eqnarray}
The relevant diagrams for the contribution of the non-local quark condensate and the non-local mixed quark-gluon condensate
are given in Figs.~\ref{nonlocalquarkcondensate}  
and \ref{nonlocalquarkgluonmixdedcondensate}, respectively.
The evaluation of the diagrams is simplified by the use of the equations of motion,
\begin{eqnarray}
\big(\!\Slash p-m\big) \, \lnp g_{s}\bar q_{\alpha}(0)F_{\mu\nu}(0)
q_{\beta}(x)\rnp_{\bar qFq} &=& 0  \,,
 \\
\big(p^{2}-m^{2}\big) \, \lnp\bar q(0)\tilde q(p)\rnp_{\bar qFq}
 &=& -\frac{\qFq}{2\,\qq}\,\lnp\bar q(0)\tilde q(p)\rnp_{\bar qq} \,,
\\
\big(\!\Slash p-m\big) \, \lnp\bar q(0)\tilde q(p)\rnp_{\bar qq} &=& 0 \,.
\end{eqnarray}

\paragraph{Gluon Condensate:}

\begin{figure}[h]
\begin{center}
\includegraphics[scale=1]{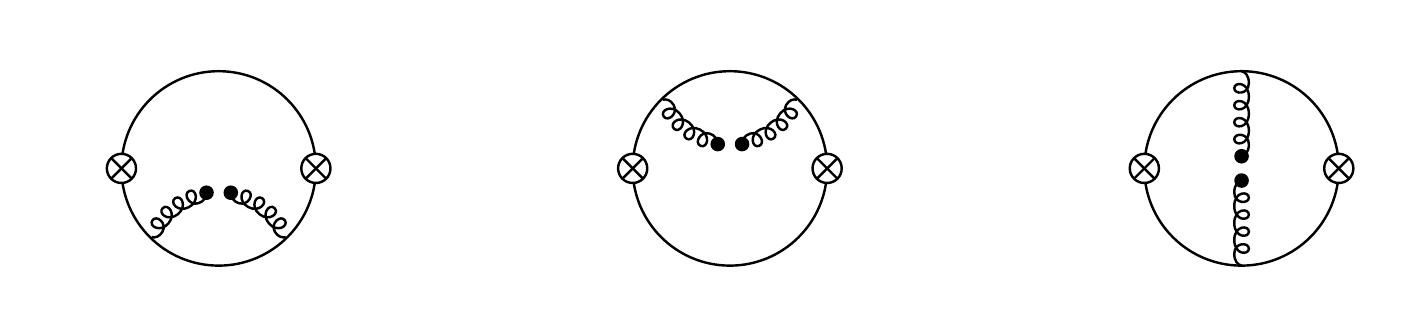}
\end{center} 
\caption{Diagrams involving the gluon condensate.}
\label{gluoncondensate}
\end{figure}

For the gluon condensate, it is more convenient to use the so-called fixed-point gauge technique, which is described in detail in \cite{Reinders:1984sr}. In the framework of the fixed-point gauge, it is possible to derive an expression for
\begin{equation}
\includegraphics[scale=1]{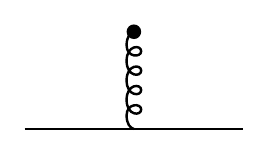} = - \frac{i}{4}gt^a G_{\kappa\lambda }^a (0)\frac{1}{{\left( {p^2  - m^2 } \right)}}\left\{ {\sigma _{\kappa \lambda } (\slashed p + m) + (\slashed p + m)\sigma _{\kappa\lambda } } \right\} \,,
\end{equation}
which is the basic building block for three lowest-order diagrams 
shown in Fig.~\ref{gluoncondensate}.


\subsubsection{Results}

\paragraph{Quark Condensate:}

The quark-condensate contribution to the coefficient for the
scalar correlation function is given by
 \begin{eqnarray}
{\chi^{S} (n = 2)} \Big|_{\bar qq} 
& = & 
\frac{\langle \bar qq\rangle\, (v+1)^3}{8 (m+M)^5 v^5} 
\ \stackrel{v\to 1}{\rightarrow}\ \frac{\langle\bar qq\rangle}{M^5} \,.
 \end{eqnarray}
The same expression (up to an overall normalization factor)
is obtained in case of the longitudinal projection
of the vector correlator,
${\chi^{V}_L (n = 1)} \big|_{\bar qq} =  (M+m)^2 v^2 \, {\chi^{S} (n = 2)} \big|_{\bar qq} $.
The transverse projection of the vector correlator leads to
 \begin{eqnarray}
{\chi ^{V}_T (n = 2)} \Big|_{\bar qq} 
& = & -\frac{\langle \bar qq\rangle\, (v+1)^3}{64 (m+M)^5 v^5} \left(7 v^2+1\right)
 \ \stackrel{v \to 1}{\rightarrow} \  -\frac{\langle\bar qq\rangle}{M^5} \,.
 \end{eqnarray}
Finally, from the relevant tensor correlator we obtain
\begin{eqnarray}
{\chi ^{T}_T (n = 3)} \Big|_{\bar qq} & = & 
-\frac{\langle \bar qq\rangle \, (v+1)^3}{32 (m+M)^5 v^5}
 \left(3 v^2+1\right)
 \ \stackrel{v\to 1}{\rightarrow} \ -\frac{\langle\bar qq\rangle }{M^5} \,.
 \end{eqnarray}

\paragraph{Gluon Condensate:}

The expressions for the gluon-condensate contributions to
the various $\chi_I^X$ coefficients read as follows:
For the scalar correlator we obtain
\begin{eqnarray}
{\chi ^{S} (n = 2)} \Big|_{G^2} 
& = &
\frac{\left\langle {\frac{\alpha}{\pi} G^2} \right\rangle }{96 \left(m+M\right)^6 v^6}
\Bigg\{
15 f_1 \, (1-v^2)^2
-15-4 v^2+27 v^4 
\cr 
&& - 6 v^2 (1-v^2) \left(  f_2 \, (1-v^2) - 4 \, \ln \left[\frac{m+M}{\mu }\right] \right)
\Bigg\}
\cr  
& \stackrel{v\to 1}{\rightarrow}&
\frac{\left\langle {\frac{\alpha}{\pi} G^2} \right\rangle }{12 M^6} \,.
\end{eqnarray}
Again, the same expression is obtained for the longitudinal projection of
the vector correlator,
${\chi^{V}_L (n = 1)} \big|_{GG} =  (M+m)^2 v^2 \, {\chi^{S} (n = 2)} \big|_{GG} $\,.
For the transverse projection of the vector correlator, one has
\begin{eqnarray}
{\chi^{V}_T (n = 2)} \Big|_{G^2} 
& = &
\frac{\left\langle {\frac{\alpha}{\pi} G^2} \right\rangle }{384 \left(m+M\right)^6 v^6}
\Bigg\{
\cr &&
45+115 v^2+3 v^4-195 v^6
-5 f_1 \, (1-v^2)^2 
\left(25 v^2+9\right) 
\cr  
 	 & &  
+ v^2  (1-v^2) \left(35 v^2+41\right)
 \left(  f_2 \, (1-v^2) - 4 \, \ln \left[\frac{m+M}{\mu }\right] \right)
\Bigg\}
 	 \cr  
	 & \stackrel{v\to1}{\rightarrow}&
-\frac{\left\langle {\frac{\alpha}{\pi} G^2} \right\rangle }{12 M^6} \,,
\end{eqnarray}
and for the tensor correlator, we get
\begin{eqnarray}
{\chi ^{T}_T (n = 3)} \Big|_{G^2} 
& = &
\frac{\left\langle {\frac{\alpha}{\pi} G^2} \right\rangle }{384 \left(m+M\right)^6 v^6}
\Bigg\{
\cr 
&& 
105+91 v^2 -17 v^4-195 v^6
-5 f_1 \, (1-v^2)^2 
\left(27 v^2+17\right) 
\cr  
 	 & &  
+2 (1-v^2) \left(15 v^4+24 v^2+5\right)
 \left(  f_2 \, (1-v^2) - 4 \, \ln \left[\frac{m+M}{\mu }\right] \right)
\Bigg\}
 	 \cr  
	 & \stackrel{v\to1}{\rightarrow}&
 -\frac{\left\langle {\frac{\alpha}{\pi} G^2} \right\rangle }{24 M^6} \,.
\end{eqnarray}

\paragraph{Mixed Condensate:}

For the mixed-condensate contributions, we finally obtain
 \begin{eqnarray}
\left. {\chi ^{S} (n = 2)} \right|_{\bar qGq} & = &  
-\frac{\big<\bar qGq\big> (1+v)}{4 (m+M)^7 v^5} \,(2+v)
\ \mathop  \to \limits^{v \to 1}  \ -\frac{3 \big<\bar qGq\big>}{2 M^7} \,,
 \end{eqnarray}
 and
 \begin{eqnarray}
\left. {\chi ^{V}_L (n = 1)} \right|_{\bar qGq} & = & 
-\frac{\big<\bar qGq\big>(1+v)}{4 (m+M)^5 v^3} \,(2+v) \
  \mathop  \to \limits^{v \to 1}  \ -\frac{3 \big<\bar qGq\big>}{2 M^5}\,,
 \end{eqnarray}
where ${\chi ^{V}_L (n = 1)} \big|_{\bar qGq}$ is again proportional
to $ {\chi ^{S} (n = 2)} \big|_{\bar qGq}$,
 as well as
 \begin{eqnarray}
\left. {\chi ^{V}_T (n = 2)} \right|_{\bar qGq} & = & 
\frac{\big<\bar qGq\big>(1+v) }{96 (m+M)^7 v^5}
\left(35 v^3+59 v^2+ 41 v+9\right)\
  \mathop  \to \limits^{v \to 1}  \ \frac{3 \big<\bar qGq\big>}{M^7} \,,
 \end{eqnarray}
and 
\begin{eqnarray}
\left. {\chi ^{TT} (n = 3)} \right|_{\bar qGq} 
& = & \frac{\big<\bar qGq\big>(1+v)}{48 (m+M)^7 v^6} 
\left(15 v^4+28 v^3+24 v^2+12 v+5\right)\cr 
  &\mathop  \to \limits^{v \to 1}  & \frac{7 \big<\bar qGq\big>}{2 M^7} \,.
 \end{eqnarray}


\section{Decomposition of the tensor-current correlator}
 
\label{app:proj}

Using the projectors
 \begin{equation}
 P _L^{\mu \nu }  = \frac{{q{}^\mu q^\nu  }}{{q^2}} \,, \qquad 
 P _T^{\mu \nu }  = \frac{{\left( {q^\mu  q^\nu   - g^{\mu \nu } q^2 } \right)}}{{(D-1)q^2 }}
\,, 
 \end{equation}
we decompose the correlator of general tensor currents,
  \begin{eqnarray}
   \Pi_{\mu\nu\alpha\beta} &=&
  i\!\int\!\! d^4x\, e^{iqx} \,\langle 0|
  T[\bar q_1(x)\sigma_{\mu\nu} q_2(x)\, \, \bar q_2(0)
  \sigma_{\alpha\beta} q_1(0)]|0\rangle\,,
  \end{eqnarray}
into the two Lorentz-invariant functions $\Pi_{TT}$ and $\Pi_{LT}$ as
follows,
\begin{eqnarray}
   \Pi_{\mu\nu\alpha\beta} &=&
   [g_{\mu\alpha}g_{\nu\beta} - g_{\mu\beta}g_{\nu\alpha}]\frac{3\Pi_{TT}(q^2)}{2}\
  \cr 
  &&{}
  + \frac{g_{\mu\beta} q_\nu q_\alpha
 +  g_{\nu\alpha} q_\mu q_\beta - g_{\mu\alpha} q_\nu q_\beta
  -  g_{\nu\beta} q_\mu q_\alpha}{q^2} \left(\frac{3\Pi_{TT}(q^2)}{2} +\Pi_{LT}(q^2)\right)\,,
  \end{eqnarray}
where 
\begin{eqnarray}
P _L^{\mu \alpha } \, P _T^{\nu \beta } \, \Pi_{\mu\nu\alpha\beta}&=&
P _T^{\mu \alpha } \, P _L^{\nu \beta }\, \Pi_{\mu\nu\alpha\beta}=\Pi_{LT}(q^2) \,,
\cr 
P _T^{\mu \alpha } \, P _T^{\nu \beta }\, \Pi_{\mu\nu\alpha\beta}&=&\Pi_{TT}(q^2) \,,
\cr 
P _L^{\mu \alpha } \, P _L^{\nu \beta }\, \Pi_{\mu\nu\alpha\beta}&=&0 \,.
\end{eqnarray}
In this notation, the correlator of the currents
\begin{equation}
 j_\mu^T =\bar q \sigma_{\mu\alpha }q^\alpha q  
\end{equation}
leads to $q^2 \, \Pi_{LT}(q^2)$.

\section{Covariance Matrices}
\label{app:covmat}

Here we give the covariance matrices as defined in (\ref{eq:U}) for the parameters corresponding to the best-fit parameters in Tables \ref{tab:AKSE} to \ref{tab:BsphiSSE}.

\paragraph{\boldmath $B\to K$ form factor fit: \unboldmath}

\begin{itemize}
 \item The fit of $B\to K$ FFs to LCSR data alone gives
the covariances matrices:
 \begin{center}
  \begin{tabular}{ccc}
\T\B & SE & SSE\\[0.3cm]
$\mathcal{A}_{V,0}$& $\left(
\begin{array}{cc}
 1.56\times 10^{{-5}} & -1.04\times 10^{{-4}} \\
 -1.04\times 10^{{-4}} & 9.59\times 10^{{-4}}
\end{array}
\right)$ & $\left(
\begin{array}{cc}
 4.39\times 10^{{-3}} & -2.91\times 10^{{-2}} \\
 -2.91\times 10^{{-2}} & 0.266
\end{array}
\right)$\\
\vspace{-.3cm}\\
$\mathcal{A}_{V,t}^{\mathrm{no\ res.}}$&$\left(
\begin{array}{cc}
 1.19\times 10^{{-4}} & -7.87\times 10^{{-4}} \\
 -7.87\times 10^{{-4}} & 6.98\times 10^{{-3}}
\end{array}
\right)$ & $\left(
\begin{array}{cc}
 7.17\times 10^{{-3}} & -4.75\times 10^{{-2}} \\
 -4.75\times 10^{{-2}} & 0.423
\end{array}
\right)$\\\vspace{-.3cm}\\
$\mathcal{A}_{V,t}$&$\left(
\begin{array}{cc}
 6.27\times 10^{{-6}} & -2.72\times 10^{{-5}} \\
 -2.72\times 10^{{-5}} & 2.19\times 10^{{-4}}
\end{array}
\right)$ & $\left(
\begin{array}{cc}
 2.61\times 10^{{-3}} & -1.08\times 10^{{-2}} \\
 -1.08\times 10^{{-2}} & 8.86\times 10^{{-2}}
\end{array}
\right)$
\vspace{.3cm}\\
$\mathcal{A}_{T,0}$&$\left(
\begin{array}{cc}
 2.1\times 10^{{-5}} & -6.55\times 10^{{-5}} \\
 -6.55\times 10^{{-5}} & 5.37\times 10^{{-4}}
\end{array}
\right)$ & $\left(
\begin{array}{cc}
 7.63\times 10^{{-4}} & 6.3\times 10^{{-4}} \\
 6.3\times 10^{{-4}} & 8.32\times 10^{{-3}}
\end{array}
\right)$\\
  \end{tabular}
\end{center}

\item For the fit of scalar/vector $B \to K$ FFs to LCSR and Lattice data,
we obtain the covariance matrices:
 \begin{center}
  \begin{tabular}{ccc}
\T\B & SE & SSE\\[0.3cm]
$\mathcal{A}_{V,0}$&
$\left(
\begin{array}{cc}
 1.48\times 10^{{-5}} & -9.81\times 10^{{-5}} \\
 -9.81\times 10^{{-5}} & 8.76\times 10^{{-4}}
\end{array}
\right)$ & $\left(
\begin{array}{cc}
 6.26\times 10^{{-3}} & -4.15\times 10^{{-2}} \\
 -4.15\times 10^{{-2}} & 0.382
\end{array}
\right)$\\\vspace{-.3cm}\\
$\mathcal{A}_{V,t}^{\mathrm{no\ res.}}$&$\left(
\begin{array}{cc}
 4.82\times 10^{{-5}} & -2.03\times 10^{{-4}} \\
 -2.03\times 10^{{-4}} & 1.6\times 10^{{-3}}
\end{array}
\right)$ & $\left(
\begin{array}{cc}
 3.08\times 10^{{-3}} & -1.39\times 10^{{-2}} \\
 -1.39\times 10^{{-2}} & 0.11
\end{array}
\right)$
\\\vspace{-.3cm}\\
$\mathcal{A}_{V,t}$&$\left(
\begin{array}{cc}
 6.21\times 10^{{-5}} & -4.11\times 10^{{-4}} \\
 -4.11\times 10^{{-4}} & 3.75\times 10^{{-3}}
\end{array}
\right)$ & $\left(
\begin{array}{cc}
 3.45\times 10^{{-3}} & -2.37\times 10^{{-2}} \\
 -2.37\times 10^{{-2}} & 0.261
\end{array}
\right)$
  \end{tabular}
\end{center}
\end{itemize}

\paragraph{\boldmath $B\to \rho$ form factor fit: \unboldmath}

\begin{itemize}
\item Fitting to LCSR data alone, the covariance matrices for the $B\to \rho$
  FFs are given by:

\begin{center}\begin{tabular}{ccc}\T\B & SE & SSE\\$\mathcal{B}_{V,0}$&$\left(
\begin{array}{cc}
 4.15\times 10^{{-6}} & -3.2\times 10^{{-5}} \\
 -3.2\times 10^{{-5}} & 7.93\times 10^{{-4}}
\end{array}
\right)$&$\left(
\begin{array}{cc}
 5.26\times 10^{{-3}} & -4.33\times 10^{{-2}} \\
 -4.33\times 10^{{-2}} & 1.57
\end{array}
\right)$\\ \vspace{-.3cm}\\$\mathcal{B}_{V,1}$&$\left(
\begin{array}{cc}
 1.57\times 10^{{-5}} & -1.28\times 10^{{-4}} \\
 -1.28\times 10^{{-4}} & 1.92\times 10^{{-3}}
\end{array}
\right)$&$\left(
\begin{array}{cc}
 3.29\times 10^{{-3}} & -2.7\times 10^{{-2}} \\
 -2.7\times 10^{{-2}} & 0.396
\end{array}
\right)$\\ \vspace{-.3cm}\\$\mathcal{B}_{V,2}$&$\left(
\begin{array}{cc}
 6.45\times 10^{{-6}} & -5.23\times 10^{{-5}} \\
 -5.23\times 10^{{-5}} & 7.98\times 10^{{-4}}
\end{array}
\right)$&$\left(
\begin{array}{cc}
 1.83\times 10^{{-3}} & -1.42\times 10^{{-2}} \\
 -1.42\times 10^{{-2}} & 0.274
\end{array}
\right)$\\ \vspace{-.3cm}\\$\mathcal{B}_{V,t}$&$\left(
\begin{array}{cc}
 1.19\times 10^{{-5}} & -9.76\times 10^{{-5}} \\
 -9.76\times 10^{{-5}} & 1.42\times 10^{{-3}}
\end{array}
\right)$ & $\left(
\begin{array}{cc}
 2.19\times 10^{{-3}} & -1.81\times 10^{{-2}} \\
 -1.81\times 10^{{-2}} & 0.258
\end{array}
\right)$\\ \vspace{-.3cm} \\$\mathcal{B}_{T,0}$&$\left(
\begin{array}{cc}
 1.01\times 10^{{-5}} & 1.29\times 10^{{-4}} \\
 1.29\times 10^{{-4}} & 1.72\times 10^{{-2}}
\end{array}
\right)$&$\left(
\begin{array}{cc}
 6.49\times 10^{{-3}} & 9.63\times 10^{{-2}} \\
 9.63\times 10^{{-2}} & 15.3
\end{array}
\right)$\\ \vspace{-.3cm}\\$\mathcal{B}_{T,1}$&$\left(
\begin{array}{cc}
 7.45\times 10^{{-7}} & -4.16\times 10^{{-6}} \\
 -4.16\times 10^{{-6}} & 5.21\times 10^{{-5}}
\end{array}
\right)$&$\left(
\begin{array}{cc}
 1.86\times 10^{{-3}} & -9.82\times 10^{{-3}} \\
 -9.82\times 10^{{-3}} & 0.13
\end{array}
\right)$\\ \vspace{-.3cm}\\$\mathcal{B}_{T,2}$&$\left(
\begin{array}{cc}
 2.7\times 10^{{-7}} & -1.41\times 10^{{-6}} \\
 -1.41\times 10^{{-6}} & 1.89\times 10^{{-5}}
\end{array}
\right)$&$\left(
\begin{array}{cc}
 8.09\times 10^{{-4}} & -2.15\times 10^{{-3}} \\
 -2.15\times 10^{{-3}} & 6.88\times 10^{{-2}}
\end{array}
\right)$\\ \vspace{-.3cm}\\\end{tabular}\end{center}


\item For the fit of vector and axial-vector
  $B\to \rho$ FFs to LCSR and Lattice data, the covariance matrices read:
\begin{center}\begin{tabular}{ccc}\T\B & SE & SSE\\$\mathcal{B}_{V,0}$&$\left(
\begin{array}{cc}
 2.62\times 10^{{-6}} & -1.35\times 10^{{-5}} \\
 -1.35\times 10^{{-5}} & 5.35\times 10^{{-4}}
\end{array}
\right)$&$\left(
\begin{array}{cc}
 2.86\times 10^{{-3}} & -5.12\times 10^{{-3}} \\
 -5.12\times 10^{{-3}} & 0.796
\end{array}
\right)$\\ \vspace{-.3cm}\\$\mathcal{B}_{V,1}$&$\left(
\begin{array}{cc}
 5.72\times 10^{{-6}} & -3.08\times 10^{{-5}} \\
 -3.08\times 10^{{-5}} & 9.15\times 10^{{-4}}
\end{array}
\right)$&$\left(
\begin{array}{cc}
 1.24\times 10^{{-3}} & -7.07\times 10^{{-3}} \\
 -7.07\times 10^{{-3}} & 0.193
\end{array}
\right)$\\ \vspace{-.3cm}\\$\mathcal{B}_{V,2}$&$\left(
\begin{array}{cc}
 1.99\times 10^{{-6}} & -3.02\times 10^{{-6}} \\
 -3.02\times 10^{{-6}} & 2.26\times 10^{{-4}}
\end{array}
\right)$&$\left(
\begin{array}{cc}
 5.21\times 10^{{-4}} & 1.52\times 10^{{-3}} \\
 1.52\times 10^{{-3}} & 6.4\times 10^{{-2}}
\end{array}
\right)$\\ \vspace{-.3cm}\\\end{tabular}\end{center}
\end{itemize}

\paragraph{\boldmath $B\to K^*$ form factor fit: \unboldmath} 

The covariance matrices for the $B\to K^*$ FFs are given by:
\begin{center}\begin{tabular}{ccc}\T\B & SE & SSE\\$\mathcal{B}_{V,0}$&$\left(
\begin{array}{cc}
 4.38\times 10^{{-6}} & -3.12\times 10^{{-5}} \\
 -3.12\times 10^{{-5}} & 1.08\times 10^{{-3}}
\end{array}
\right)$&$\left(
\begin{array}{cc}
 4.85\times 10^{{-3}} & -3.26\times 10^{{-2}} \\
 -3.26\times 10^{{-2}} & 1.85
\end{array}
\right)$\\ \vspace{-.3cm}\\$\mathcal{B}_{V,1}$&$\left(
\begin{array}{cc}
 1.94\times 10^{{-5}} & -1.63\times 10^{{-4}} \\
 -1.63\times 10^{{-4}} & 3.06\times 10^{{-3}}
\end{array}
\right)$&$\left(
\begin{array}{cc}
 3.\times 10^{{-3}} & -2.54\times 10^{{-2}} \\
 -2.54\times 10^{{-2}} & 0.467
\end{array}
\right)$\\ \vspace{-.3cm}\\$\mathcal{B}_{V,2}$&$\left(
\begin{array}{cc}
 1.01\times 10^{{-5}} & -8.52\times 10^{{-5}} \\
 -8.52\times 10^{{-5}} & 1.59\times 10^{{-3}}
\end{array}
\right)$&$\left(
\begin{array}{cc}
 2.42\times 10^{{-3}} & -1.87\times 10^{{-2}} \\
 -1.87\times 10^{{-2}} & 0.456
\end{array}
\right)$\\ \vspace{-.3cm}\\$\mathcal{B}_{V,t}$&
$\left(
\begin{array}{cc}
 1.79\times 10^{{-5}} & -1.53\times 10^{{-4}} \\
 -1.53\times 10^{{-4}} & 2.75\times 10^{{-3}}
\end{array}
\right)$ & $\left(
\begin{array}{cc}
 2.24\times 10^{{-3}} & -1.93\times 10^{{-2}} \\
 -1.93\times 10^{{-2}} & 0.34
\end{array}
\right)$\\ \vspace{-.3cm}\\$\mathcal{B}_{T,0}$&$\left(
\begin{array}{cc}
 1.63\times 10^{{-5}} & 3.6\times 10^{{-4}} \\
 3.6\times 10^{{-4}} & 2.41\times 10^{{-2}}
\end{array}
\right)$&$\left(
\begin{array}{cc}
 9.38\times 10^{{-3}} & 0.246 \\
 0.246 & 17.7
\end{array}
\right)$\\ \vspace{-.3cm}\\$\mathcal{B}_{T,1}$&$\left(
\begin{array}{cc}
 1.17\times 10^{{-6}} & -6.23\times 10^{{-6}} \\
 -6.23\times 10^{{-6}} & 9.82\times 10^{{-5}}
\end{array}
\right)$&$\left(
\begin{array}{cc}
 2.26\times 10^{{-3}} & -1.12\times 10^{{-2}} \\
 -1.12\times 10^{{-2}} & 0.191
\end{array}
\right)$\\ \vspace{-.3cm}\\$\mathcal{B}_{T,2}$&$\left(
\begin{array}{cc}
 4.04\times 10^{{-7}} & -2.07\times 10^{{-6}} \\
 -2.07\times 10^{{-6}} & 3.4\times 10^{{-5}}
\end{array}
\right)$&$\left(
\begin{array}{cc}
 1.12\times 10^{{-3}} & -2.27\times 10^{{-3}} \\
 -2.27\times 10^{{-3}} & 0.11
\end{array}
\right)$\\ \vspace{-.3cm}\\\end{tabular}\end{center}

\paragraph{\boldmath $B_s\to \phi$ form factor fit: \unboldmath}  

The covariance matrices for the $B_s \to \phi$ FFs are given by:
\begin{center}\begin{tabular}{ccc}\T\B & SE & SSE\\$\mathcal{B}_{V,0}$&$\left(
\begin{array}{cc}
 1.16\times 10^{{-6}} & -8.32\times 10^{{-6}} \\
 -8.32\times 10^{{-6}} & 3.47\times 10^{{-4}}
\end{array}
\right)$&$\left(
\begin{array}{cc}
 4.56\times 10^{{-3}} & -2.81\times 10^{{-2}} \\
 -2.81\times 10^{{-2}} & 1.98
\end{array}
\right)$\\ \vspace{-.3cm}\\$\mathcal{B}_{V,1}$&$\left(
\begin{array}{cc}
 8.44\times 10^{{-6}} & -7.79\times 10^{{-5}} \\
 -7.79\times 10^{{-5}} & 1.63\times 10^{{-3}}
\end{array}
\right)$&$\left(
\begin{array}{cc}
 3.37\times 10^{{-3}} & -3.15\times 10^{{-2}} \\
 -3.15\times 10^{{-2}} & 0.643
\end{array}
\right)$\\ \vspace{-.3cm}\\$\mathcal{B}_{V,2}$&$\left(
\begin{array}{cc}
 3.6\times 10^{{-6}} & -3.26\times 10^{{-5}} \\
 -3.26\times 10^{{-5}} & 7.08\times 10^{{-4}}
\end{array}
\right)$&$\left(
\begin{array}{cc}
 2.91\times 10^{{-3}} & -2.38\times 10^{{-2}} \\
 -2.38\times 10^{{-2}} & 0.662
\end{array}
\right)$\\  \vspace{-.3cm}\\$\mathcal{B}_{V,t}$& $\left(
\begin{array}{cc}
 6.61\times 10^{{-6}} & -6.07\times 10^{{-5}} \\
 -6.07\times 10^{{-5}} & 1.28\times 10^{{-3}}
\end{array}
\right)$ & $\left(
\begin{array}{cc}
 2.05\times 10^{{-3}} & -1.9\times 10^{{-2}} \\
 -1.9\times 10^{{-2}} & 0.394
\end{array}
\right)$ \\  \vspace{-.3cm}\\$\mathcal{B}_{T,0}$&$\left(
\begin{array}{cc}
 7.03\times 10^{{-6}} & 1.75\times 10^{{-4}} \\
 1.75\times 10^{{-4}} & 1.04\times 10^{{-2}}
\end{array}
\right)$&$\left(
\begin{array}{cc}
 1.41\times 10^{{-2}} & 0.406 \\
 0.406 & 25.2
\end{array}
\right)$\\ \vspace{-.3cm}\\$\mathcal{B}_{T,1}$&$\left(
\begin{array}{cc}
 6.39\times 10^{{-7}} & -3.91\times 10^{{-6}} \\
 -3.91\times 10^{{-6}} & 6.67\times 10^{{-5}}
\end{array}
\right)$&$\left(
\begin{array}{cc}
 3.37\times 10^{{-3}} & -1.93\times 10^{{-2}} \\
 -1.93\times 10^{{-2}} & 0.35
\end{array}
\right)$\\ \vspace{-.3cm}\\$\mathcal{B}_{T,2}$&$\left(
\begin{array}{cc}
 1.63\times 10^{{-7}} & -8.97\times 10^{{-7}} \\
 -8.97\times 10^{{-7}} & 1.69\times 10^{{-5}}
\end{array}
\right)$&$\left(
\begin{array}{cc}
 1.64\times 10^{{-3}} & -3.86\times 10^{{-3}} \\
 -3.86\times 10^{{-3}} & 0.187
\end{array}
\right)$\\ \vspace{-.3cm}\\\end{tabular}\end{center}




\providecommand{\href}[2]{#2}\begingroup\raggedright\endgroup

\end{document}